%% file: lsa.tex
\documentclass[preprint2]{aastex61}    

\usepackage{graphicx}                 
\usepackage{lineno}                   
\usepackage{amsmath}                  
\usepackage{dcolumn}                  
\usepackage{bm}                       
\usepackage{xspace}                   
\usepackage{hyperref}                 
\usepackage{yfonts} 

\usepackage{tabularx}

\begin{document}

\title{Observation of Anisotropy\\of TeV Cosmic Rays with Two Years of HAWC}

\input authors

\begin{abstract}
After two years of operation, the High-Altitude Water Cherenkov (HAWC) Observatory has analyzed the TeV 
cosmic-ray sky over an energy range between $2.0$ and $72.8$ TeV.
\deleted{The HAWC detector is a ground-based air-shower array located at high altitude in the state of Puebla, Mexico.
Using 300 light-tight water tanks, it collects the Cherenkov light from the particles of extensive air showers 
from primary gamma rays and cosmic rays.
This detection method allows for uninterrupted observation of the entire overhead sky 
(2~sr instantaneous, 8.5~sr integrated) in the energy range from a few TeV to hundreds of TeV.}
\explain{Request to shorten abstract text.}
Like other detectors in the northern and southern hemispheres, HAWC observes an energy-dependent anisotropy 
in the arrival direction distribution of cosmic rays.
\replaced{The observed cosmic-ray}{This} anisotropy is dominated by a dipole moment with phase \added{in right ascension}\explain{Defining alpha.} 
$\alpha\approx40^{\circ}$ and amplitude that slowly rises 
in relative intensity from 
$8\times10^{-4}$ at 2 TeV to $14\times10^{-4}$ around 30\deleted{.3} TeV \explain{Using "around" to qualify curtails the precise energy estimate energy of that bin.}, above which the dipole decreases in strength.
A significant large-scale ($>60^{\circ}$ in angular extent) signal is also observed in the quadrupole and octupole moments, 
and significant small-scale features are also present, with locations and shapes consistent with previous observations.
Compared to previous measurements in this energy range, the HAWC cosmic-ray sky maps improve on the energy resolution 
and fit precision of the anisotropy.
These data can be used in an effort to better constrain local cosmic-ray accelerators and the intervening magnetic fields. 
\end{abstract}


\section{Introduction}
\label{sec:Introduction}
The study of the anisotropy in the arrival direction of cosmic rays
has entered an era of precision measurement.
Combined with improved modeling of the local interstellar medium, these measurements are maturing into an
important way to understand simultaneously cosmic-ray acceleration and propagation. 
For a recent review see \citep{Ahlers:2016rox}.

Anisotropy is a well-studied consequence of standard propagation models where cosmic rays diffuse \replaced{in}{due to} random magnetic 
fields and their sources are distributed inhomogeneously \citep{Erlykin:2006ri, Blasi:2011fm, Pohl:2012xs, Sveshnikova:2013ui}.
Anisotropy can also arise from motion relative to the rest frame of the cosmic rays \citep{Compton}.
Both scenarios result in a dominantly dipolar anisotropy, yet \added{the predicted dipole amplitude is at least
an order of magnitude larger than the observed value \citep{kumar:2014apj, mertsch:2015prl}, and} 
the measured dipole orientation can not be explained by these simple models.
Recent studies have included the effects of regular magnetic fields to probe the origins of the dipole 
direction \citep{Ahlers:2016njl} in hopes of identifying the locations of\deleted{dominant cosmic-ray} accelerators
\added{contributing most to the locally observed cosmic-ray flux}.

While the observed TeV cosmic-ray anisotropy is primarily dipolar with amplitude $\sim10^{-3}$, 
it also contains smaller scale structure $O(\lesssim45^{\circ})$ with strength $\sim10^{-4}$.
It is likely that \replaced{the}{an} initial dipolar \replaced{signal}{distribution} is distorted 
as it passes through the interstellar medium.
For example, isotropic magnetic turbulence can create a feed down of angular power to higher multipole moments
\citep{Giacinti:2011mz, Ahlers:2013ima,Ahlers:2015dwa,Giacinti:2016tld}, 
and pitch-angle scattering \citep{Giacinti:2016tld} alters the shape of the large-scale multipoles.
\added{It is possible that heliospheric effects perturb the anisotropy,
manifesting in the observed small-scale structure \citep{Desiati:2011xg, Schwadron:2014}.}
Anisotropy can also result from non-standard diffusion such as strong regularities in 
local magnetic field lines \citep{Drury:2013uka,Harding:2015pna}.  
Thus, divergence from a pure dipole anisotropy provides a probe into the bulk properties of the interstellar medium.

Measuring anisotropy signals of $O(10^{-3} -10^{-4})$ at significant levels requires several key detector attributes.
A large instantaneous sky-coverage and long, uninterrupted observation periods are 
needed to achieve statistical uncertainties below the signal strength and to resolve features with large angular extent ($180^{\circ}$). 
Only earthbound air-shower detectors fit these requirements, combining large fields-of-view, 
effective areas of $\sim10^4$~m$^2$, and high duty cycles with long-term stability. 

Cosmic-ray anisotropy has been observed in the energy
range $\sim500$ GeV - $100$ PeV by air shower arrays such as
Tibet-AS$\gamma$ \citep{Amenomori:2005dy,Amenomori:2006bx,Amenomori:2017jbv}, 
Milagro \citep{Abdo:2008kr,Abdo:2008aw},
EAS-TOP \citep{Aglietta:2009mu}, 
IceCube/IceTop \citep{Abbasi:2010mf,Abbasi:2011ai,Abbasi:2011zka,Aartsen:2012ma,Aartsen:2016ivj}, 
ARGO-YBJ \citep{DiSciascio:2013cia,ARGO-YBJ:2013gya,Bartoli:2015ysa}, 
and HAWC \citep{Abeysekara:2014sna}. 
Below these energies, cosmic rays begin to follow geomagnetic field lines and no longer probe interstellar scales.
Only the Pierre Auger Observatory has a significant measurement \citep{Aab:2016ban,Aab:2017tyv} above EeV energies. 
The most recent experimental overviews are given in \citep{DiSciascio:2014jwa} and in \citep{Ahlers:2016rox}. 

Air shower arrays must use the observations themselves to determine intrinsic detector acceptances,
which limits sensitivity to the anisotropy component along the direction of the Earth's rotation,
i.e. along the right ascension $\alpha$ in equatorial coordinates.
These detectors are thus unable to recover a dipolar signal aligned with the equatorial poles, 
bounding the maximally recoverable dipole strength according to its orientation in declination.
Direct modeling of the detector acceptance, as done in  \citep{Aab:2016ban}
can eliminate this bias but has not been demonstrated for arrays operating at TeV energies,
as it requires an agreement between simulation and data over the full zenith angle range
at the level of $10^{-5}$ or better, which is currently not achieved.

In this paper, we describe the results of an analysis of the cosmic-ray anisotropy on all angular scales and as a function 
of energy using the first two years of data recorded by the HAWC experiment.
In a previous paper \citep{Abeysekara:2014sna}, we used 113 days of data to study the small-scale anisotropy of 
cosmic rays\deleted{at energies} above 1.7 TeV.
Here, we extend on this analysis by also studying the large-scale anisotropy and using an improved energy estimator \citep{crpaper} 
to study the energy-dependence of both the small- and large-scale structures.

In addition, we apply a new iterative method \citep{Ahlers:2016njl} to reconstruct the maximally recoverable strength of the anisotropy.
This method compensates for the reduction in the measured dipole strength caused by the fact that mid-latitude detectors only see a fraction 
of the cosmic-ray dipole at any given time.

With the first two years of data taking, the HAWC array can currently study the cosmic-ray anisotropy up to energies of about 70 TeV.
In future studies, we will \replaced{extent}{extend} the energy range to higher energies, using the same methods described in this paper.

This paper is organized as follows: we first describe the HAWC detector in Section~\ref{sec:detector}, then
the event selection and data set used for the measurement in Section~\ref{sec:dataset}.
An explanation of the analysis methods is provided in Section~\ref{sec:analysis}.
The results of the observed anisotropy are presented and discussed in section \ref{sec:results}.
Section \ref{sec:conclusion} summarizes the main conclusions of this work.

\section{The HAWC Detector}
\label{sec:detector}

The High-Altitude Water Cherenkov (HAWC) Gamma-Ray Observatory is an
extensive air-shower array located at 4100\,m a.s.l. on the slopes
of Volcan Sierra Negra at $19^{\circ}$N in the state of Puebla, Mexico.
While HAWC is designed to study the sky in gamma rays between 500\,GeV and 100\,TeV,  
it is also sensitive to showers from primary cosmic rays up to multi-PeV energies.

The detector consists of a 22,000\,m$^2$ array of 300 close-packed water Cherenkov detectors (WCDs), 
each containing 200 kiloliters of purified water and four upward-facing photomultiplier tubes (PMTs).
As secondary air shower particles pass through the WCDs, the Cherenkov light produced is collected
by the PMTs, permitting the reconstruction of primary particle properties including the local arrival direction,
core location, and the energy.
Further details on the HAWC detector can be found in \citep{Abeysekara:2017mjj}.

The light-tight nature of the WCDs allows the detector to operate at nearly 100\% up-time efficiency,
with the data acquisition system recording air showers at a rate of $\sim$ 25 kHz.
With a resulting daily sky coverage of $8.4$\,sr, \replaced{HAWC is an ideal 
instrument for measuring the cosmic-ray arrival direction distribution with unprecedented precision}
{HAWC is an instrument well-suited for measuring the cosmic-ray arrival direction distribution}.

\section{The Data Set}
\label{sec:dataset}

\begin{figure*}[t]
  \centering
    \includegraphics[width=\textwidth]{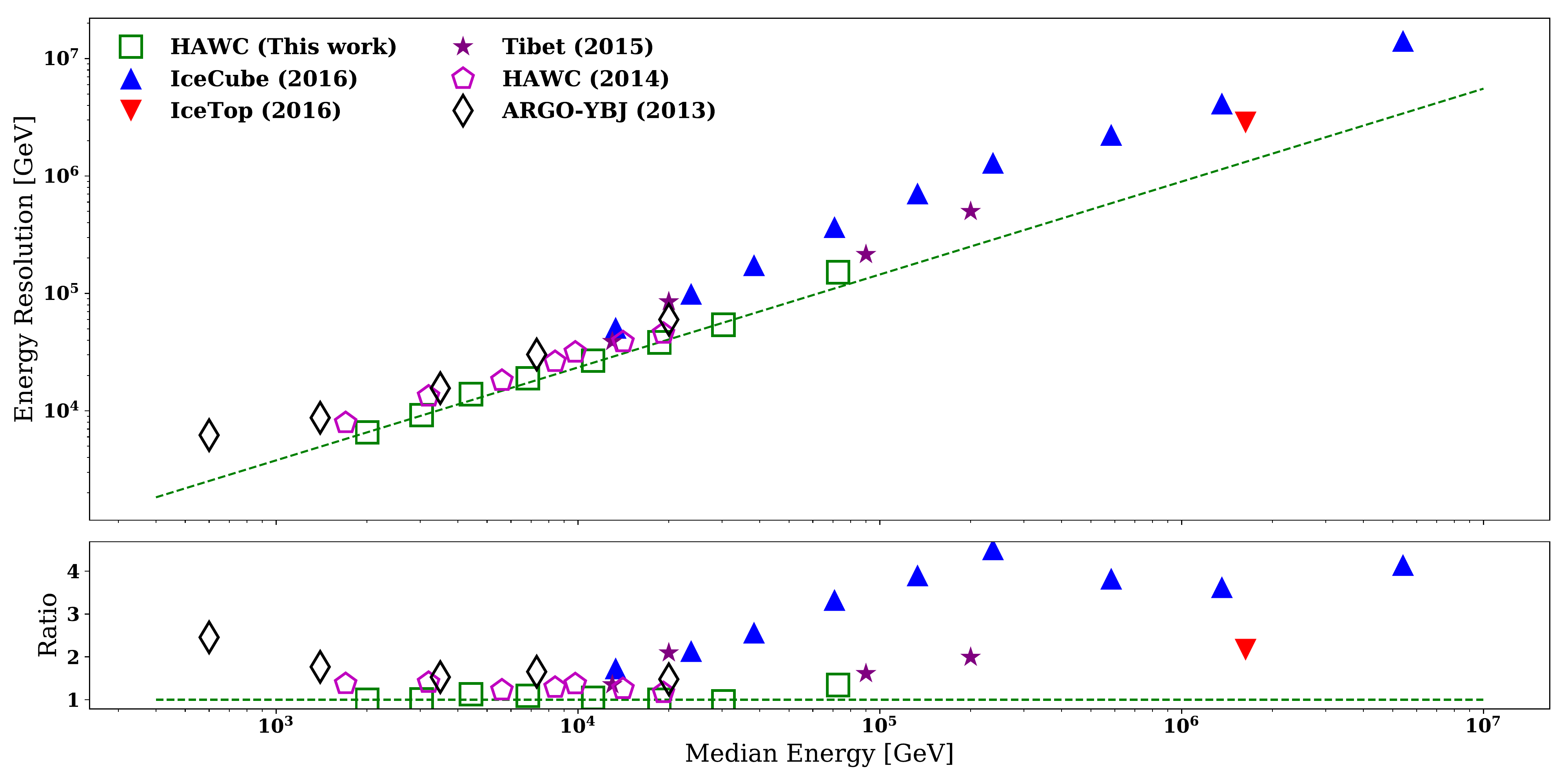}
     \caption{
               The energy resolution for previous cosmic-ray anisotropy measurements (using the multiplicity method)
               compared to the maximum-likelihood energy estimator used in this analysis.
               The improvement is about 30\% below 10 TeV from previous HAWC results.
               A dashed green line connecting the first and sixth HAWC energy bins is shown to guide the eye. 
               Ratios relative to that green line are shown in the lower panel. 
               The energy resolution values for HAWC \citep{Abeysekara:2014sna} as well as IceCube and IceTop  \citep{Aartsen:2016ivj}
               are given in their publications. 
               For ARGO-YBJ \citep{DiSciascio:2013cia} and Tibet \citep{Amenomori:2017jbv}, the values were estimated as the 
               full-width at half-maximum of the provided energy distributions.
              }
    \label{fig:eres}
\end{figure*}

The data set for this study consists of 508 uninterrupted sidereal days 
between 1 May, 2015 and 1 May, 2017.
The detector operated with 294 WCDs, recording about $1.2\times{10}^{12}$ air shower triggers.
To determine the energy of the primary air shower particle, we apply a maximum likelihood-based estimator  
that uses the lateral distribution of measured PMT signals as a function of simulated  
primary proton energy \citep{crpaper}.
To improve the estimated energy resolution, poorly reconstructed showers are removed from the data set
by application of moderate event selection.
The selection criteria are:
\begin{enumerate}
   \item Air shower events must pass a minimum multiplicity threshold of $\ge75$ PMTs. 
         This improves angle and energy reconstruction accuracy.
 
  \item At least 1 PMT within 40 meters of the core position ($N_{\text{r40}} \ge 1$) must record a signal.
        This criterion selects air showers landing on or near the array,
        resulting in a core resolution of better than 15 meters above 10 TeV.
        
   \item The zenith angle acceptance range is $0^\circ-60^\circ$.
   
   \item  The data set is composed of periods covering complete sidereal days,
          hence events from incomplete days are not included.
          This removes non-uniformities in sky exposure along right ascension, 
          reducing systematics in the estimation of the reference map (see Section~\ref{subsec:referencemap}).

\end{enumerate}
Requiring a multiplicity of $\ge 75$ PMTs reduces the trigger rate to $23\%$.
The remaining selection criteria further reduce the number of events by $52\%$,
leaving a total of 123 billion air shower events.

Using the selection criteria and the likelihood energy estimator, 
we achieve $\sim30\%$ improvement in energy resolution compared to the 
multiplicity energy-proxy method from previous HAWC results \citep{Abeysekara:2014sna},
and a $\sim50\%$ improvement over ARGO-YBJ \citep{DiSciascio:2013cia} below 10 TeV.
This also permits more energy bins, as well as an increase in the median energy of the highest-energy bin.
The median energy and $68\%$ containment for the eight analysis bins are listed in Table~\ref{tab:fits}
and shown in Figure~\ref{fig:eres} along with the energy bins from comparable experimental results.

\begin{figure}[tb]
   \centering
     \includegraphics[width=0.48\textwidth]{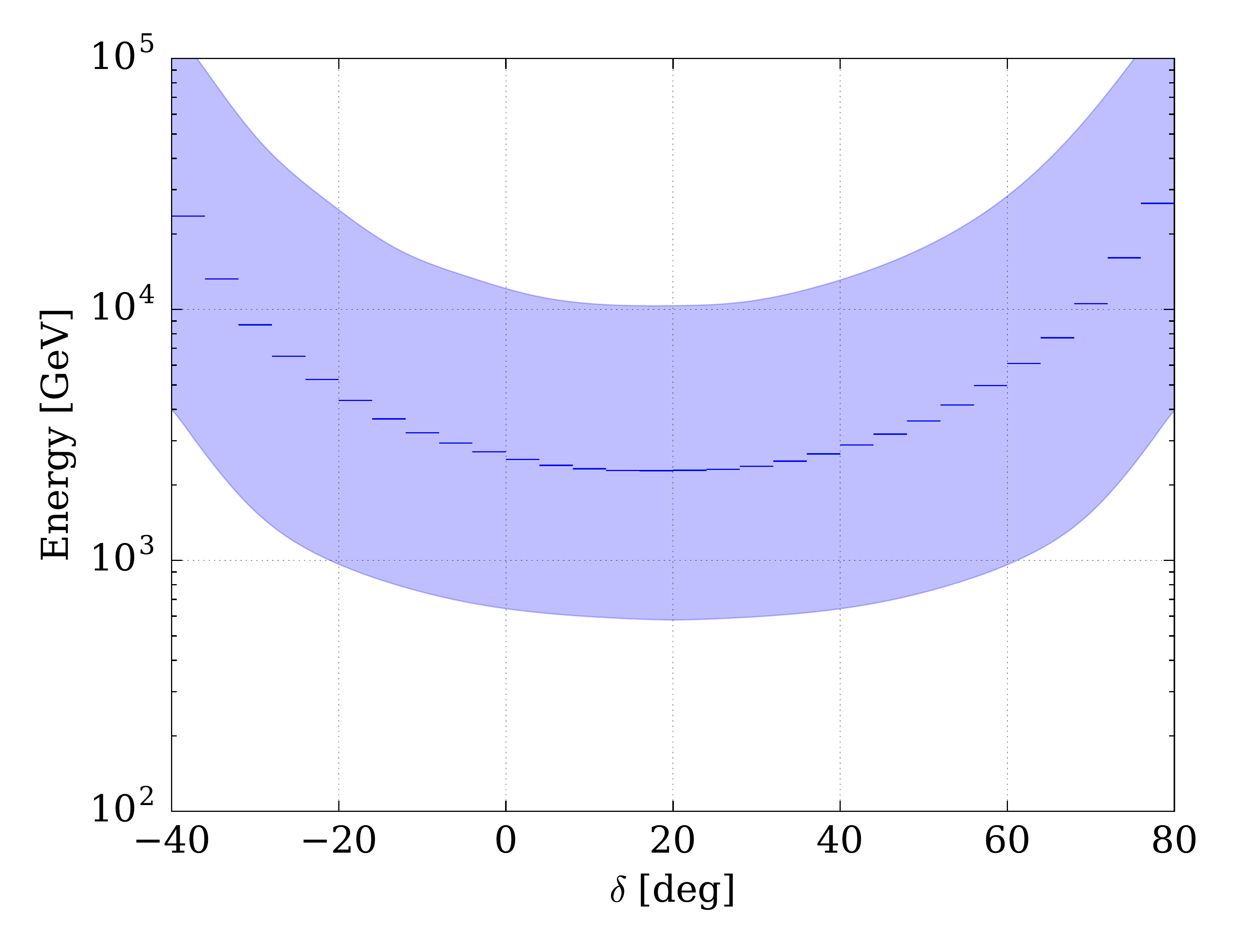}
     \caption{
              Median energy as a function of declination for the combination of all energy bins.
              The blue band indicates the 68\% central containment region.
             }
     \label{fig:energy_vs_declination}
 \end{figure}

The estimated energy exhibits a slight dependence on declination as determined by simulation, 
shown in Figure \ref{fig:energy_vs_declination}.
This is attributed to an increasing energy threshold with increasing zenith angle, 
as air showers must traverse more atmospheric overburden.
The values presented in Figure~\ref{fig:eres} and Table~\ref{tab:fits} were calculated for the 
overhead sky ($\delta=19^{\circ}$), 
thus this declination-dependent energy shift must be considered when viewing the resulting sky maps.

Furthermore, for all maps in the lowest energy bin ($2.0$ TeV), there is a decreased range in 
declination used for analysis, corresponding to the first two zenith bins $(\theta < 35.2^{\circ})$ 
of the energy estimation described in \citep{crpaper}.
This is due to the limited number of selected events having both large zenith angles 
and low reconstructed energies available to measure the anisotropy via the methods described 
in Section \ref{sec:analysis}.

The estimated angular resolution given the selection criteria improves from $0.8^{\circ}$ to 
$0.5^{\circ}$ between 1 and 10 TeV, and plateaus at $0.5^{\circ}$ above 10 TeV. 
A full description of the \textit{in-situ} angular resolution and energy-scale verifications with 
the cosmic-ray Moon shadow is presented in \citep{crpaper}.

\section{Analysis}
\label{sec:analysis}
The measured anisotropy requires comparison of the observed data event distribution $\mathcal{D}$ with 
a background distribution or ``reference map'' $\mathcal{B}$, which represents the detector response to an isotropic flux of cosmic rays.
This reference map is not in itself isotropic because of effects of the detector exposure and geometry.
In principle, the reference map can be obtained from a complete simulation of the detector response to an isotropic flux of cosmic rays, 
but as previously mentioned, measuring anisotropy at the $10^{-4}$ level requires an accuracy of the detector simulation 
that can currently not be achieved for detectors like HAWC.
The reference map is therefore estimated from the data themselves.

We report the anisotropy distribution using two-dimensional sky maps, represented via the 
equal-area pixellation scheme provided by the HEALPix~\citep{Gorski:2004by} package.
Maps are tessellated with 12 base pixels, and each pixel is further partitioned into $N_{\text{side}}$ subdivisions. 
We chose a fine pixellation of $N_{\text{side}} = 256$, corresponding to a pixel width of $0.23^{\circ}$ and \replaced{area}{solid angle} of $1.6 \times 10^{-5}$~sr.

\subsection{Reference Map (Isotropic Expectation)}
\label{subsec:referencemap}

The reference map is determined using the method described in  \citep{Atkins:2003ep},
due \replaced{on}{to} its minimal assumptions on the data compared to similar background estimation methods  \citep{Amenomori:2005a,Aglietta:2009mu}.
This technique uses the number of events in local angular coordinates (the local detector acceptance) 
and in sidereal time (all-sky rate) to construct the expected counts map from an isotropic flux expectation.

\begin{figure*}[t]
  \centering
    \includegraphics[width=\textwidth]{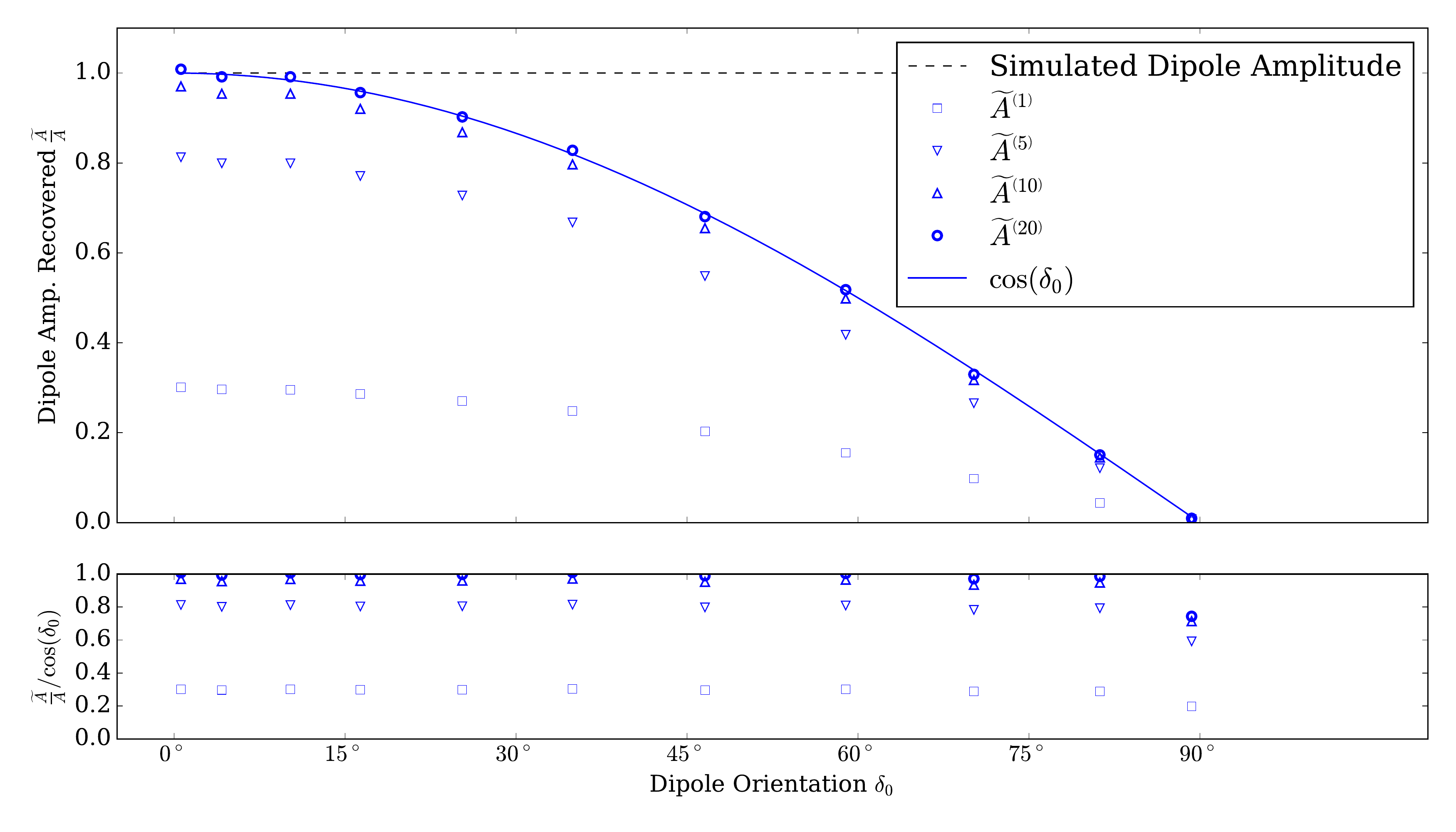}
      \caption{
         Fit to $\delta I$ \added{at 1, 5, 10, and 20 iterations (given by the superscript on $\tilde{A}$)} for 10 simulated HAWC 
         datasets with an injected dipole (oriented with $\delta_0$ between $0^\circ$ and $90^\circ$) plus Poisson noise. 
         This demonstrates the necessity of the convergence of iterative background estimation (light to dark blue points) 
         in order to reach the maximally recoverable signal \replaced{(blue line)}{shown by the blue curve}.
         For our method, the maximally recoverable signal is the projection of the dipole onto the right ascension axis. 
         The black \added{dashed} line at $\frac{\tilde{A}}{A}=1$ shows the relative true strength of the simulated dipole.
         \added{For comparison, the amplitudes given by $\tilde{A}^{(1)}$ (squares) are equivalent to performing direct integration with $\Delta t = 24$hr
         \citep{Ahlers:2016njl}.}
         }
    \label{fig:simdipole}
\end{figure*}

For the calculation of the relative intensity of the cosmic-ray anisotropy, 
this paper \added{uses} for the first time\deleted{uses} a new analysis technique developed by
\citep{Ahlers:2016njl} which mitigates a common artifact of previous methods.
For mid-latitude detectors like HAWC, methods like \citep{Atkins:2003ep} 
severely underestimate the relative intensity of any large-scale structure, 
in particular the strength of the dipole component.
The reason for this underestimation is the fact that these detectors
have an instantaneous field of view that is much smaller
than the size of the large-scale anisotropy structure.
As a consequence, these large-scale structures are attenuated.
\deleted{, so the detector
observes only a small fraction of the anisotropy pattern at any given time}

Over the course of a sidereal day, as the Earth rotates, the detector eventually 
accumulates an event distribution that shows the entire large-scale structure.
However, a consequence of observing different parts of the anisotropy at different 
times is that the observed event distribution is a function of \added{both} the \added{instantaneous}
detector exposure\deleted{to different parts of the sky} as well as the cosmic-ray anisotropy itself.

The new method overcomes the effect of the limited instantaneous exposure
by simultaneously fitting for the\deleted{cosmic-ray} anisotropy and the detector
exposure, using a maximum likelihood technique.
The resulting equations cannot be solved \replaced{in explicit form}{explicitly}, but best-fit solutions 
can be attained by \replaced{using an iterative process}{iteratively} adjusting the local acceptance and all-sky rate.
The convergence criterion depends on the likelihood value of the calculated $\mathcal{B}$ 
provided $\mathcal{D}$ compared to the previous iteration. 
Typically, the calculation requires less than 20 iterations.  

We demonstrate the stable convergence of the process with simulated dipoles of various 
orientations on the sky using a set of $10^{10}$ simulated events drawn from a HAWC-like sky exposure.
The rate as a function of sidereal time was varied by a simple sinusoid of amplitude $5\%$. 
On top of the simulated events, dipoles of strength $10^{-3}$ with ten different orientations
in declination ($\delta_0$, from 0--90 degrees) were added, providing ten fake data sets.
For each simulated data set, the differential relative intensity map was created using the iterative method,
and the maximally recoverable dipole amplitude was then obtained. 
Figure~\ref{fig:simdipole} shows the stable convergence of the fit results after 1, 5, 10, and 20 iterations.     
We also simulated various sky coverages confirming that the fit obtains the maximally 
recoverable dipole amplitude.

The all-sky rate varies by $\sim5\%$ due to diurnal pressure cycles in the upper atmosphere, 
and the local detector acceptance is verified to be stable for each sidereal day by evaluating
the $\chi^2$-difference of local angular distributions in 2 minute intervals compared to 
the mean calculated over the entire sidereal day.
With such minimal variations over the\deleted{entire} data set,
we chose to sum the local detector acceptance and all-sky rate for all 508
sidereal days prior to the background calculation.
Since the all-sky rate is binned in sidereal time bins of $1^{\circ}$, 
much smaller than the 
\replaced{large and small features of the anisotropy}{large- and small-scale features} ($>10^{\circ}$),
variations within a single sidereal time bin have negligible impact on the observed features.

\subsection{Relative Intensity and Significance Map}
\label{subsec:map}

The amplitude of the measured anisotropy and its statistical strength are given by the sky maps in differential relative intensity $\delta I$ 
and significance $S$, respectively. 
For a given pixel, $\delta I$ is the fractional difference between the observed counts in that pixel, $\mathcal{D}$, 
and the expected counts from the reference map, $\mathcal{B}$:

\begin{equation}
\footnotesize{ \delta I= \frac{\mathcal{D}}{\mathcal{B}} - 1~~~.}
\end{equation}
The significance of $\delta I$ is conservatively estimated via
\begin{equation}
\footnotesize{ S \simeq \sqrt{\frac{\mathcal{D}}{1+\alpha_{\text{exp}}}}\delta I~~~,}
\end{equation}
based on  \citep{Li:1983fv}, where $\alpha_{\text{exp}}$ is the relative exposure of the data map compared to the reference map. 
The reference map is overexposed compared to the data because it uses information from all local pixels to calculate its values.
\replaced{For regular direct integration, this value is analytical}
{The value of $\alpha_{\text{exp}}$ is found analytically via the method in \citep{Atkins:2003ep}},
but a direct calculation was not determined for the iterative method.
A conservative value of $\alpha_{\text{exp}}=1$ underestimates the statistical significance by at most 70\%.

\begin{figure*}[t]

  \begin{center}
     \includegraphics[width=0.4\textwidth,trim={0 2.7cm 0 0},clip]{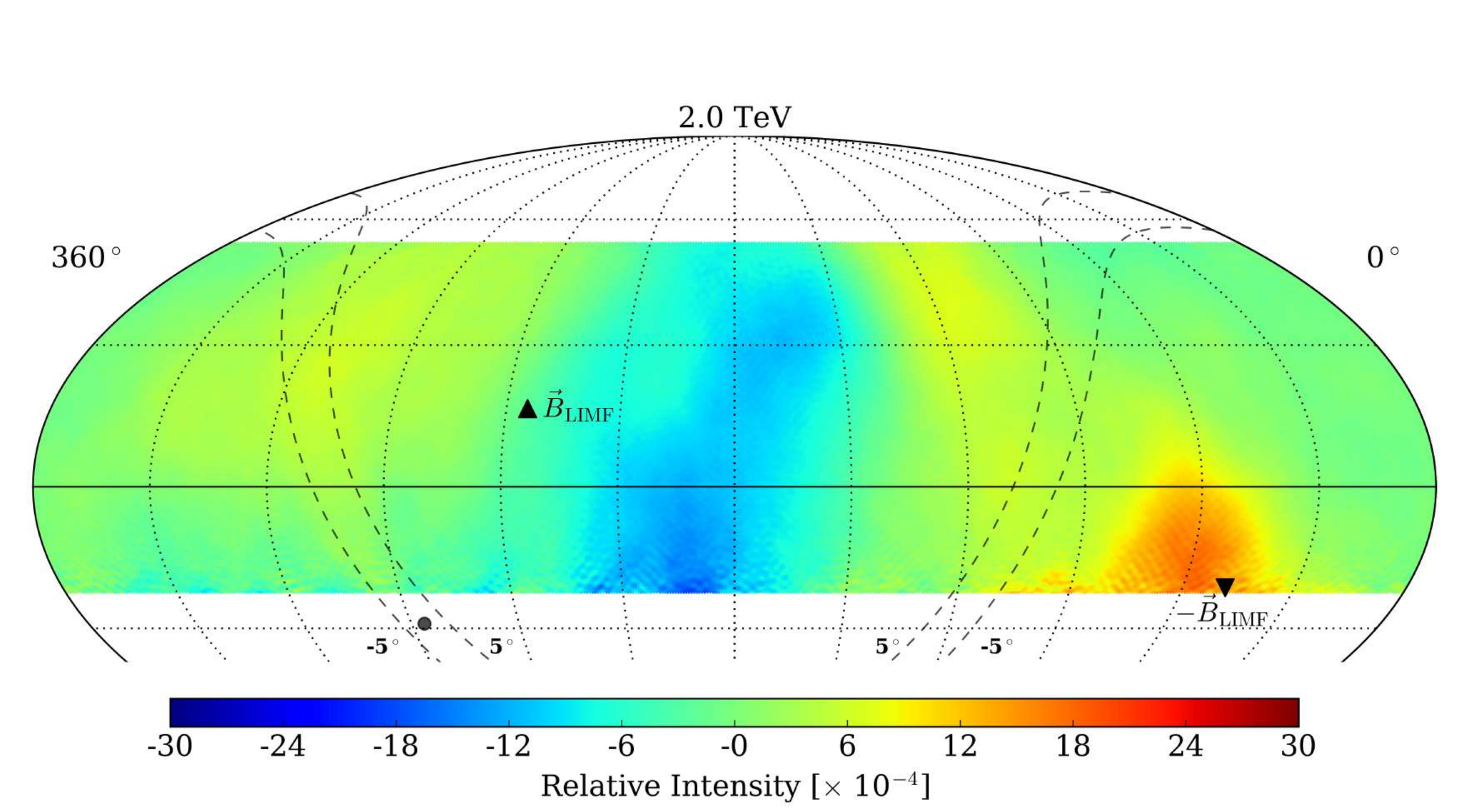}
     \includegraphics[width=0.4\textwidth,trim={0 2.7cm 0 0},clip]{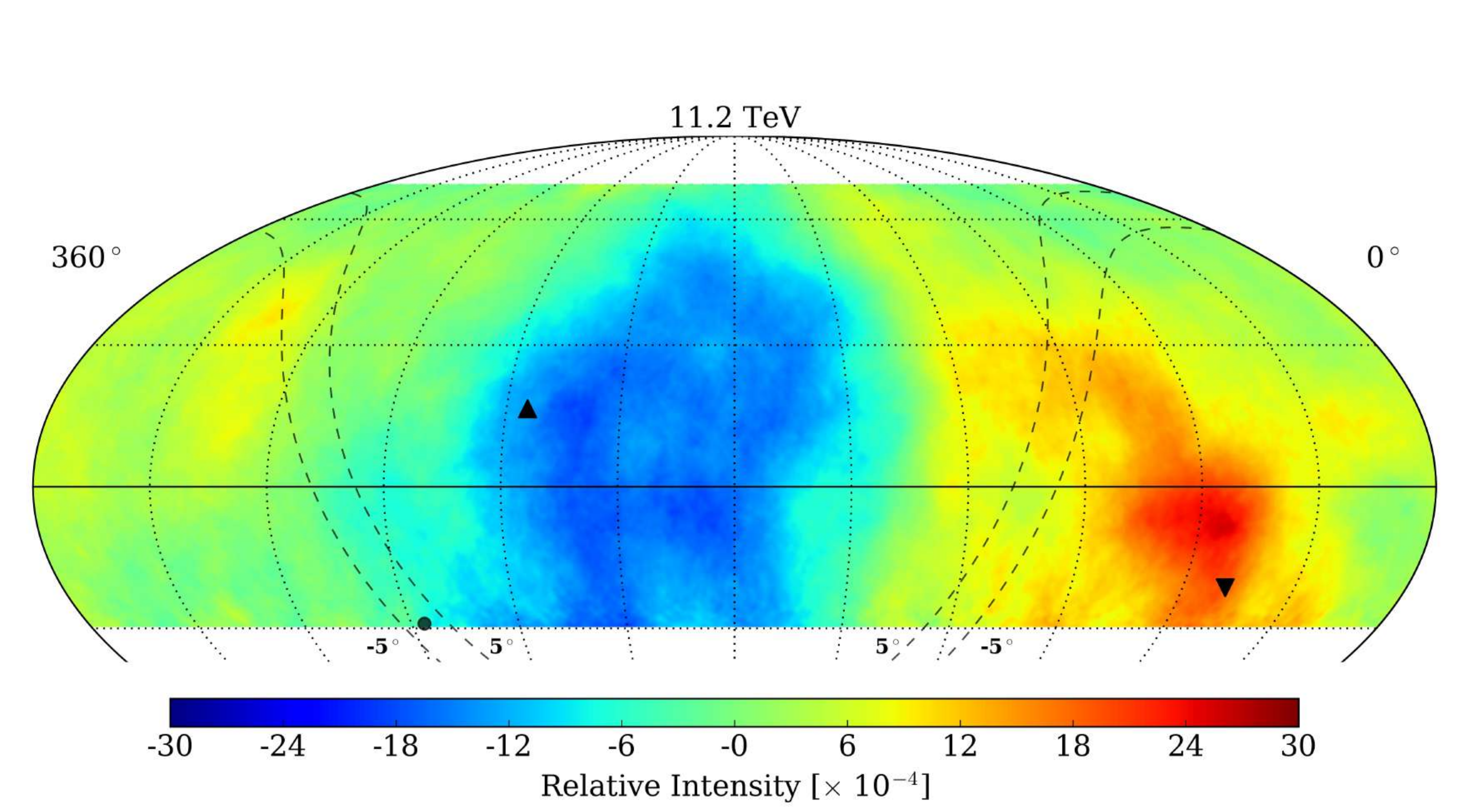}
     \includegraphics[width=0.4\textwidth,trim={0 2.7cm 0 0},clip]{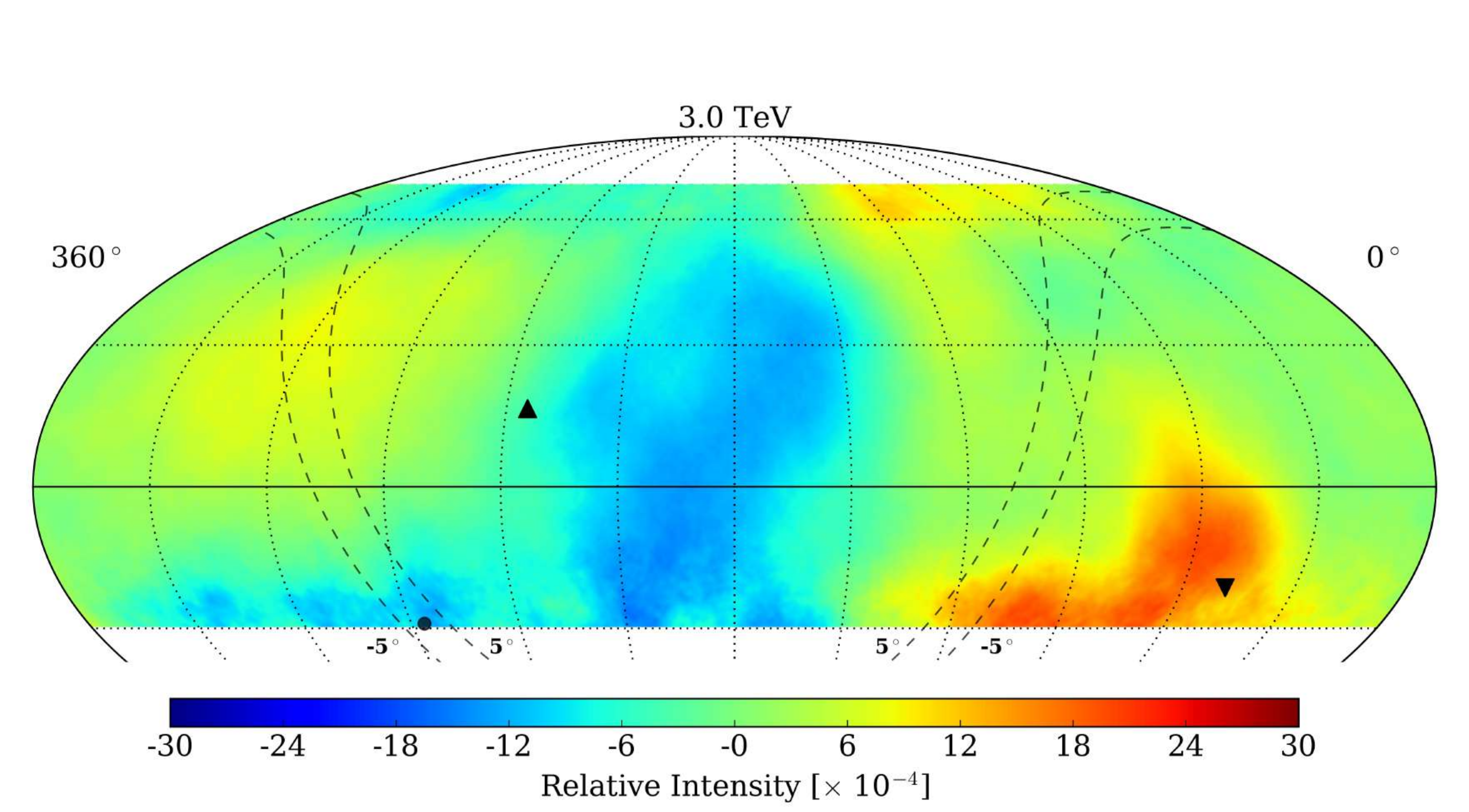}
     \includegraphics[width=0.4\textwidth,trim={0 2.7cm 0 0},clip]{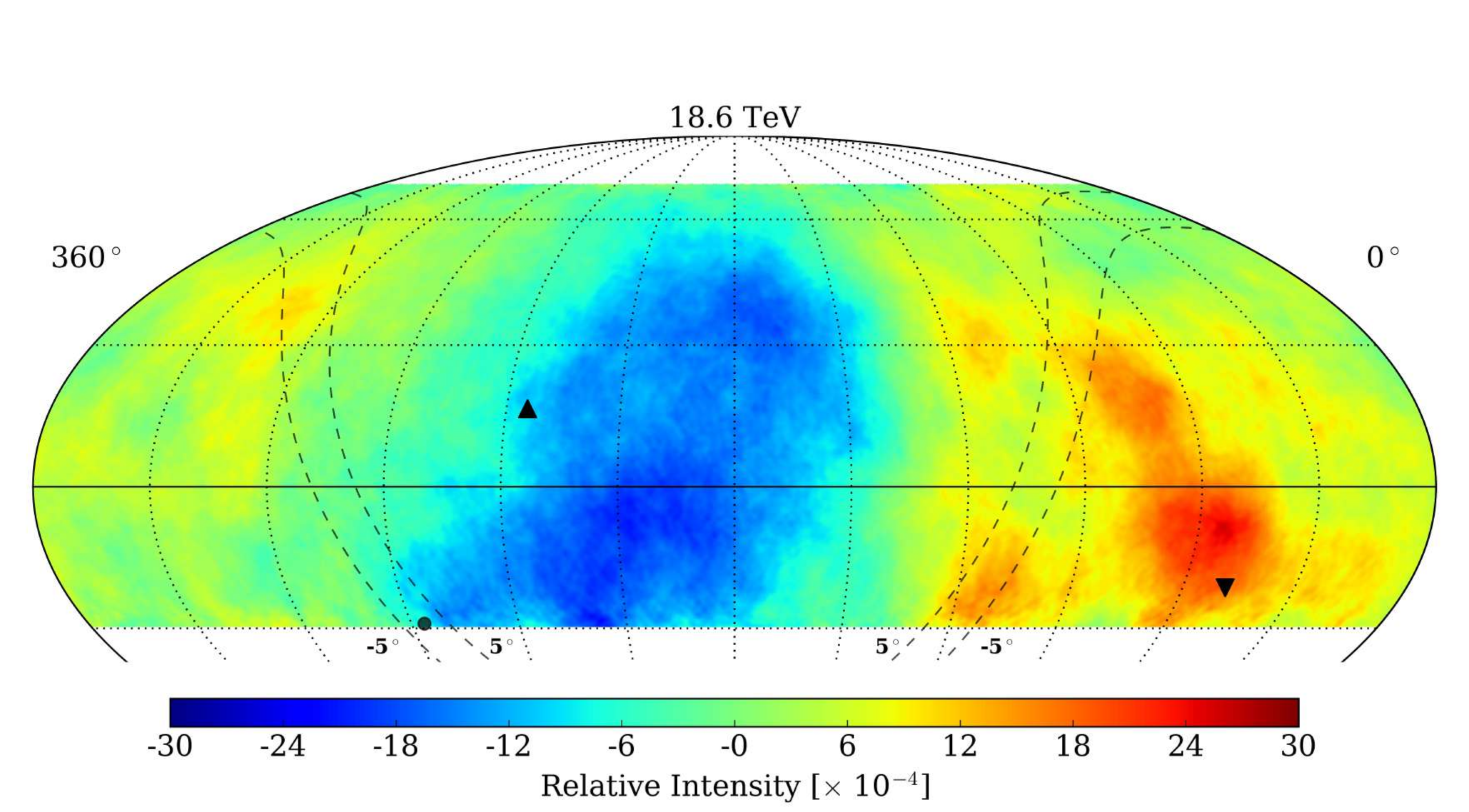}
     \includegraphics[width=0.4\textwidth,trim={0 2.7cm 0 0},clip]{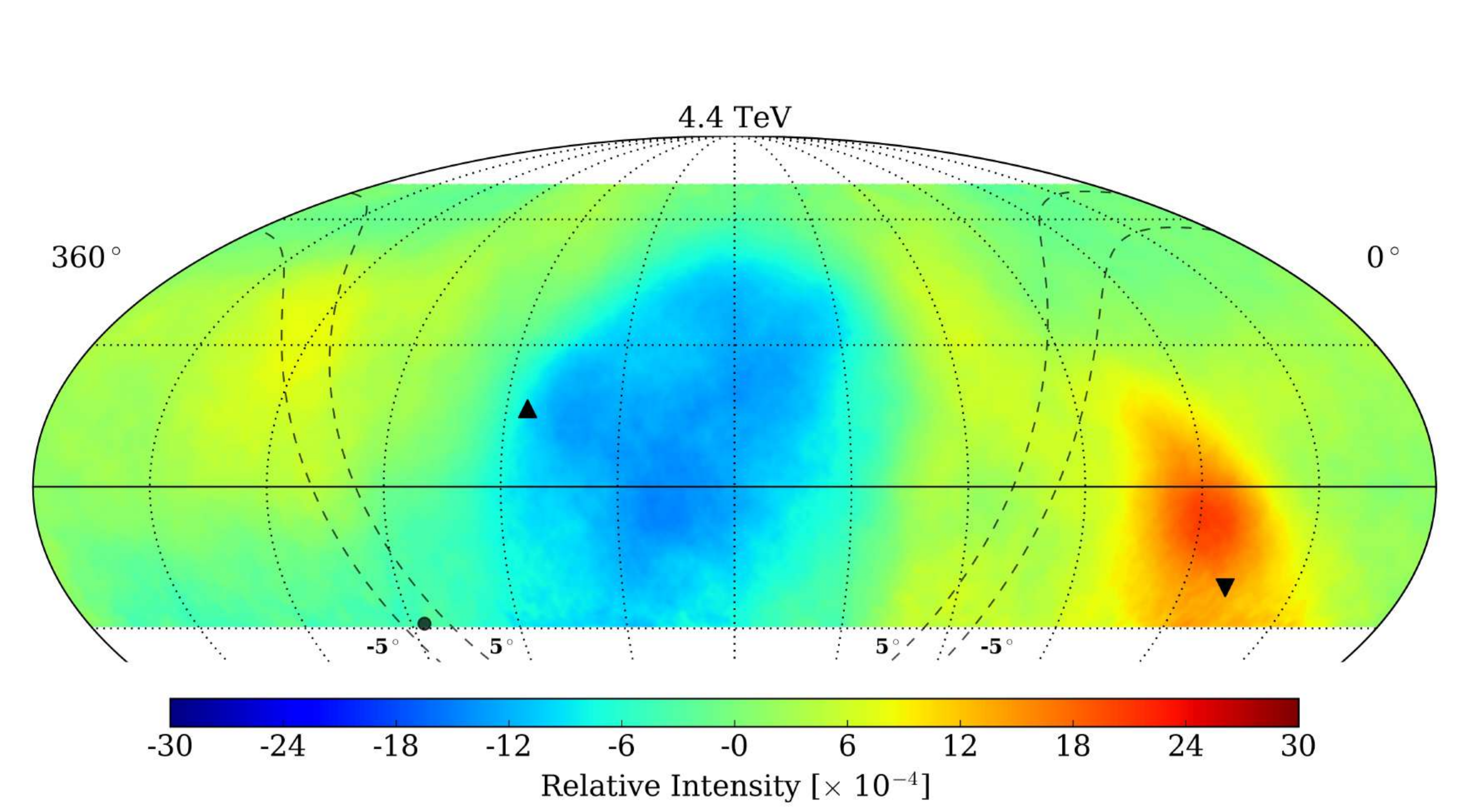}
     \includegraphics[width=0.4\textwidth,trim={0 2.7cm 0 0},clip]{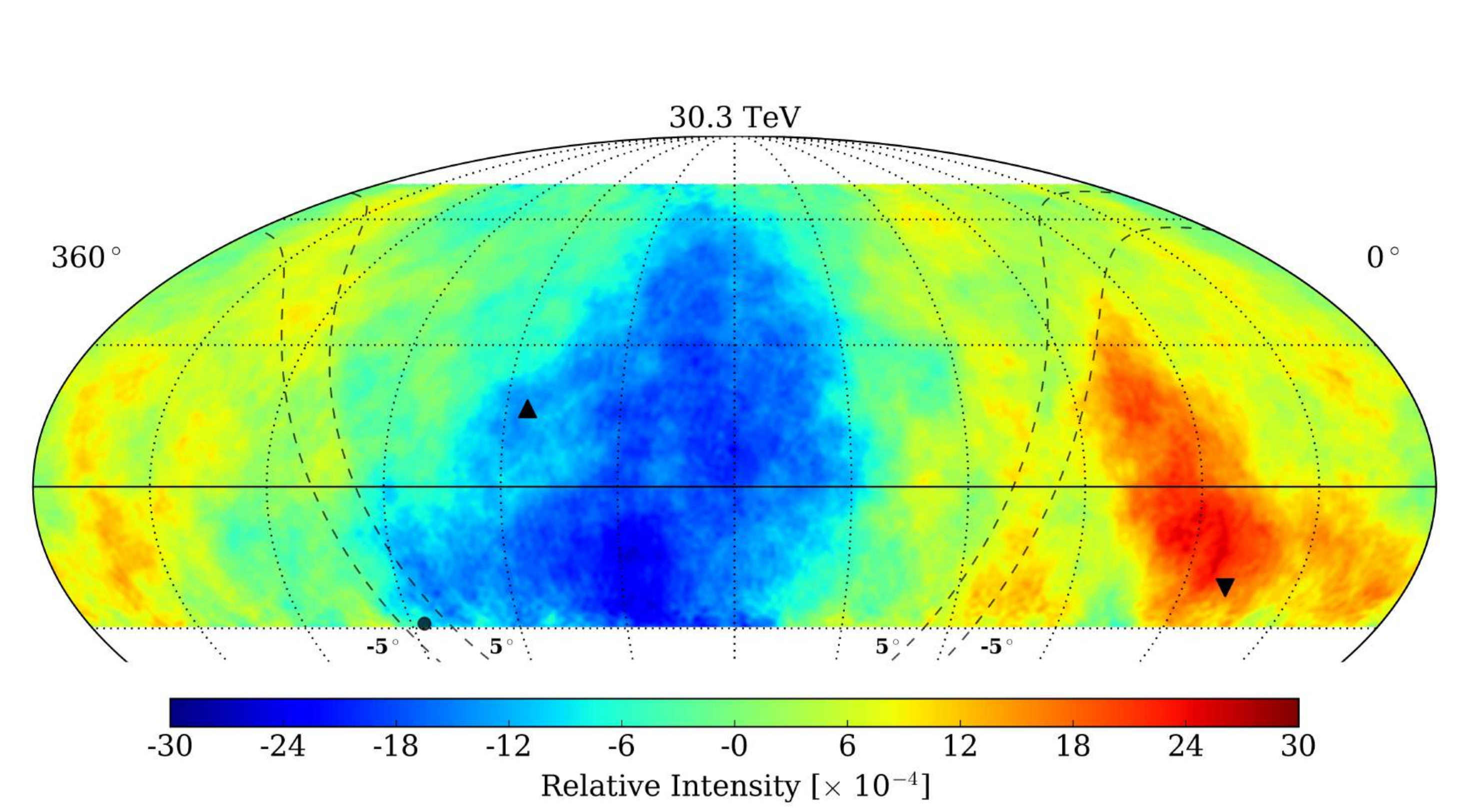}
     \includegraphics[width=0.4\textwidth,trim={0 2.7cm 0 0},clip]{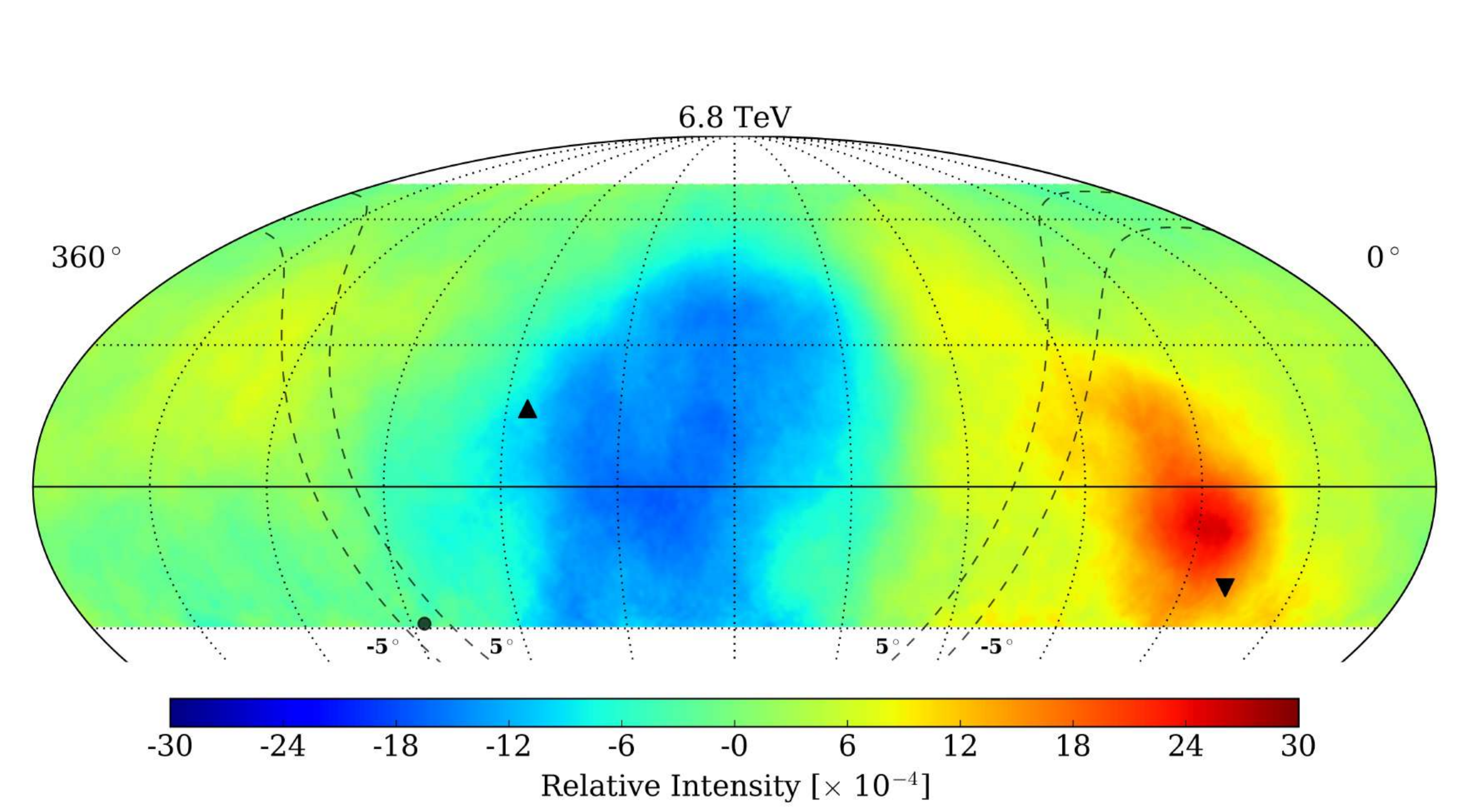}
     \includegraphics[width=0.4\textwidth,trim={0 2.7cm 0 0},clip]{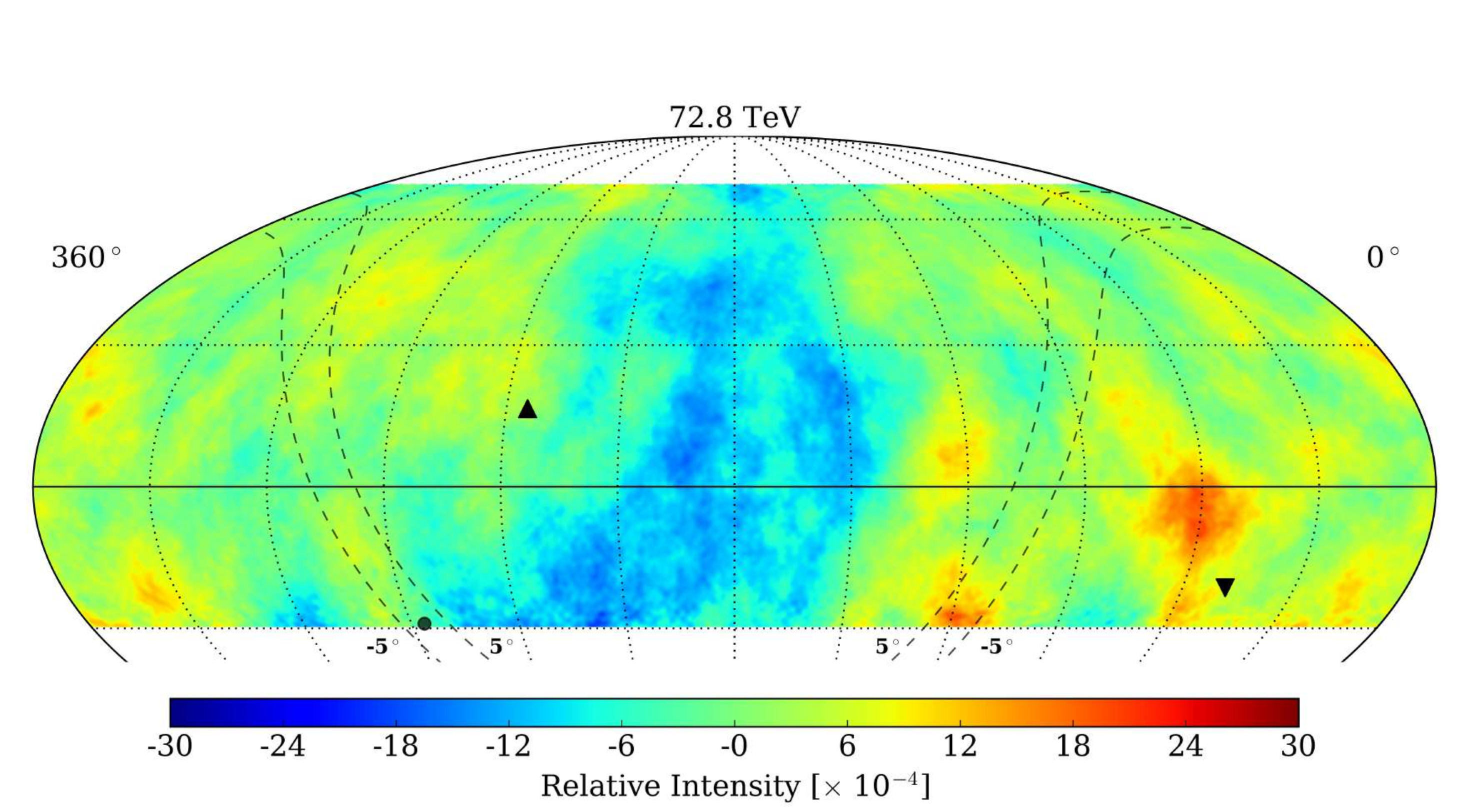}
\includegraphics[width=0.85\textwidth,trim={0 0 0 13.475cm},clip]{bin1_N256_S10p000_relint.pdf}
  \end{center}
    \caption{
             Anisotropy maps in differential relative intensity $\delta I$ (smoothed $10^\circ$) for 
             eight\deleted{independent} energy bins separated by a likelihood-based reconstructed energy variable.
             Energies are reported as the median with the 68\% containment of the bin according to \replaced{our Monte Carlo}{simulation}.
             \added{The two triangle markers indicate the positive (upward-pointing) and negative (downward-pointing)
             directions of the local interstellar magnetic field $\vec{B}_{\mathrm{LIMF}}$ as inferred from
             Interstellar Boundary Explorer (IBEX) observations \citep{Zirnstein_2016}.}
             The Galactic Plane is shown with lines at $+5^\circ$ and $-5^\circ$ in Galactic latitude,
             \added{and the Galactic Center is demarcated by the solid circle}.
             }
  \label{fig:relint}
\end{figure*}

\begin{figure*}[t]

  \begin{center}
     \includegraphics[width=0.4\textwidth,trim={0 2.7cm 0 0},clip]{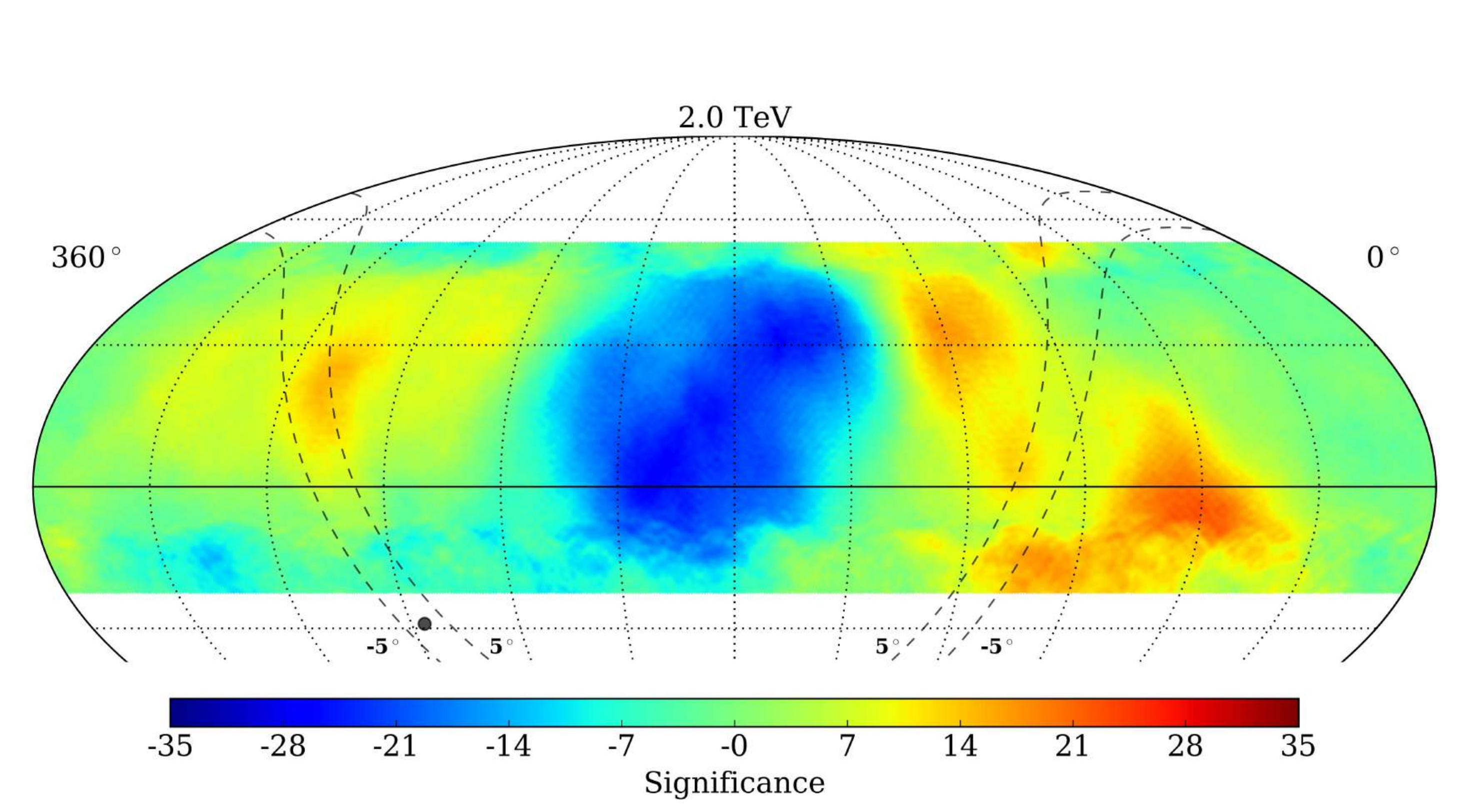}
     \includegraphics[width=0.4\textwidth,trim={0 2.7cm 0 0},clip]{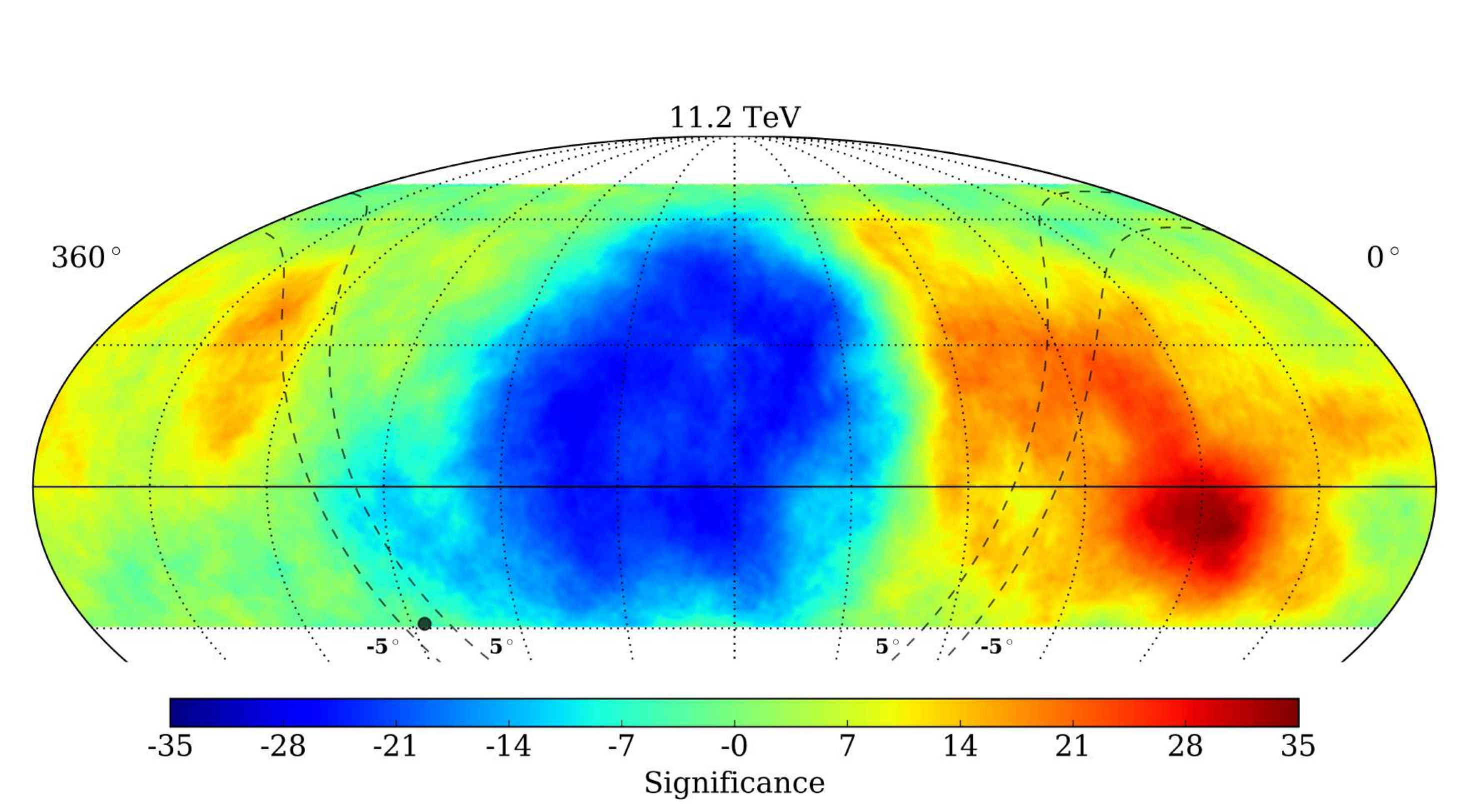}
     \includegraphics[width=0.4\textwidth,trim={0 2.7cm 0 0},clip]{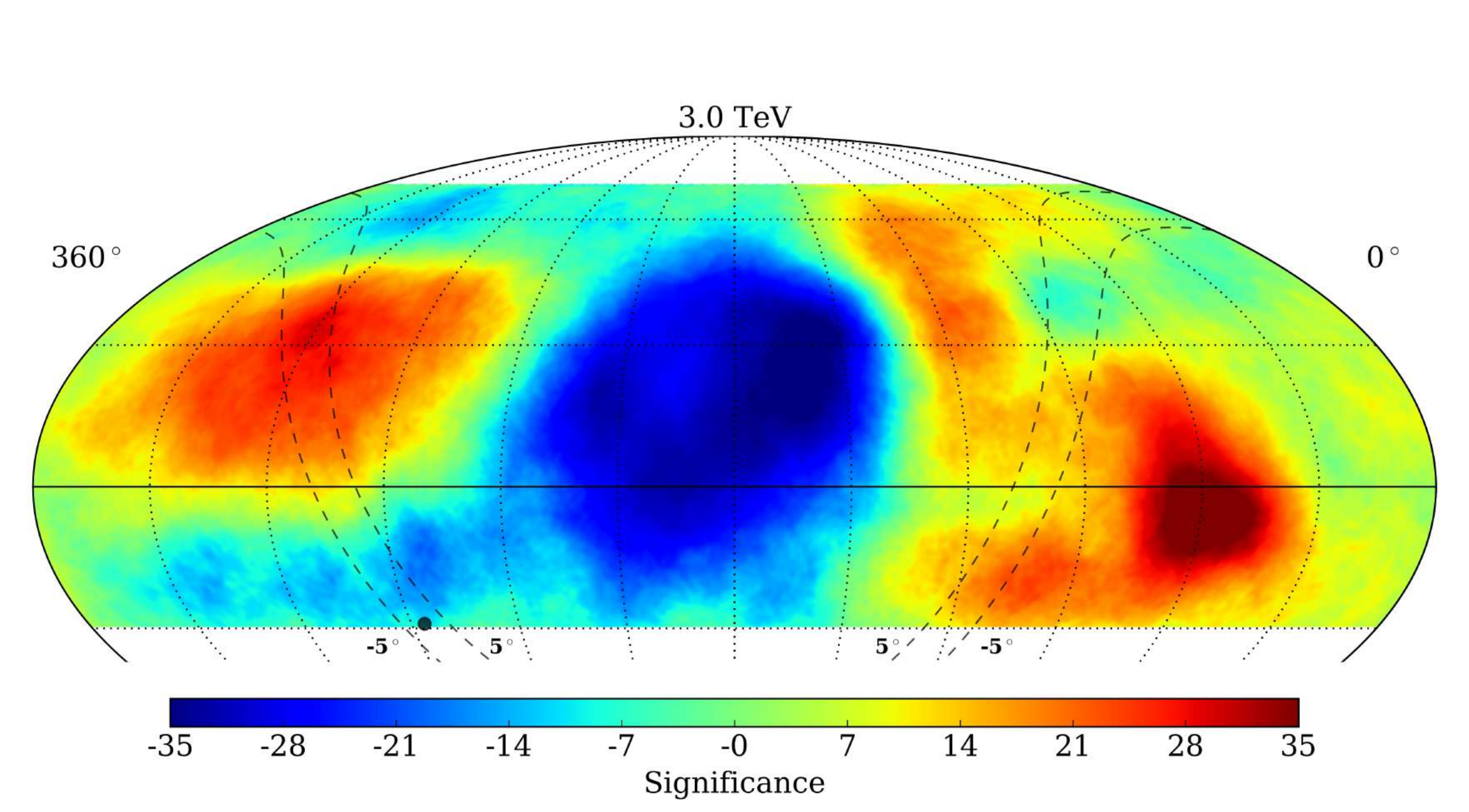}
     \includegraphics[width=0.4\textwidth,trim={0 2.7cm 0 0},clip]{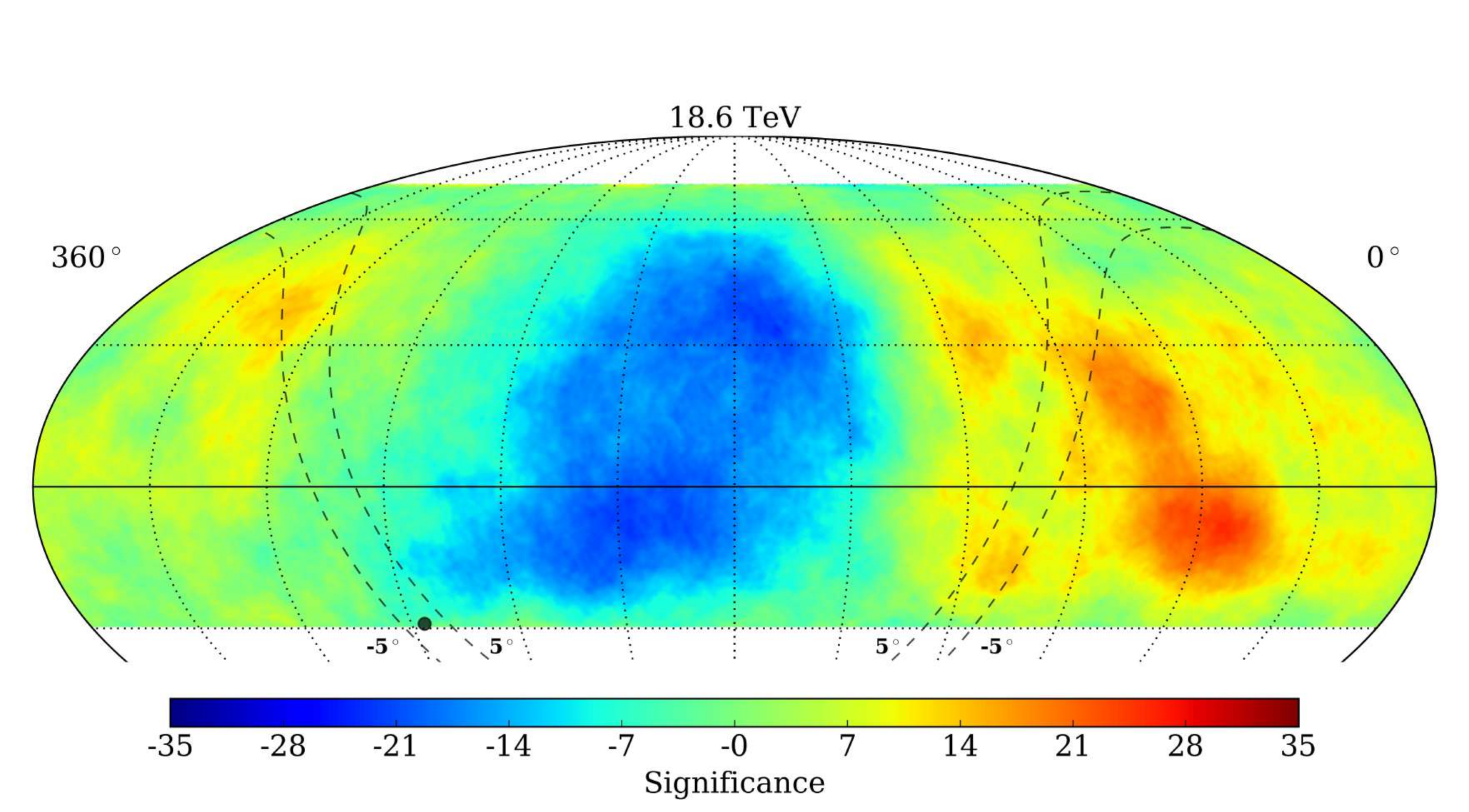}
     \includegraphics[width=0.4\textwidth,trim={0 2.7cm 0 0},clip]{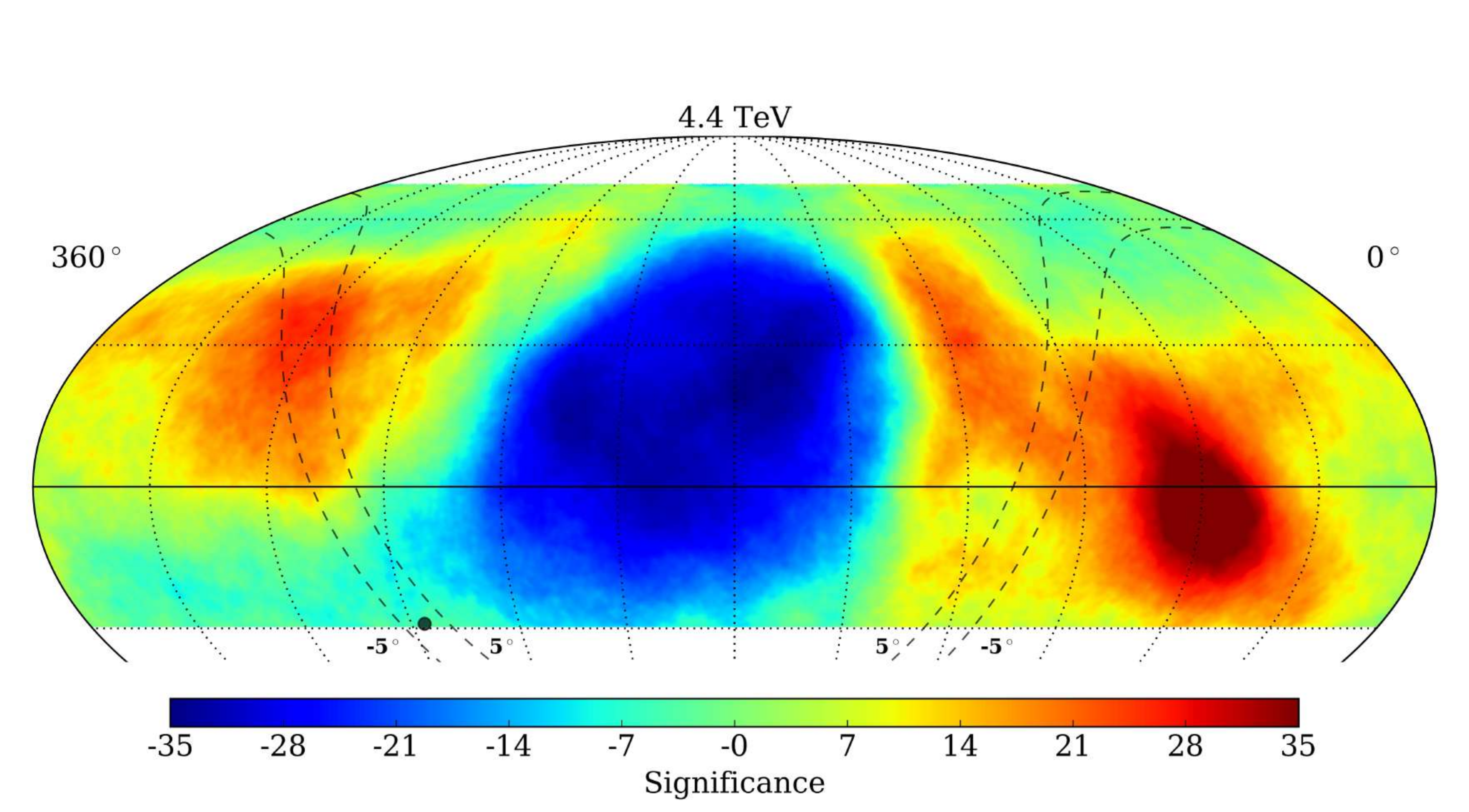}
     \includegraphics[width=0.4\textwidth,trim={0 2.7cm 0 0},clip]{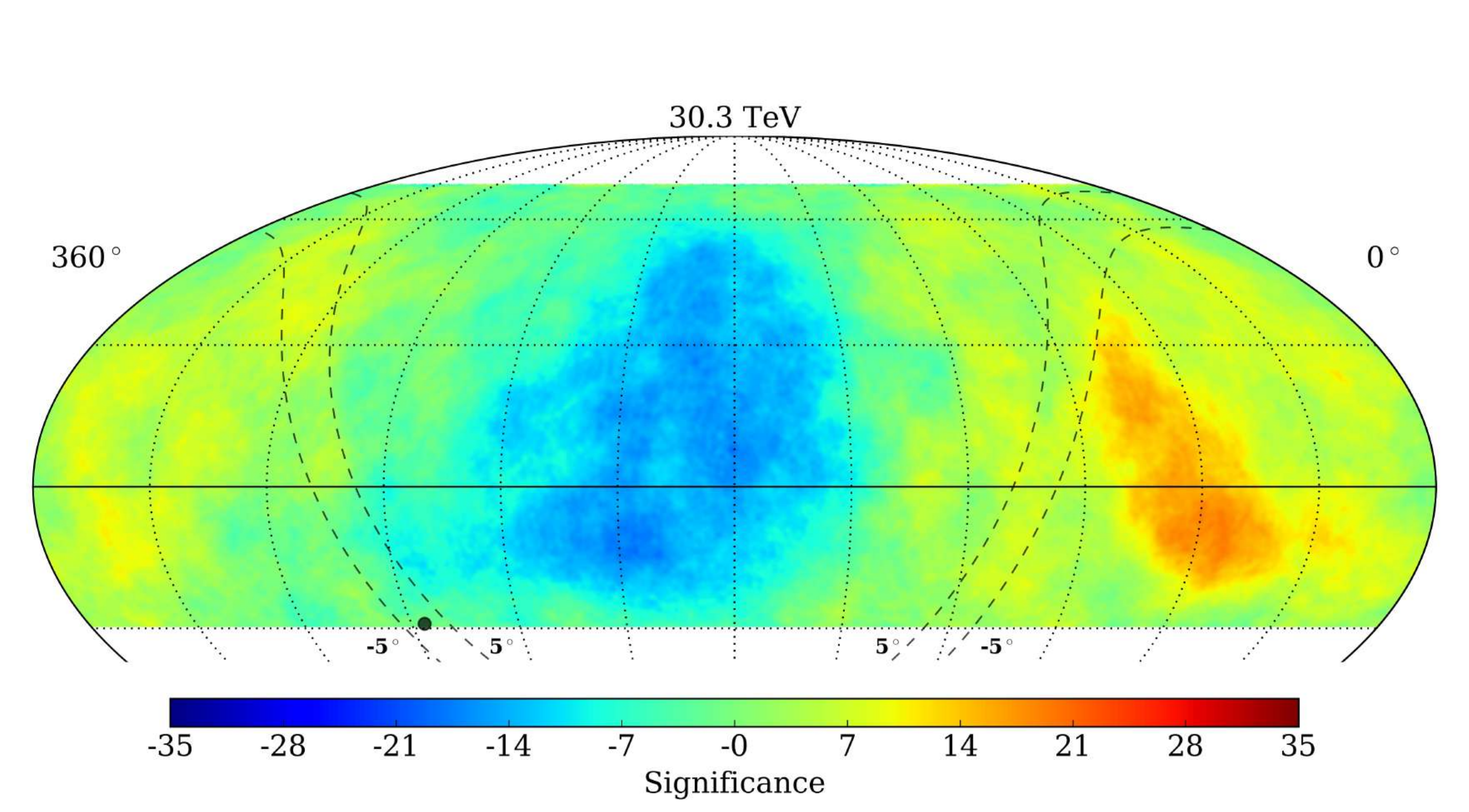}
     \includegraphics[width=0.4\textwidth,trim={0 2.7cm 0 0},clip]{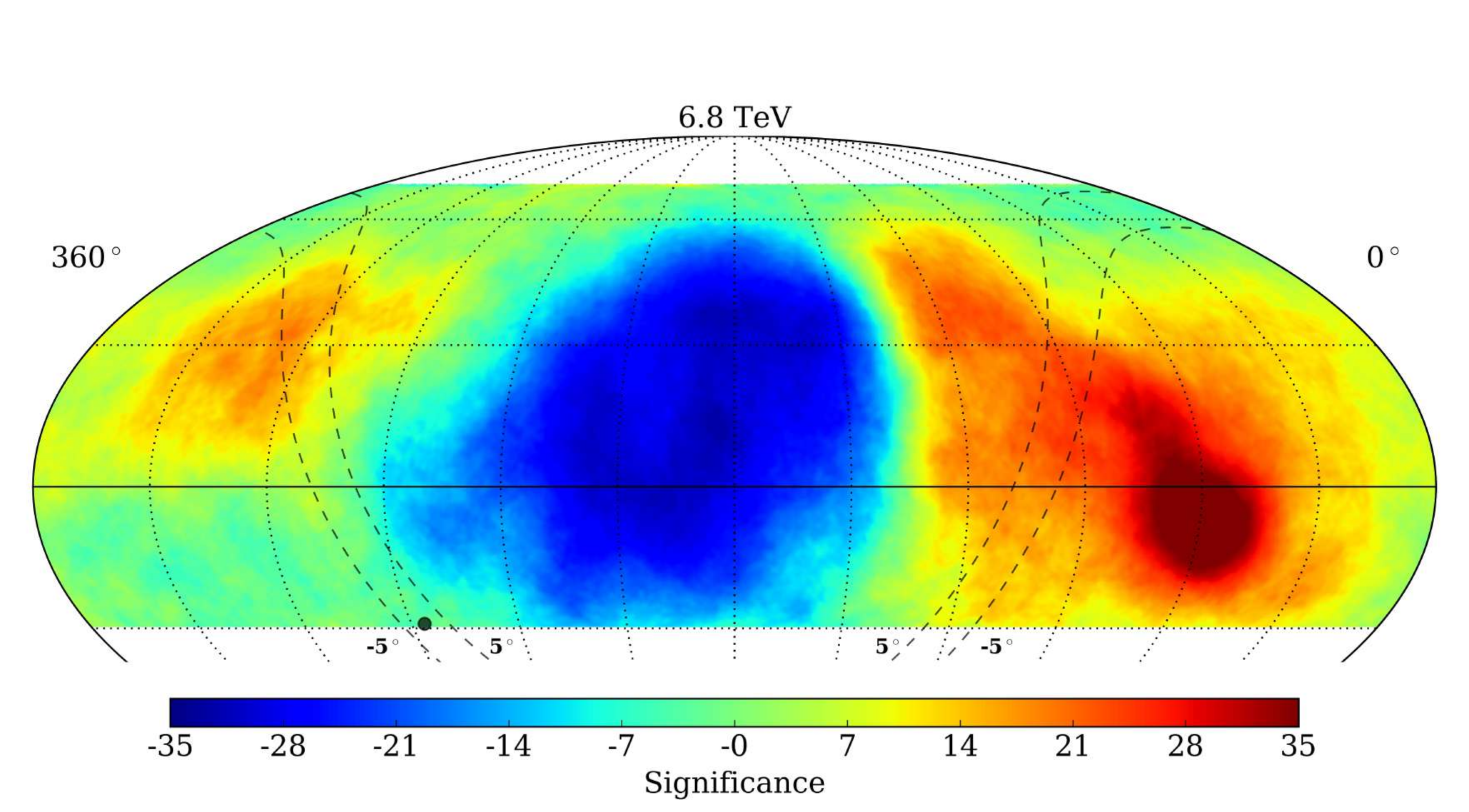}
     \includegraphics[width=0.4\textwidth,trim={0 2.7cm 0 0},clip]{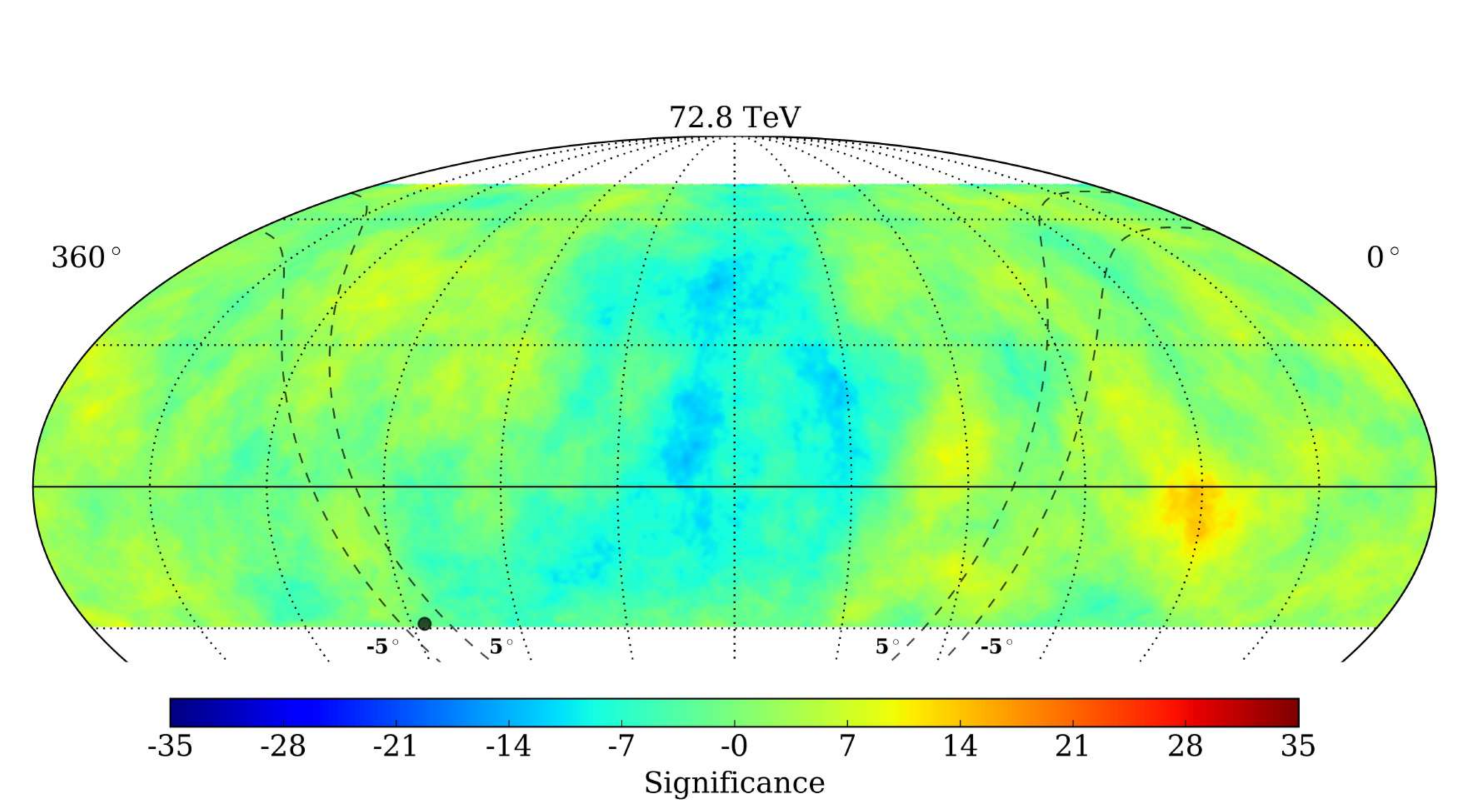}
     \includegraphics[width=0.85\textwidth,trim={0 0 0 13.475cm},clip]{bin1_N256_S10p000_sig_corr.pdf}

  \end{center}
  \caption{
           Anisotropy maps in statistical significance (smoothed $10^\circ$) for 
           eight\deleted{independent} energy bins separated by a likelihood-based reconstructed energy variable.
           \deleted{The Galactic Plane is shown with lines at $+5^\circ$ and $-5^\circ$ in Galactic latitude.}
           }
  \label{fig:signif}
\end{figure*}

The results of the HAWC analysis are shown in Figure \ref{fig:relint}, which depicts the relative intensity of the arrival 
direction distribution of cosmic rays in equatorial coordinates for eight \added{independent} energy bins ranging from 2.0 to 72.8 TeV.
Figure \ref{fig:signif} shows the significance of the deviation from isotropy 
for the same energy bins, where negative values of significance correspond to pixels with $\delta I < 1$.
The maps show significant deviation of the cosmic-ray flux from isotropy, 
dominated by a \replaced{dipole moment}{dipolar feature} which increases in strength
up to 30.3 TeV while maintaining a nearly constant phase.

\subsection{Multipole Fitting}
\label{subsec:fit}

To better quantify the observed large-scale features, the relative intensity map $\delta I$ is fit
to the following truncated series of spherical harmonics:
\replaced{
\begin{equation}\label{eq:sph_harm}
\footnotesize{   \sum^{\ell_\text{max}}_{\ell=1} \sum^{\ell}_{m=-\ell} a_{\ell m} Y_{\ell m}~~~(m\neq0),}
\end{equation}
}
{
\begin{equation}\label{eq:sph_harm}
\footnotesize{  \delta I = \sum^{\ell_\text{max}}_{\ell=1} \sum^{\ell}_{m=-\ell} a_{\ell m} Y_{\ell m}(\theta, \phi)~~~(m\neq0),}
\end{equation}
}
where $\ell_\text{max}$ is chosen to distinguish between large-scale and small-scale features.
The choice of $\ell_\text{max}$ will be discussed in Section~\ref{sec:results}.

Since the local detector acceptance is estimated from the data, features in the anisotropy which only depend on declination 
can not be disentangled from declination-dependent asymmetries in the detector acceptance.
This results from the fact that the HAWC detector only samples the sky in the Earth's rotational direction.
Thus, we set $a_{\ell,0}=0$ for all $\ell$, i.e. the $m=0$ terms,
to reflect the knowledge that the analysis method is insensitive to anisotropy \replaced{solely oriented}{oriented solely}
along the declination direction.

In the fit, we chose to use the real-valued (tesseral) spherical harmonics,
\replaced{
\begin{equation}\label{eq:tess_sph}
\footnotesize{  Y_{\ell m} =
\begin{cases}
 \sqrt{{2(2\ell+1)\over 4\pi}{(\ell-|m|)!\over (\ell+|m|)!}} P_\ell^{|m|}(x) \sin |m|\varphi  & {\mbox{if } m<0} \\
 \sqrt{{(2\ell+1)\over 4\pi}} P_\ell^m(x) & \mbox{if } m=0\\
 \sqrt{{2(2\ell+1)\over 4\pi}{(\ell-m)!\over (\ell+m)!}} P_\ell^m(x) \cos m\varphi & \mbox{if } m>0
\end{cases}}
\end{equation}
}
{
\begin{equation}\label{eq:tess_sph}
\footnotesize{  Y_{\ell m}(\theta, \phi) =
\begin{cases}
 \sqrt{{2(2\ell+1)\over 4\pi}{(\ell-|m|)!\over (\ell+|m|)!}} P_\ell^{|m|}(x) \sin |m|\varphi  & {\mbox{if } m<0} \\
 \sqrt{{2(2\ell+1)\over 4\pi}{(\ell-m)!\over (\ell+m)!}} P_\ell^m(x) \cos m\varphi & \mbox{if } m>0
\end{cases}}
\end{equation}
}
where $x =\cos \theta$, again discarding the $m=0$ terms which are symmetric in right ascension.
The $a_{\ell m}$ are then determined by a $\chi^2$-minimisation fit of $\delta I$ to equation \ref{eq:sph_harm},
and the variance for each pixel is calculated via propagation of uncertainties of the quantities $\mathcal{D}$ and $\mathcal{B}$ 
which comprise $\delta I$.

The strongest feature of the measured anisotropy is the dipole ($l=1$), 
which can be more conveniently expressed as an amplitude and phase by projection onto right ascension. 
The amplitude $\widetilde{A}_1$ can be expressed as the sum of the $a_{1,m}$ terms added in quadrature:
\begin{equation}
  \footnotesize{\widetilde{A}_1 = \sqrt{\frac{3}{8\pi}\sum_{m=1,-1} |a_{1m}|^2}}
\label{eq:Amp}
\end{equation}
with variance 
\begin{equation}
   \footnotesize{\sigma^2_{\left(\widetilde{A}_1\right )} = \frac{3}{8\pi \displaystyle\sum_{m=1,-1} |a_{1m}|^2}
\sum_{m=1,-1} \left( |a_{1m}|^2 \sigma^2_{\left ( a_{1m} \right ) } \right ).}
\end{equation}
The maximally recoverable dipole amplitude $\widetilde{A}_1$ obtained via the iterative method 
is related to the true amplitude $A_1$ through the original declination position of the maximum $\delta_0$,
\begin{equation}
\footnotesize{A_1=\frac{\widetilde{A}_1}{\cos\delta_0} }.
\label{eq:amp}
\end{equation}
The term $a_{10}=0$ leaves one term which scales with the cosine ($a_{1,1}$) and one that scales with 
the sine ($a_{1,-1}$) of the dipole phase $\widetilde{\phi_1}$.
This constrains the maximum amplitude to a declination of $0^\circ$,
simplifying the measured phase to
\begin{equation}
   \footnotesize{\widetilde{\phi_1}=\tan^{-1}\left ( \frac{a_{1,-1}}{a_{1,1}} \right),}
\end{equation}
with variance
\begin{equation}
   \footnotesize{\sigma^2_{\left (\widetilde{\phi_1} \right )}=
\frac{1}{\left(\displaystyle\sum_{m=1,-1}|a_{1m}|^2 \right)^2} \sum_{m=1,-1} 
\left( |a_{1m}|^2 \sigma^2_{\left (a_{1m} \right ) } \right ).}
\end{equation}

\begin{figure*}[t]
  \centering
      \begin{center}
     \includegraphics[width=0.375\textwidth]{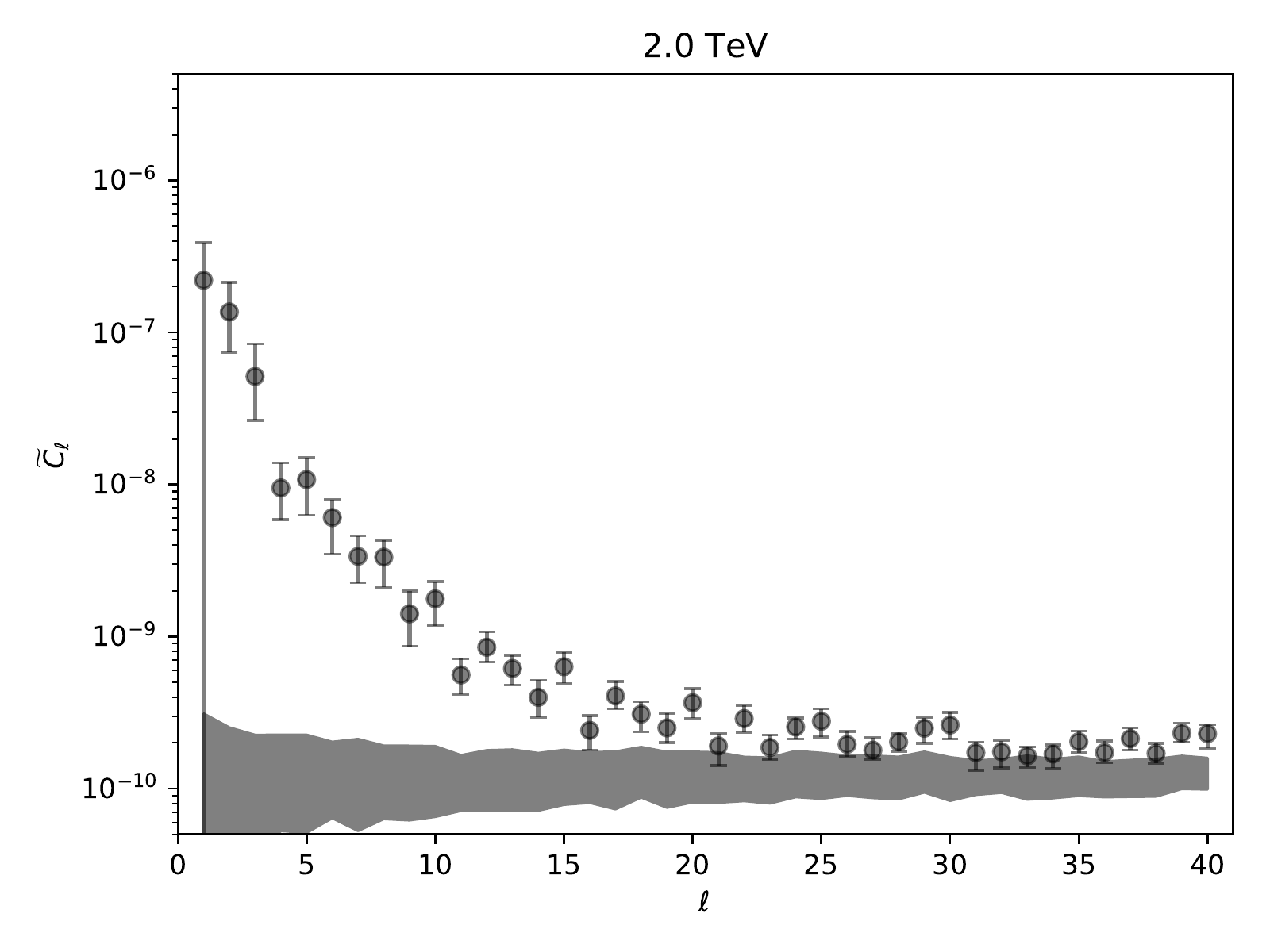}
     \includegraphics[width=0.375\textwidth]{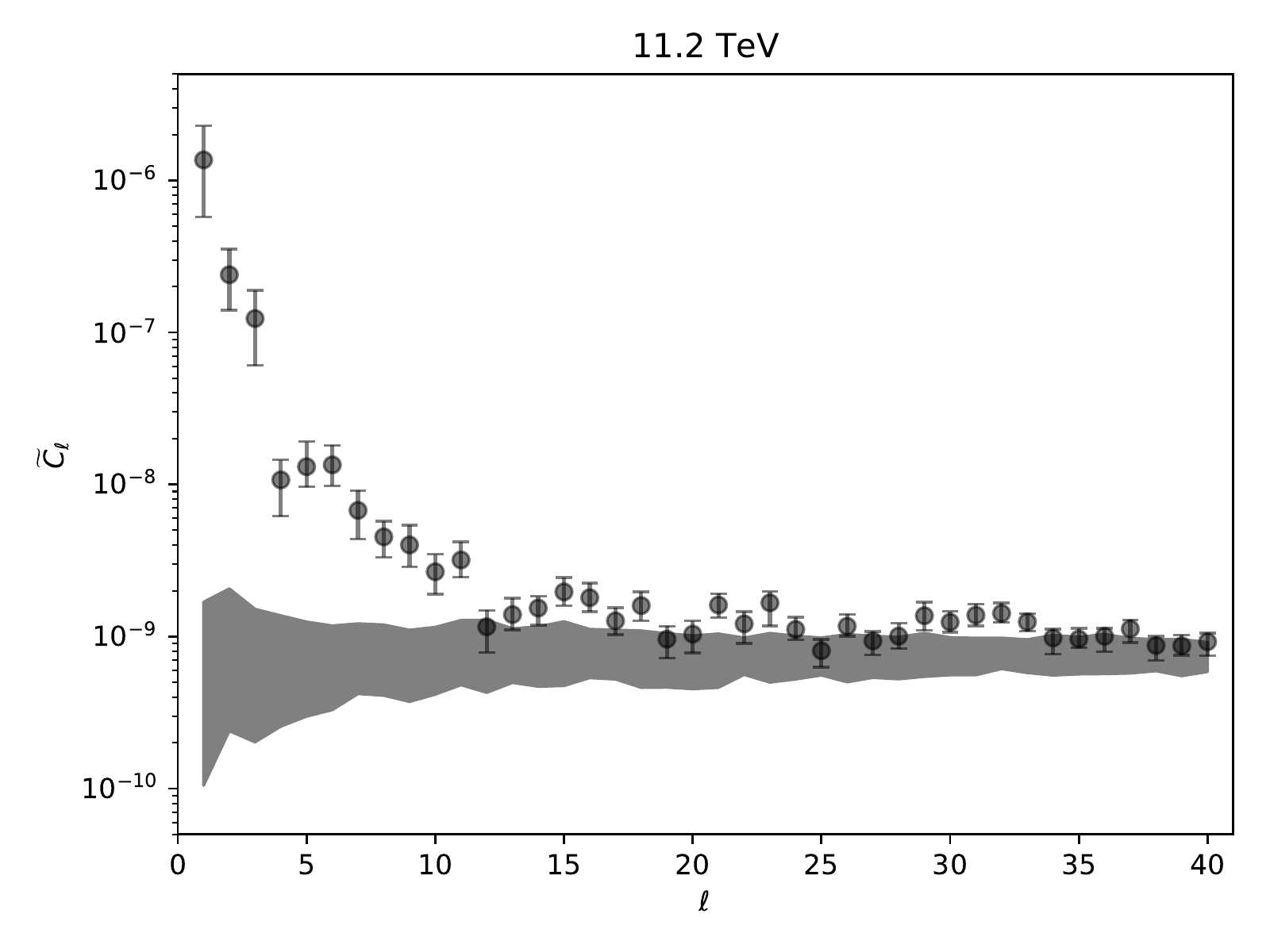}
     \includegraphics[width=0.375\textwidth]{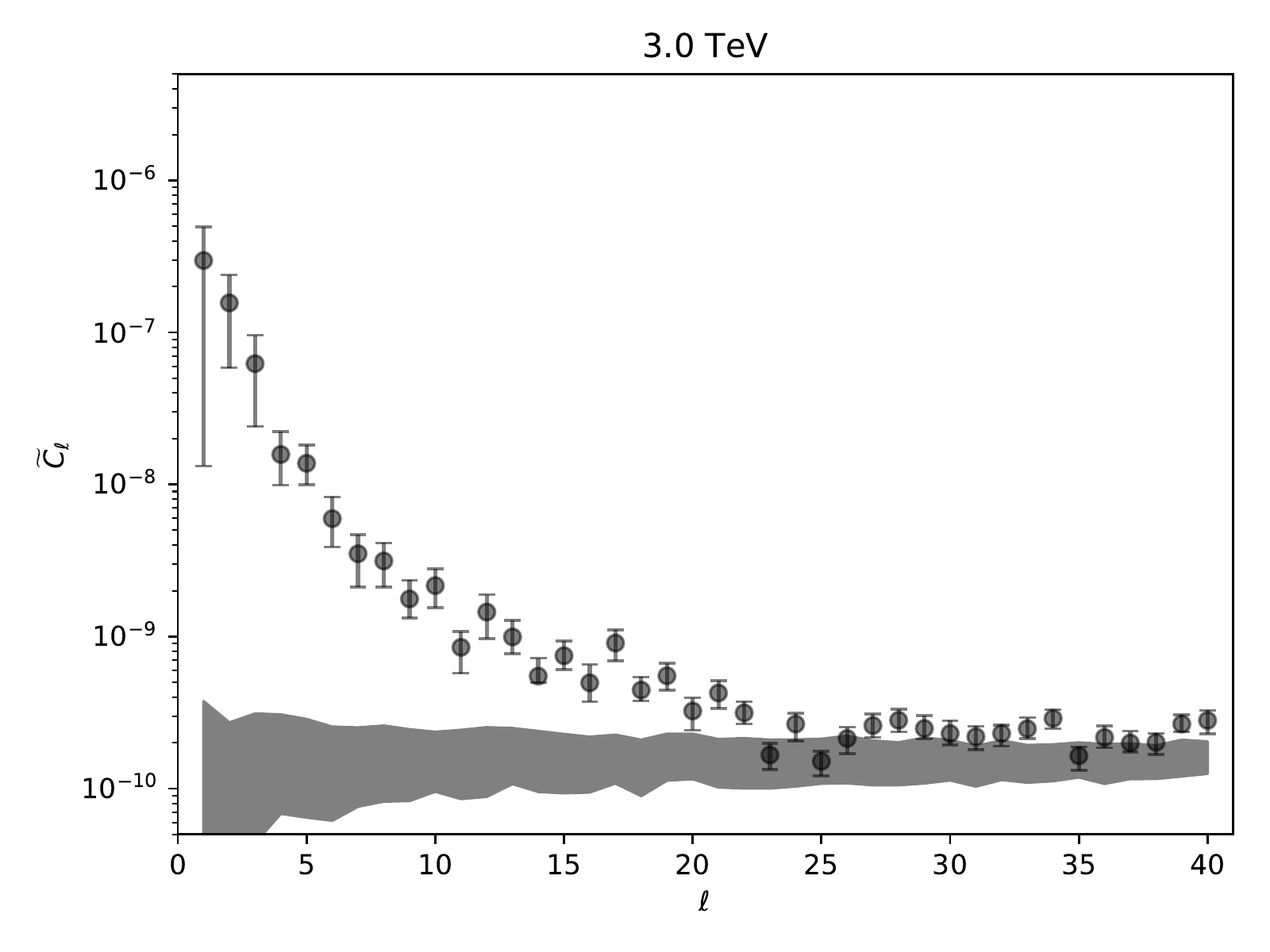}
     \includegraphics[width=0.375\textwidth]{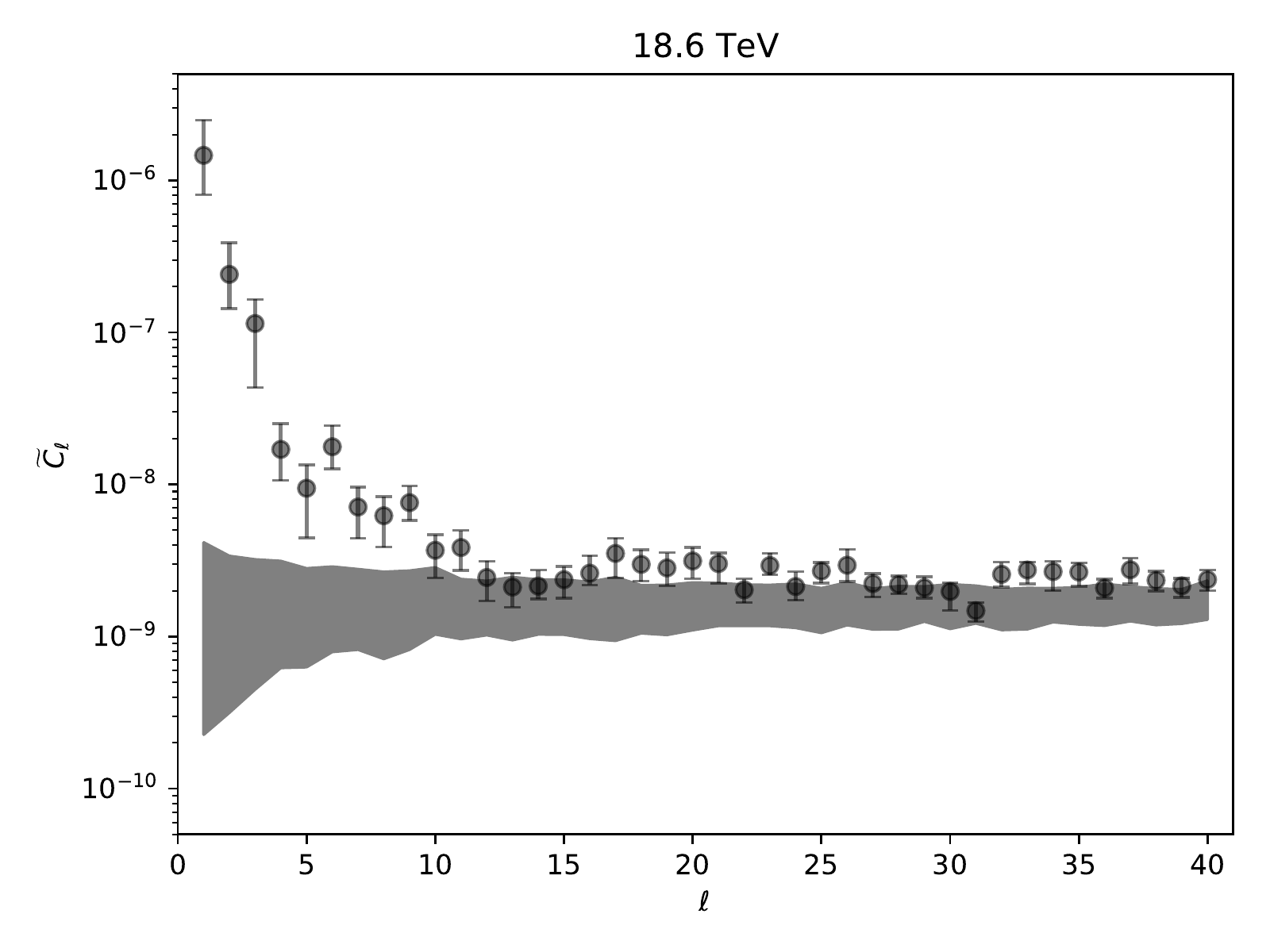}
     \includegraphics[width=0.375\textwidth]{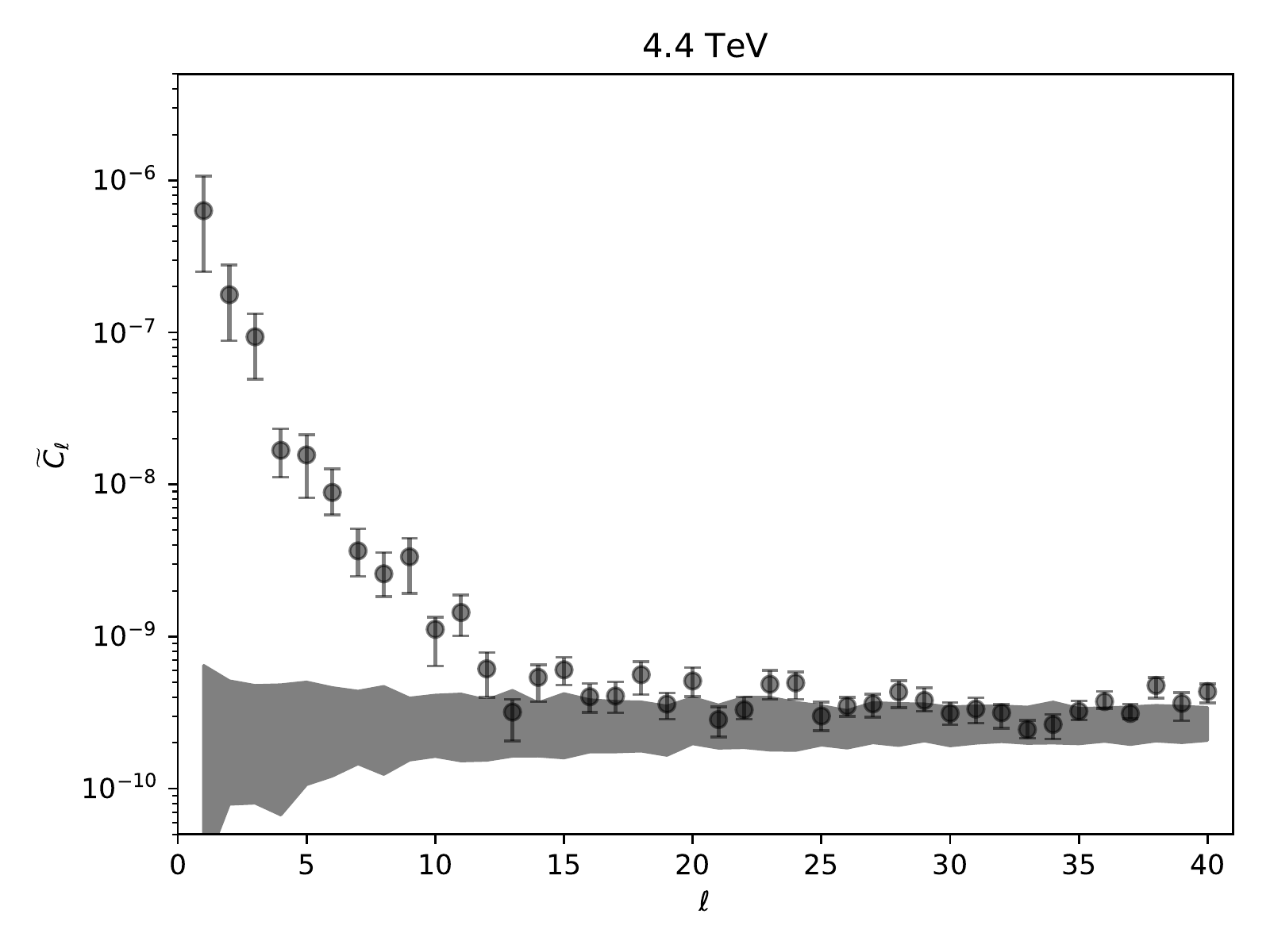}
     \includegraphics[width=0.375\textwidth]{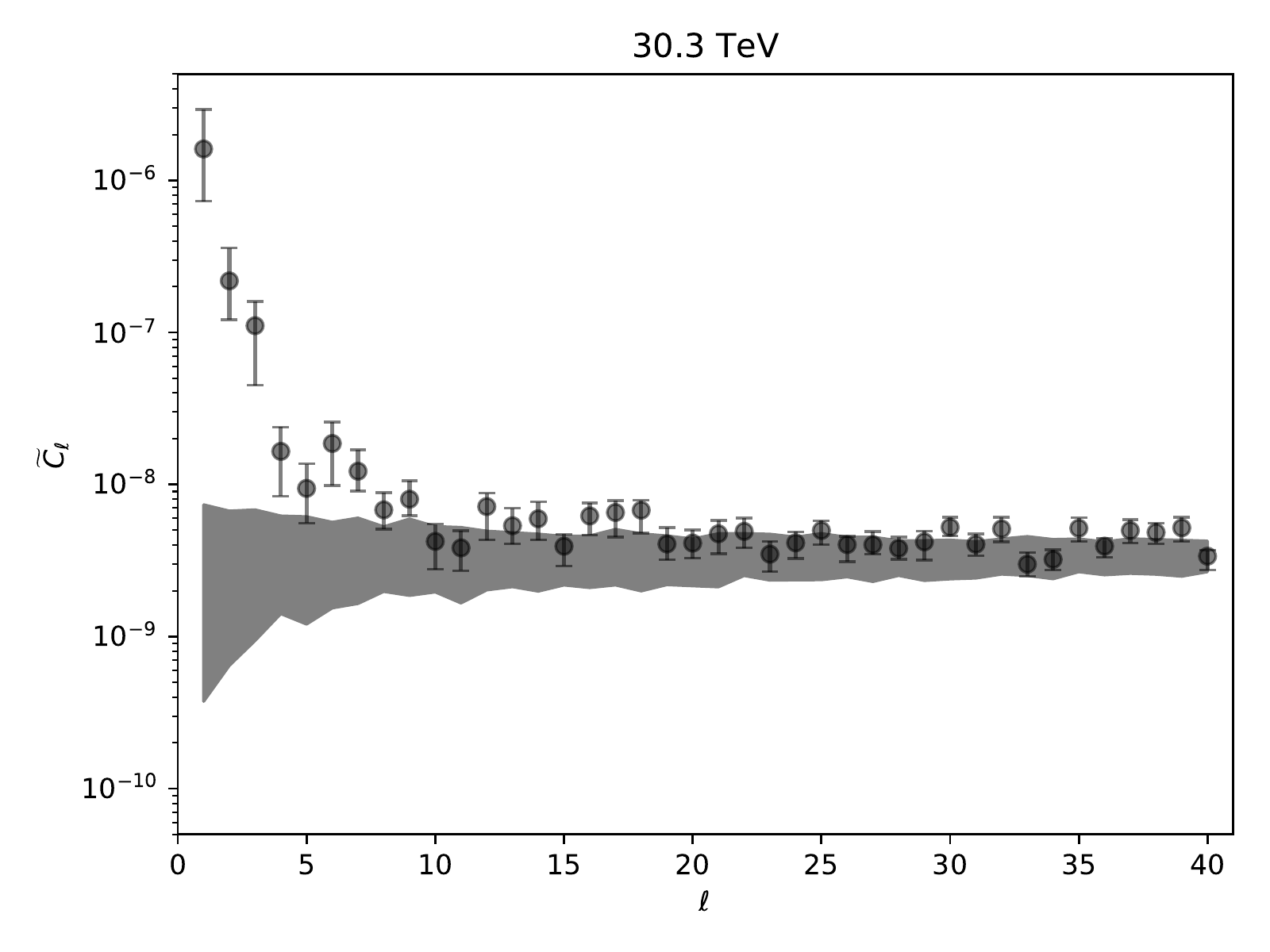}
     \includegraphics[width=0.375\textwidth]{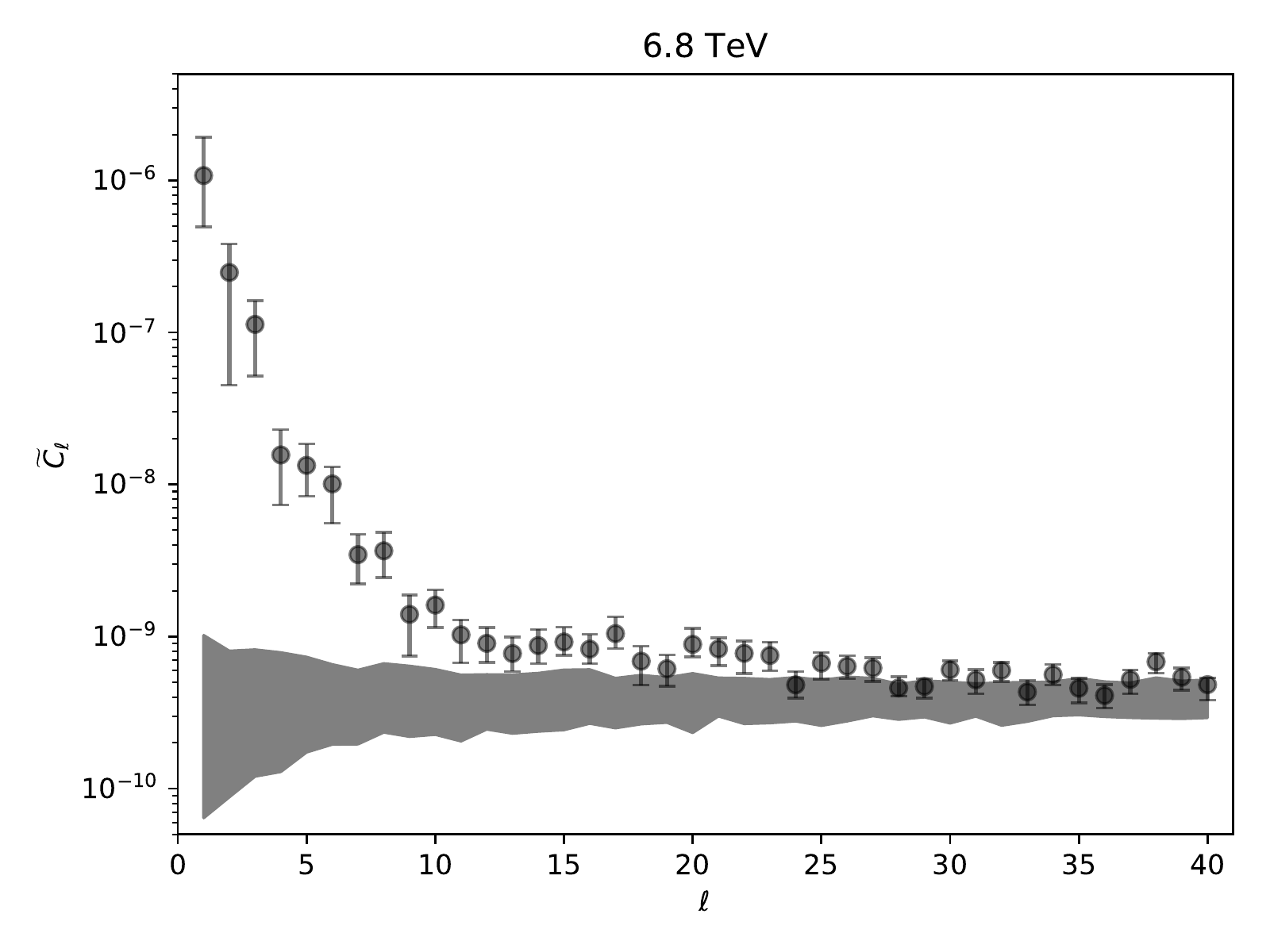}
     \includegraphics[width=0.375\textwidth]{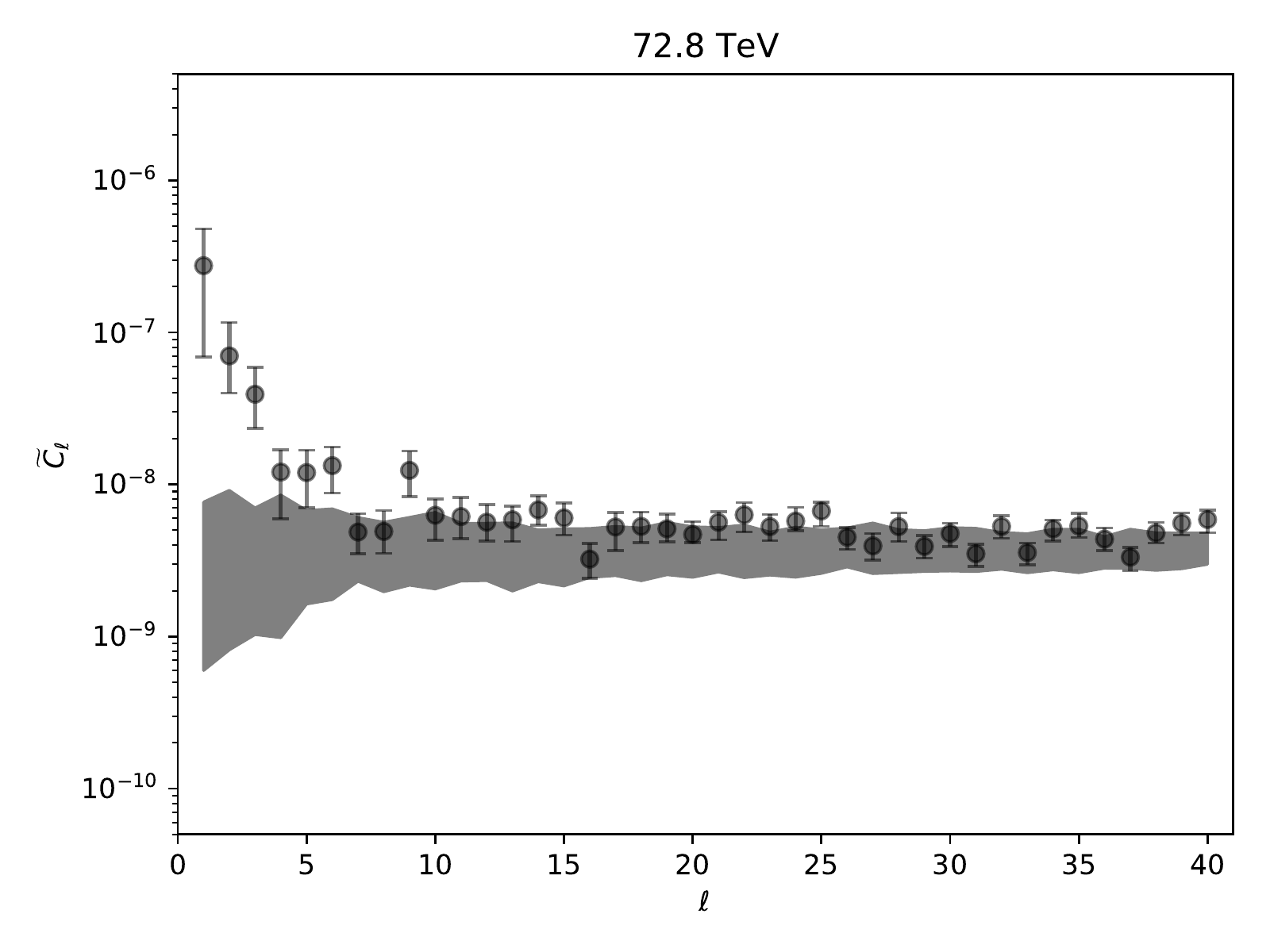}

    \end{center}
    \caption{
             Angular power spectra for each of the eight independent energy bins.
             \added{The gray bands represent the power spectra for isotropic sky maps at the 90\% confidence level}.
             The uncertainties on the data points are systematic, representing the $68\%$ containment
             of the measured angular power spectrum for maps with the same true power spectrum as the data.
             The statistical uncertainties are smaller than the data points.
             The multipole moments $\ell$ correspond to angular scales $180^\circ/\ell$. 
            }
    \label{fig:angps}
\end{figure*}

\subsection{Angular Power Spectrum}
\label{subsec:powerspectrum}

The cosmic-ray anisotropy is not a pure dipole, and its full angular power spectrum
reveals the strength of correlations at various angular scales. 
Figure~\ref{fig:angps} shows for each energy bin the pseudo-angular power spectra 
as derived from the anafast routine in HEALPix,
\added{where $\displaystyle \tilde{C}_{\ell} = \frac{1}{2 \ell+1} \sum_{m=-\ell}^{\ell} | a_{\ell m} |^2 $ }. 
The infinite series of multipoles was truncated at $\ell_\text{max}=40$ for quicker computation.
Truncation at the maximum multipole that can be calculated for a HEALPix grid ($\ell_\text{max}=3\, N_{\text{side}}-1$) was done for comparison, 
with no noticeable effect on the reported power spectra.

The uncertainties shown in Figure~\ref{fig:angps} are systematic, representing the $68\%$ central containment region
of the measured power spectrum for maps generated to have the same power spectrum as the data. 
The statistical uncertainties are smaller than the data points. 
These were determined by the $68\%$ containment of angular power spectra derived 
from random sky maps drawn from a Poisson distribution using the true map as the mean. 

The gray band shows the $90\%$ confidence interval for the expected angular power spectrum 
for an isotropic map containing the same event statistics.
The isotropic power band is flat across the multipole moments, and its magnitude level is determined by the number of events in the map.
This band was derived in the same way as the statistical uncertainties, except the data fluctuated by a Poisson distribution was 
compared to the true data instead of the reference map.
The deviation of the spectral points from this gray band represents the measured signal strength compared to isotropy.

\subsection{Systematics}
\label{subsec:Systematics}

As previously described\deleted{ and depicted in Figure~\ref{fig:simdipole}}, 
the measured dipole orientation in declination is unconstrained due to the limitations in estimating the reference map.
For example, \replaced{for a true dipole (black dashed line ) oriented along $\delta_0>65^\circ$, the measured amplitude is smaller by
more than a factor of two}{the measured amplitude (blue solid line in Figure~\ref{fig:simdipole}) of a dipole with
orientation $\delta_0>65^\circ$ is decreased by more than a factor of two from its true value (black dashed line)}.
By modeling the detector acceptance with functional forms, 
it is possible to recover the declination orientation as done in \citep{Aab:2017tyv}.
This systematic has not been studied for HAWC in this capacity, as it requires an inordinate amount of simulation
that matches the data to a precision below the observed level of anisotropy. 
However, the degeneracies in the angular power spectrum caused by the lack of full sky coverage
are taken into account in the systematic uncertaintes presented in Figure~\ref{fig:angps}.

The known solar dipole signal from the motion of the Earth around the Sun is present in the data,
and can in principle contaminate the dipole signal in equatorial coordinates.
However, in this analysis, where  an integer number of years of data taken at a constant rate is used, 
the influence of the solar dipole cancels\deleted{ out} and can be neglected.
We estimate a maximum residual signal of $2\times 10^{-5}$ from the solar dipole, 
$\sim1\%$ of the sidereal signal.

A thorough verification of the absolute energy scale has been performed\deleted{in \citep{crpaper}} using 
the energy dependence of the cosmic-ray Moon shadow \added{\citep{crpaper}}.
The uncertainty in the scale of the reported median energies for the eight analysis bins
is estimated to be $\sim5\%$.

\section{Results and Discussion}
\label{sec:results}

 \begin{figure}[tb]
   \centering
     \includegraphics[width=0.48\textwidth]{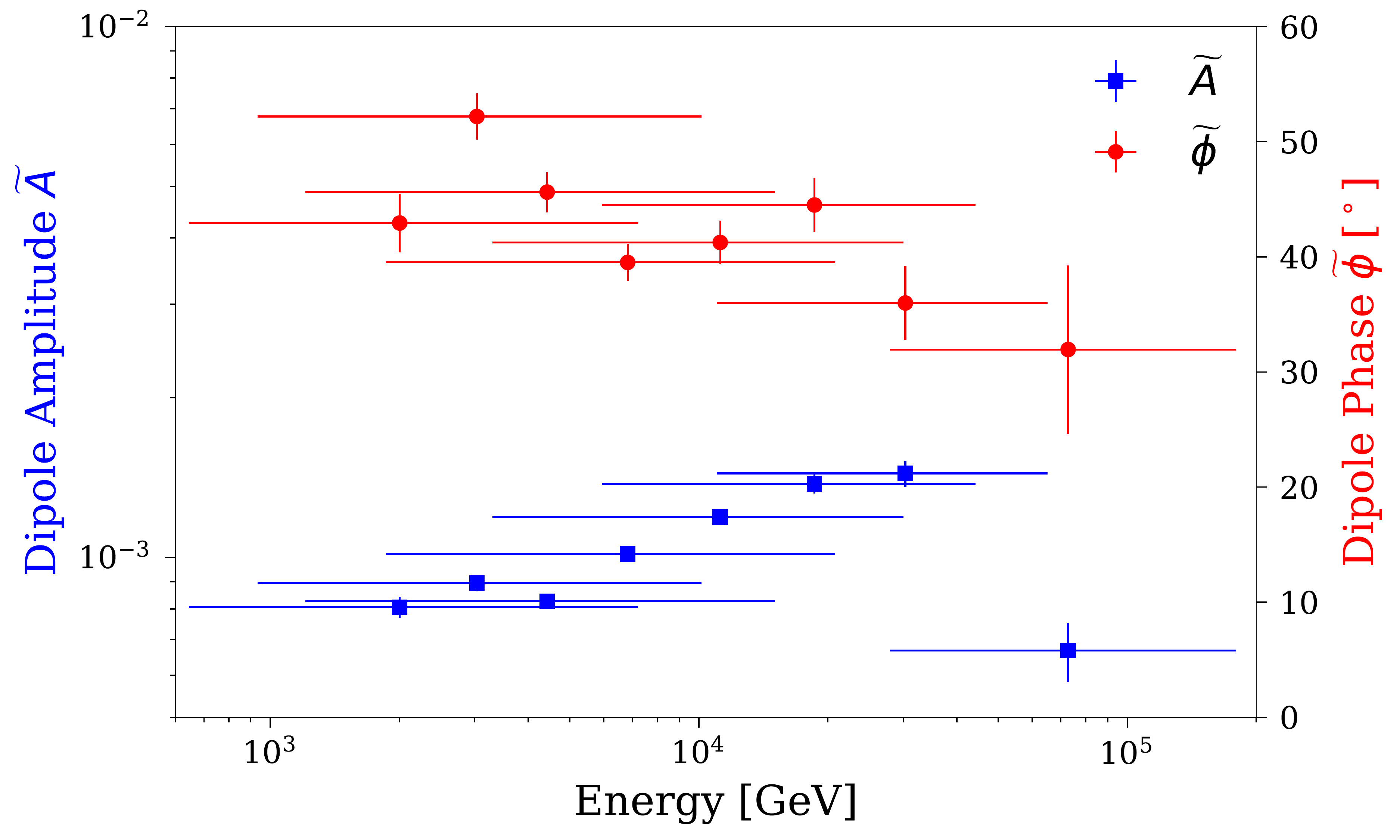}
     \caption{
              The best fit dipole amplitude (blue squares) and phase (red circles) as functions
              of energy.
             }
     \label{fig:dipole_amp_phase}
 \end{figure}

  \begin{deluxetable*}{r r r r r r r}
    \tablecaption{Reported median energy (with 68\% central containment region) and fit of two-dimensional dipole anisotropy (amplitude and phase) for each independent energy bin. \label{tab:fits}}
    \tablehead{ \colhead{Energy} &\colhead{Events}  & \colhead{Amplitude } & \colhead{Phase} & \colhead{$a_{1,1}$  } & \colhead{$a_{1,-1}$ } & \colhead{$\chi^2/N_{\text{dof}}$}  \\ 
     \colhead{[TeV]} &  & \colhead{\footnotesize{$[\times 10^{-4}]$}}  & & \colhead{\footnotesize{$[\times 10^{-4}]$}} & \colhead{\footnotesize{$[\times 10^{-4}]$}} } 

    \startdata
      $2.0   \left ( ^{-1.4}_{+5.2} \right )$& $4.0\times 10^{10}$  & $8.1 \pm 0.4$& $42.9^\circ \pm 2.5^\circ$& $ -17.1 \pm 1.0$& $-15.9 \pm 1.0$ & $13.34$ \\
      $3.0   \left ( ^{-2.1}_{+7.1} \right )$& $2.9\times 10^{10}$  & $8.9 \pm 0.3$& $52.2^\circ \pm 2.0^\circ$& $ -15.9 \pm 0.9$& $-20.5 \pm 0.9$ & $1.24$ \\
      $4.4   \left ( ^{-3.2}_{+10.6} \right )$& $2.4\times 10^{10}$  & $8.3 \pm 0.3$& $45.6^\circ \pm 1.8^\circ$& $ -16.7 \pm 0.7$& $-17.1 \pm 0.7$ & $0.79$ \\
      $6.8   \left ( ^{-5.0}_{+14.0} \right )$& $1.6\times 10^{10}$  & $10.1 \pm 0.3$& $39.5^\circ \pm 1.6^\circ$& $ -22.7 \pm 0.8$& $-18.7 \pm 0.8$ & $0.85$ \\
      $11.2   \left ( ^{-7.9}_{+18.8} \right )$& $7.9\times 10^9$  & $11.9 \pm 0.4$& $41.3^\circ \pm 1.9^\circ$& $ -25.9 \pm 1.1$& $-22.7 \pm 1.1$ & $0.81$ \\
      $18.6   \left ( ^{-12.7}_{+25.6} \right )$& $3.8\times 10^9$  & $13.8 \pm 0.6$& $44.5^\circ \pm 2.4^\circ$& $ -28.4 \pm 1.6$& $-27.9 \pm 1.6$ & $1.07$ \\
      $30.3   \left ( ^{-19.3}_{+34.8} \right )$& $1.8\times 10^9$  & $14.4 \pm 0.8$& $36.0^\circ \pm 3.2^\circ$& $ -33.7 \pm 2.3$& $-24.5 \pm 2.3$ & $1.25$ \\
      $72.8   \left ( ^{-44.9}_{+106.7} \right )$& $1.6\times 10^9$  & $6.7 \pm 0.9$& $31.9^\circ \pm 7.3^\circ$& $ -16.4 \pm 2.5$& $-10.2 \pm 2.5$ & $1.03$ \\
    \enddata
  \end{deluxetable*}

The resulting relative intensity maps, significance maps, and power spectra for each of the eight energy bins 
from 2.0 TeV to 72.8 TeV are shown in Figures~\ref{fig:relint}, \ref{fig:signif}, and \ref{fig:angps}, respectively.
The fit dipole amplitudes and phases obtained from these large-scale maps are shown in Figure~\ref{fig:dipole_amp_phase}.
In order to enhance regional correlations, the sky maps have been smoothed by a circular top hat function of radius $10^{\circ}$, 
in accordance with other studies \citep{Abdo:2008kr,Abeysekara:2014sna}.

Each map in Figure~\ref{fig:relint} shares the common significant features of having a broad region of 
deficit around $\alpha=150^\circ$ to $\alpha=240^\circ$
and a broad but more sharply-peaked excess around $\alpha=30^\circ$ to $\alpha=90^\circ$.
The deficit grows in intensity with energy until the final bin at 72.8 TeV, where the feature diminishes.
The center of the excess starts low in the HAWC field of view near $\delta=-15^\circ$ and rises to about $\delta=-5^\circ$ 
by 4.4 TeV, where it remains for the remaining energy bins. 
Its strength increases until 11.2 TeV, slowly diminishing until nearly disappearing in the 72.8 TeV bin.
Starting at 6.8 TeV the excess develops an extension higher in declination and slightly higher in right ascension.
This extension is strongest in the 30.3 TeV bin before also diminishing by 72.8 TeV.

\replaced{The angular power spectra in Figure~\ref{fig:angps} 
depict that}{As shown in Figure~\ref{fig:angps},} 
the dipole moments possess the most angular power for each bin,
also reflected in the evolution of the broad deficit.
The second strongest moment is the quadropole, whose power remains fairly constant
at around $1.5\times10^{-7}$ for all energies save the highest energy bin.
The octupole moment is typically half of the quadrupole moment.
In most bins there is a rapid decrease in power from these first three moments to the sextupole ($\ell=4$),
at which point we differentiate between the large and small angular scales.
Significant anisotropy is seen up to $\ell=10$ (characteristic angular scale of $18^{\circ}$) 
until statistics dip below 5 billion events above 18.6 TeV.

\begin{figure*}[t]
  \centering
    \includegraphics[width=\textwidth]{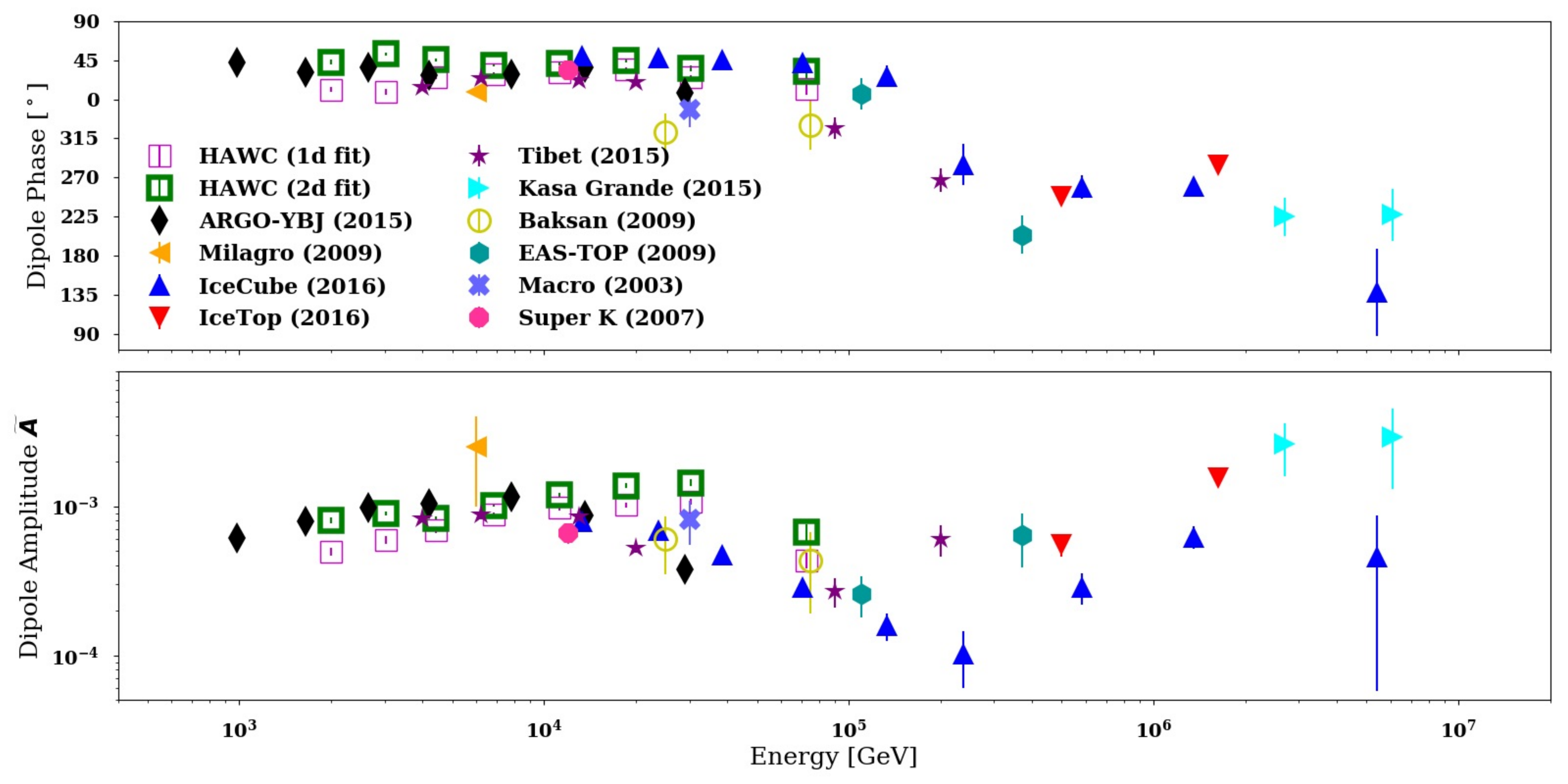}
    \caption{
             Comparison of the fit dipole phase (top) and amplitude (bottom) of HAWC and previously reported \replaced{dipole amplitudes}{results} 
             for all-sky cosmic-ray anisotropy maps.
             All previous measurements fit the dipole projected in the right-ascension axis.
             The median energy for the energy bin is reported.
             The HAWC data is shown for both the method described in the paper (2d fit) and using the projection method (1d fit).
             The most notable feature across energies is the abrupt decrease in amplitude around 100 TeV.
             Above 100 TeV, \added{the} dipole reappears with a similar amplitude, but the phase has changed by nearly $180^\circ$ \citep{Aartsen:2016ivj, Amenomori:2017jbv}.}
    \label{fig:dipolefits}
\end{figure*}

\added{
The decrease in $\tilde{C}_{\ell}$ as a function of $\ell$, especially for $\ell>3$ 
becomes more rapid with increasing energy,
and the anisotropy becomes less significant due to the rising noise floor of maps with fewer data.
This noise level is represented as the expected power spectrum of an isotropic cosmic-ray distribution,
shown by the gray bands in Figure~\ref{fig:angps}.
The large uncertainty for the $\ell=1$ term in the lowest energy bin results from the reduced
declination range available for the multipole fit.
As the integrated sky coverage decreases, angular power from lower multipoles becomes increasingly degenerate
with power from higher multipoles.
}

A summary of the dipole fit parameters obtained per the methods of Section~\ref{subsec:fit}, and
the median cosmic ray energies and numbers of events for each bin are given in Table~\ref{tab:fits}.
The dipole component is detected at a significant level in all bins, 
and as shown in Figure~\ref{fig:dipole_amp_phase}, its amplitude steadily increases with energy 
from $8.1\times10^{-4}$ to $14.4\times10^{-4}$ until the final bin at 72.8 TeV where its value 
drops to $6.7\times10^{-4}$.
As shown in Figure~\ref{fig:signif}, this bin also has the most significant excess 
near $\alpha=300^\circ$, the declination of the Cygnus region which has more than one extended TeV gamma-ray emission features. 
Determining the precise contribution of gamma-ray contamination will be considered in future studies.

The resulting phases and amplitudes from the dipole fits are compared with measurements from other 
experiments in Figure~\ref{fig:dipolefits}, with the HAWC measurements (green squares) being 
in fair agreement with the observed trends.
Though not all phases are consistent within statistical errors,
systematics which are not accounted for in the estimation of the reference maps may
contribute to the $5^{\circ}-10^{\circ}$ differences between HAWC and ARGO-YBJ (black diamonds) at energies below 10 TeV.
The phases measured by HAWC and IceCube are consistent within the overlapping energy range,
noting that the reference map methods used here and in the IceCube study are nearly identical. 

The evolution with energy of the fit amplitudes matches well with previous results,
showing
a steady rise until a sudden decrease between $50-100$ TeV.
For the HAWC measurement, the highest energy bin at 72.8 TeV has nearly the same 
number of events as at 30.3 TeV, yet its dipole amplitude is reduced by more than half.
The energy scale of the amplitude behavior is in slight disagreement with several other experiments. 
This tension could be resolved by shifting along the abscissa, as the experiments's energy scales may be offset relative to one another.
For example, the energy scale reported for HAWC in this work may \replaced{be adjusted}{vary} by 5\% \citep{crpaper},
while the ARGO-YBJ proton energy is reported to within $13\%$ \citep{Bartoli:2015ysa}.

\begin{figure*}[t]

  \begin{center}
     \includegraphics[width=0.4\textwidth,trim={0 2.7cm 0 0},clip]{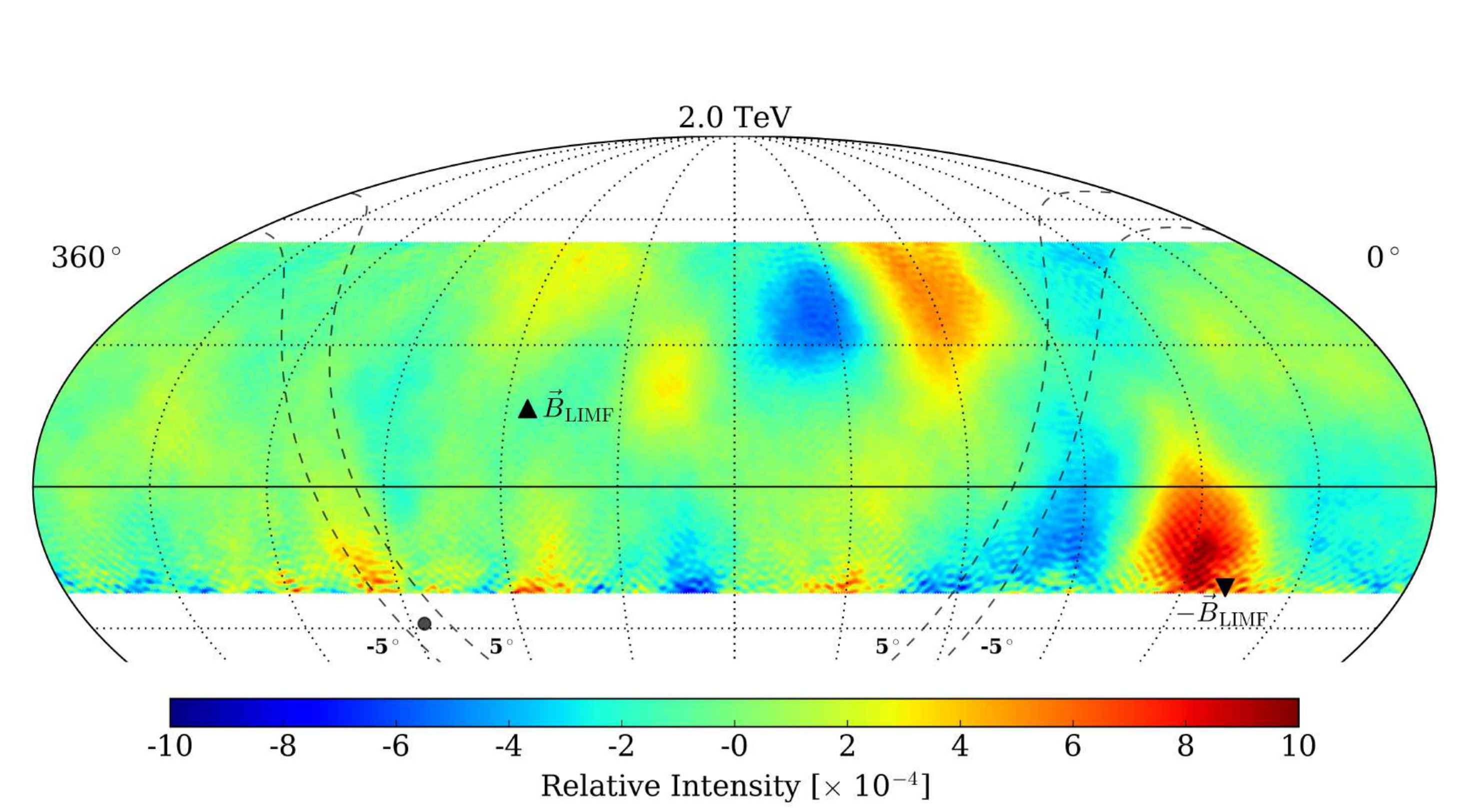}
     \includegraphics[width=0.4\textwidth,trim={0 2.7cm 0 0},clip]{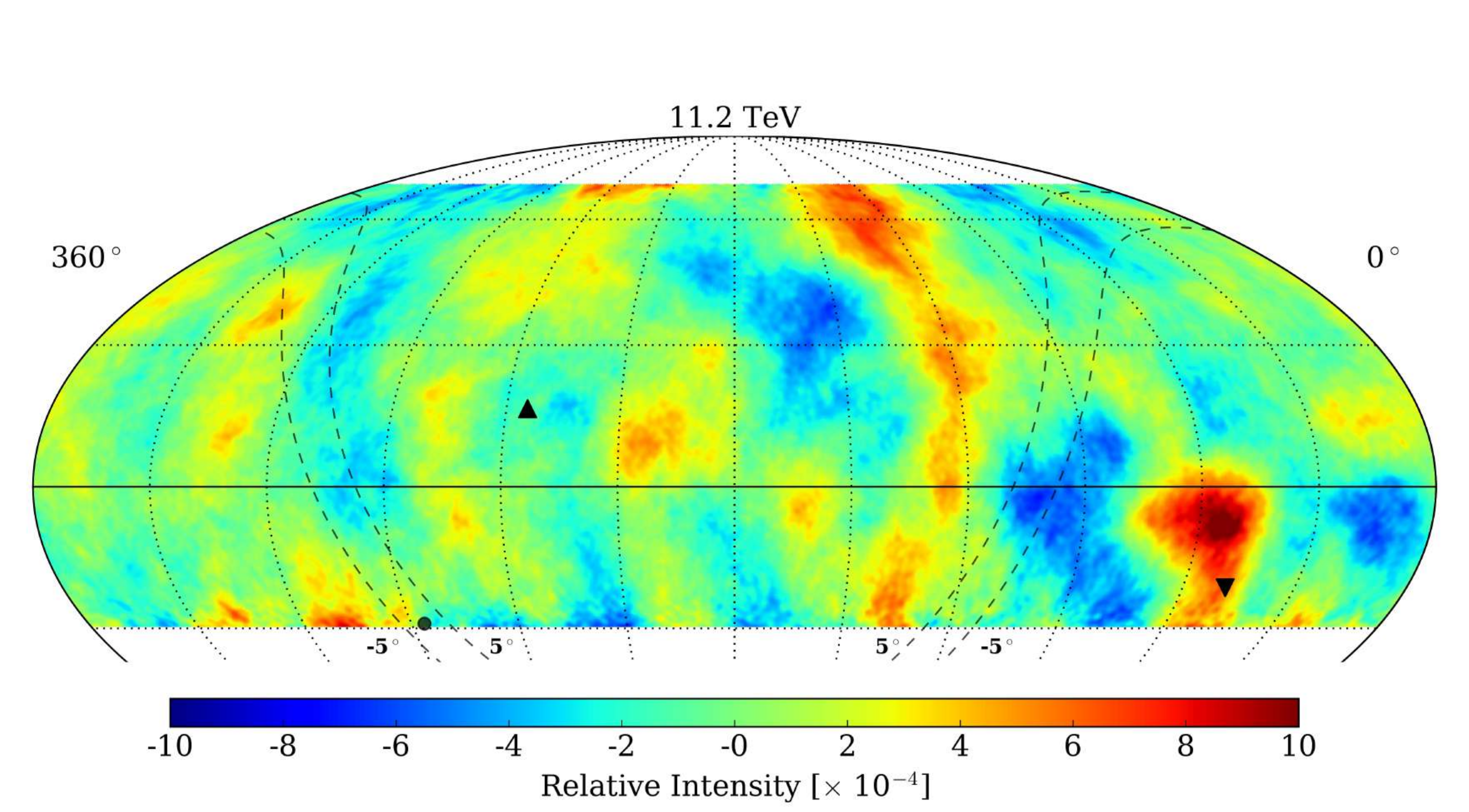}
     \includegraphics[width=0.4\textwidth,trim={0 2.7cm 0 0},clip]{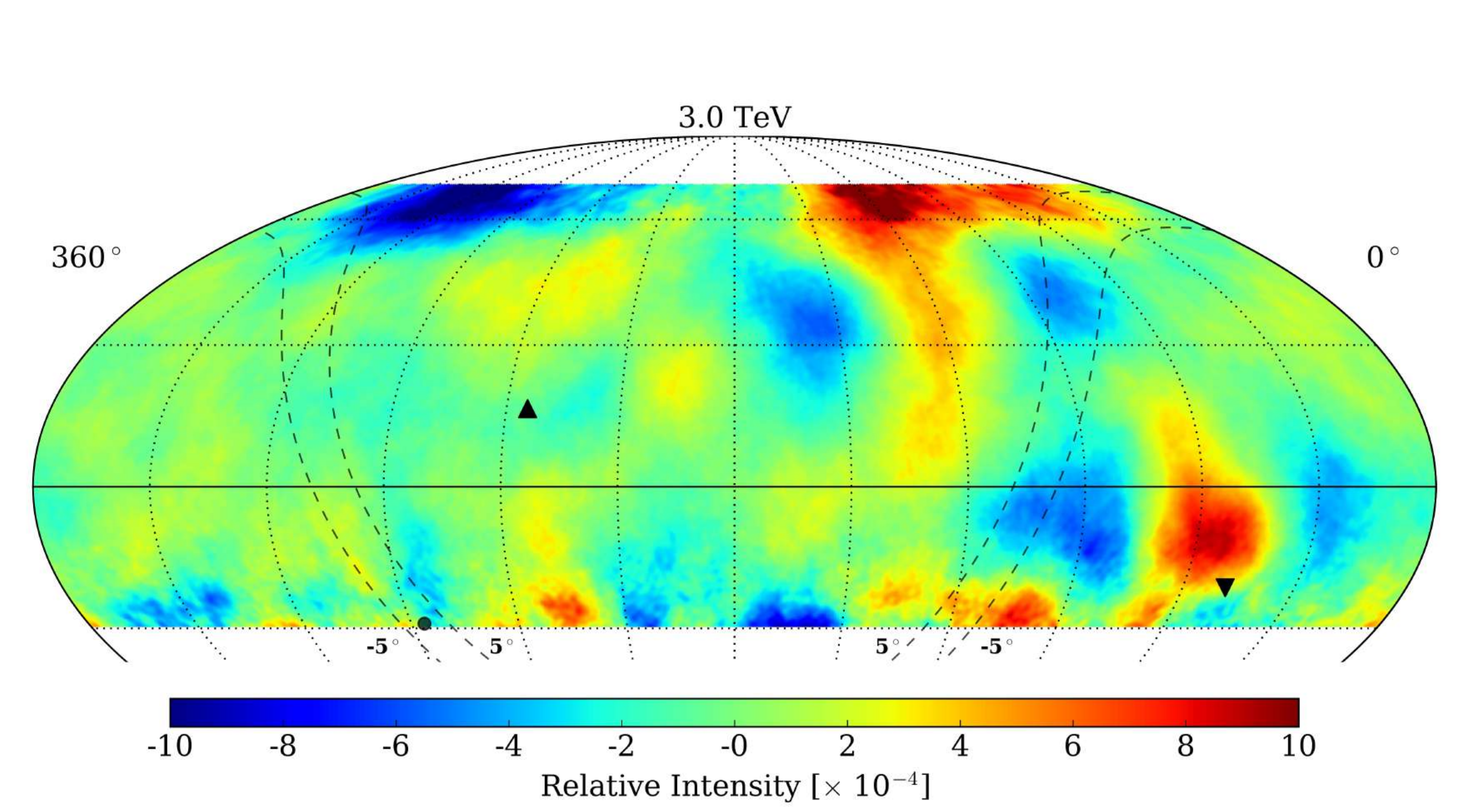}
     \includegraphics[width=0.4\textwidth,trim={0 2.7cm 0 0},clip]{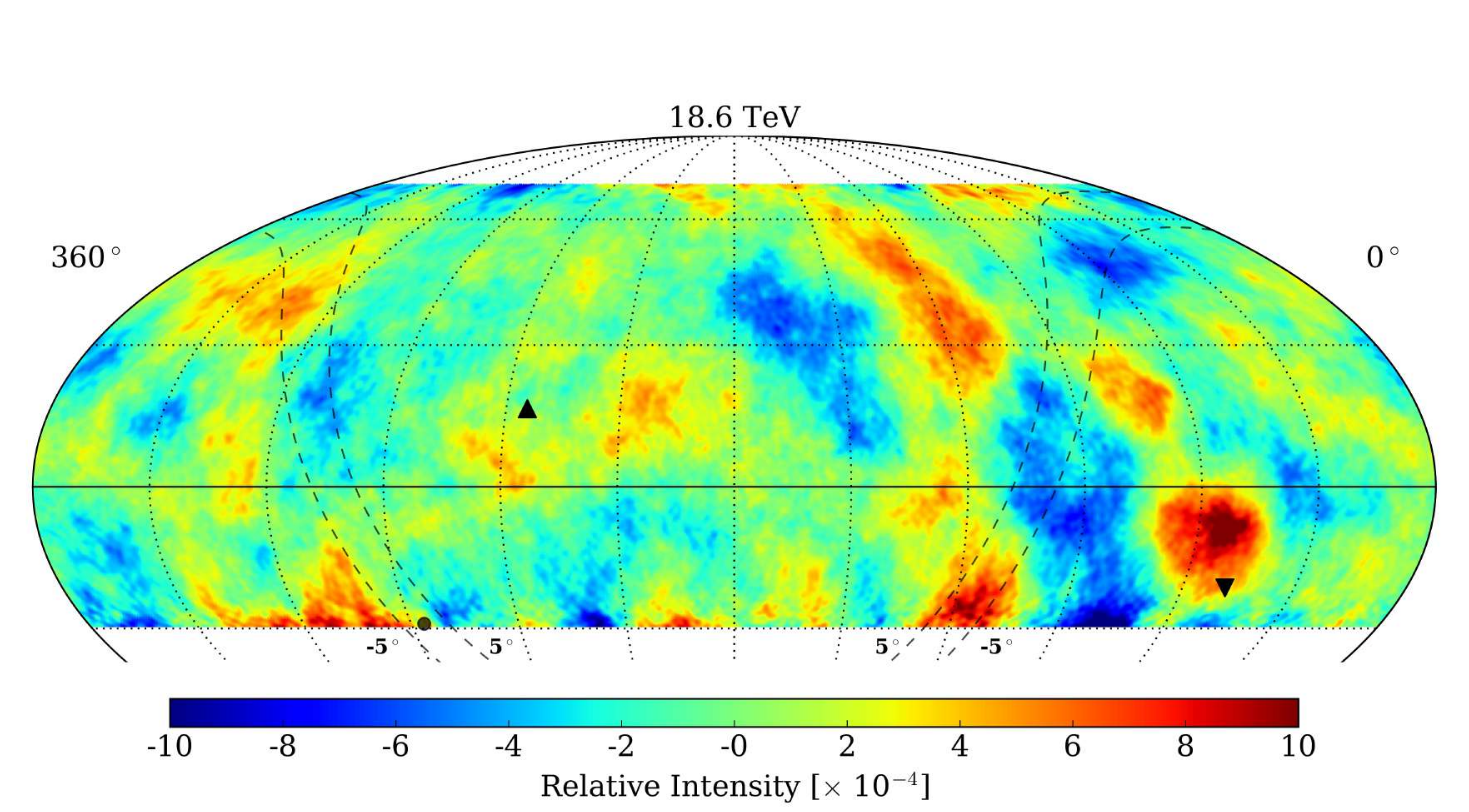}
     \includegraphics[width=0.4\textwidth,trim={0 2.7cm 0 0},clip]{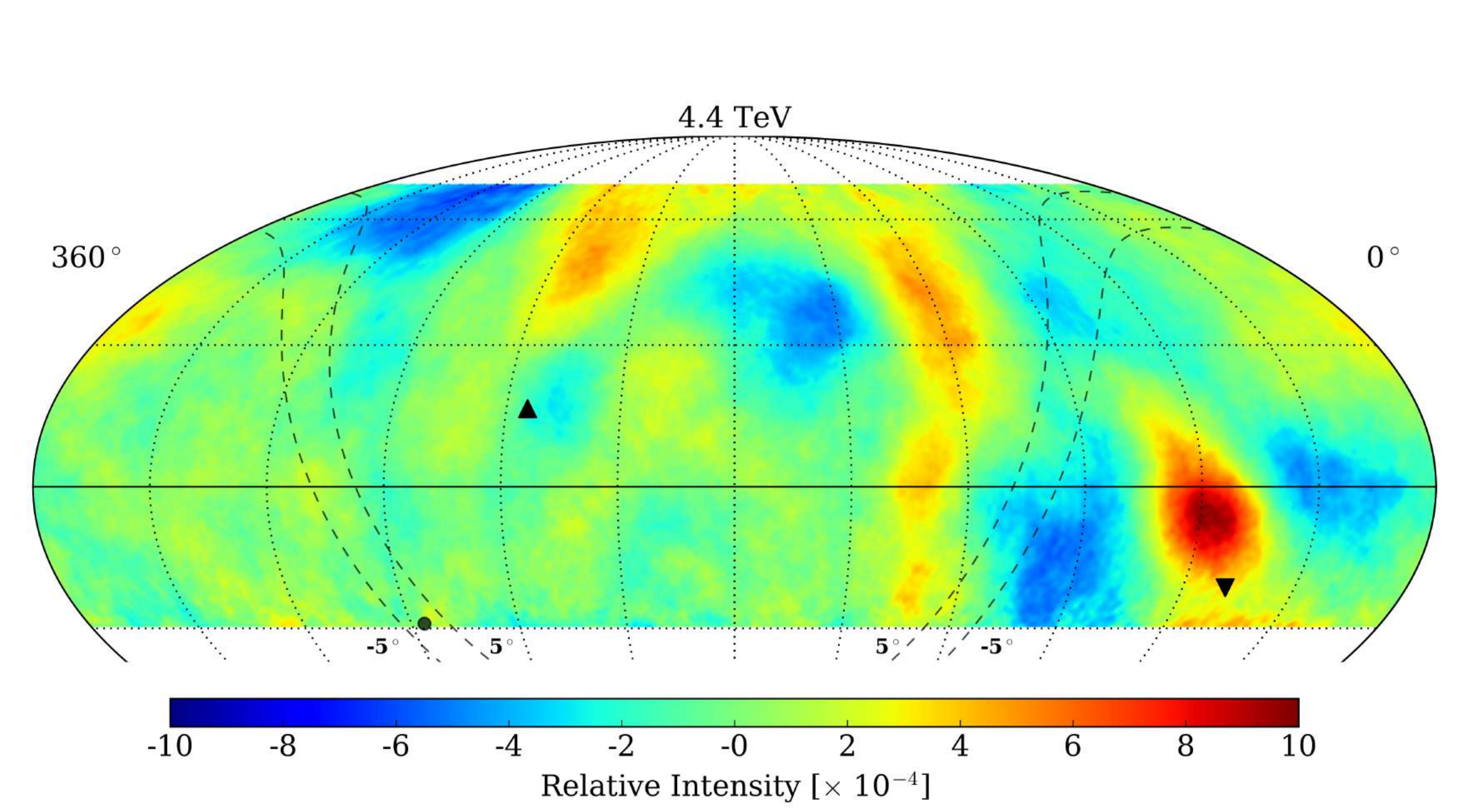}
     \includegraphics[width=0.4\textwidth,trim={0 2.7cm 0 0},clip]{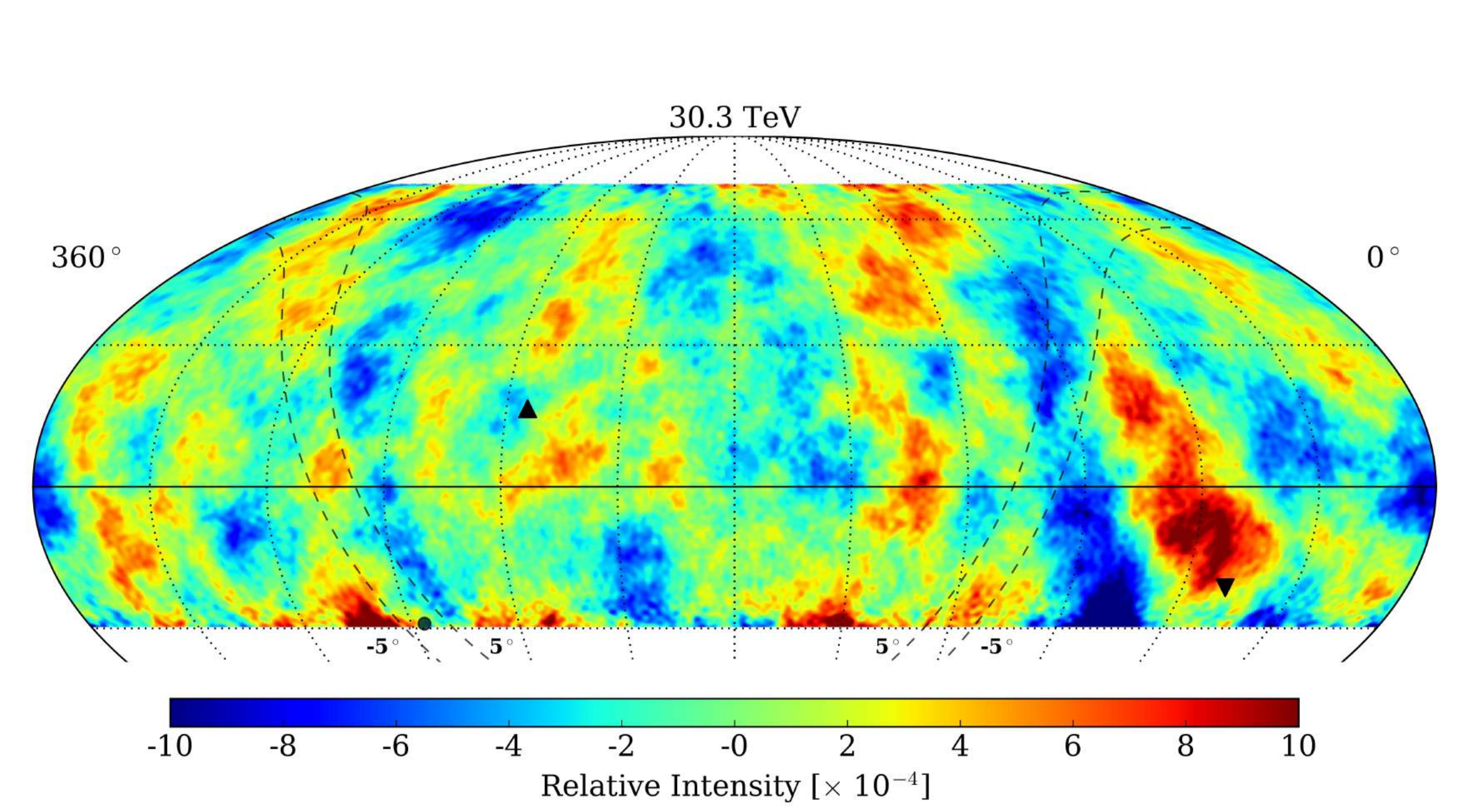}
     \includegraphics[width=0.4\textwidth,trim={0 2.7cm 0 0},clip]{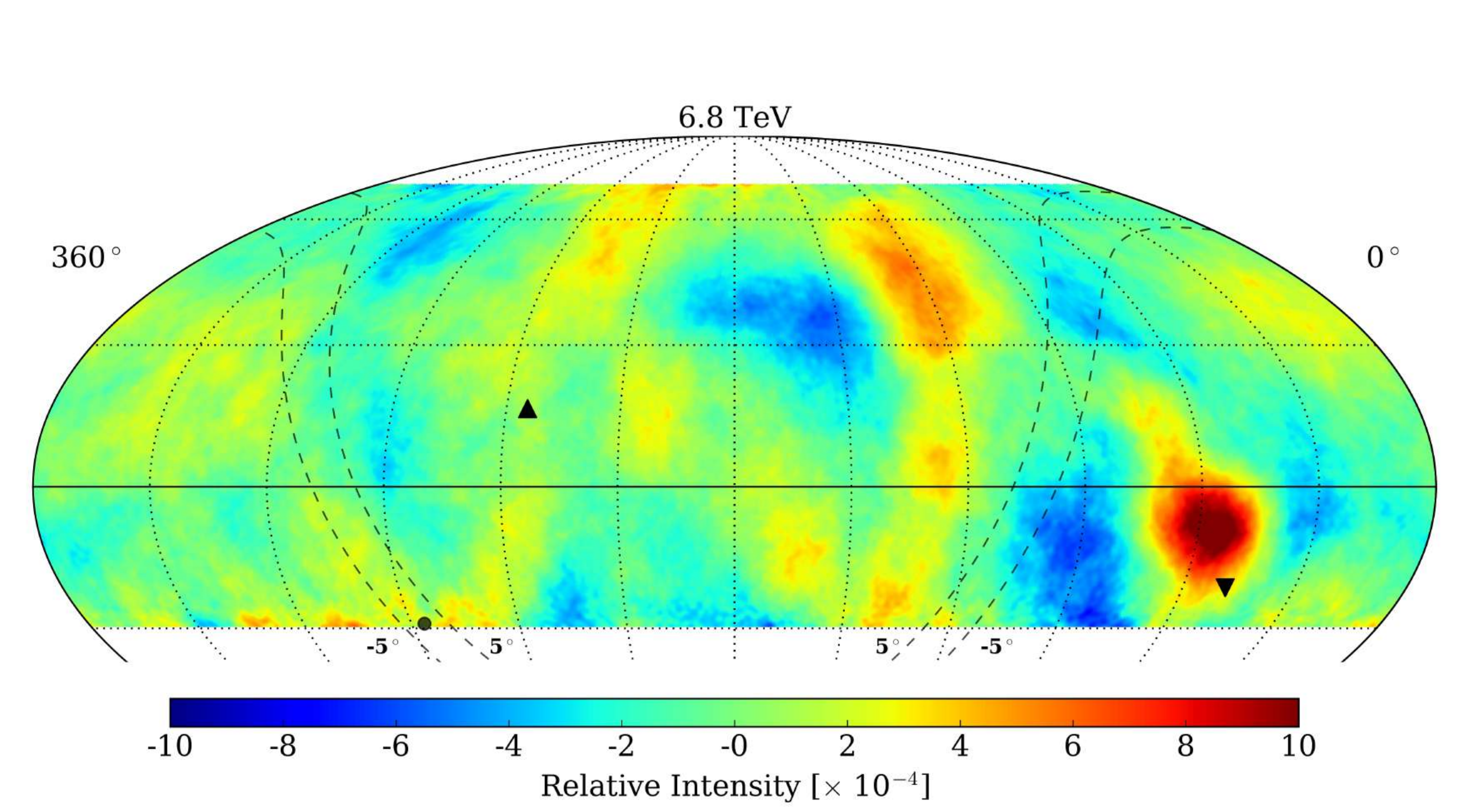}
     \includegraphics[width=0.4\textwidth,trim={0 2.7cm 0 0},clip]{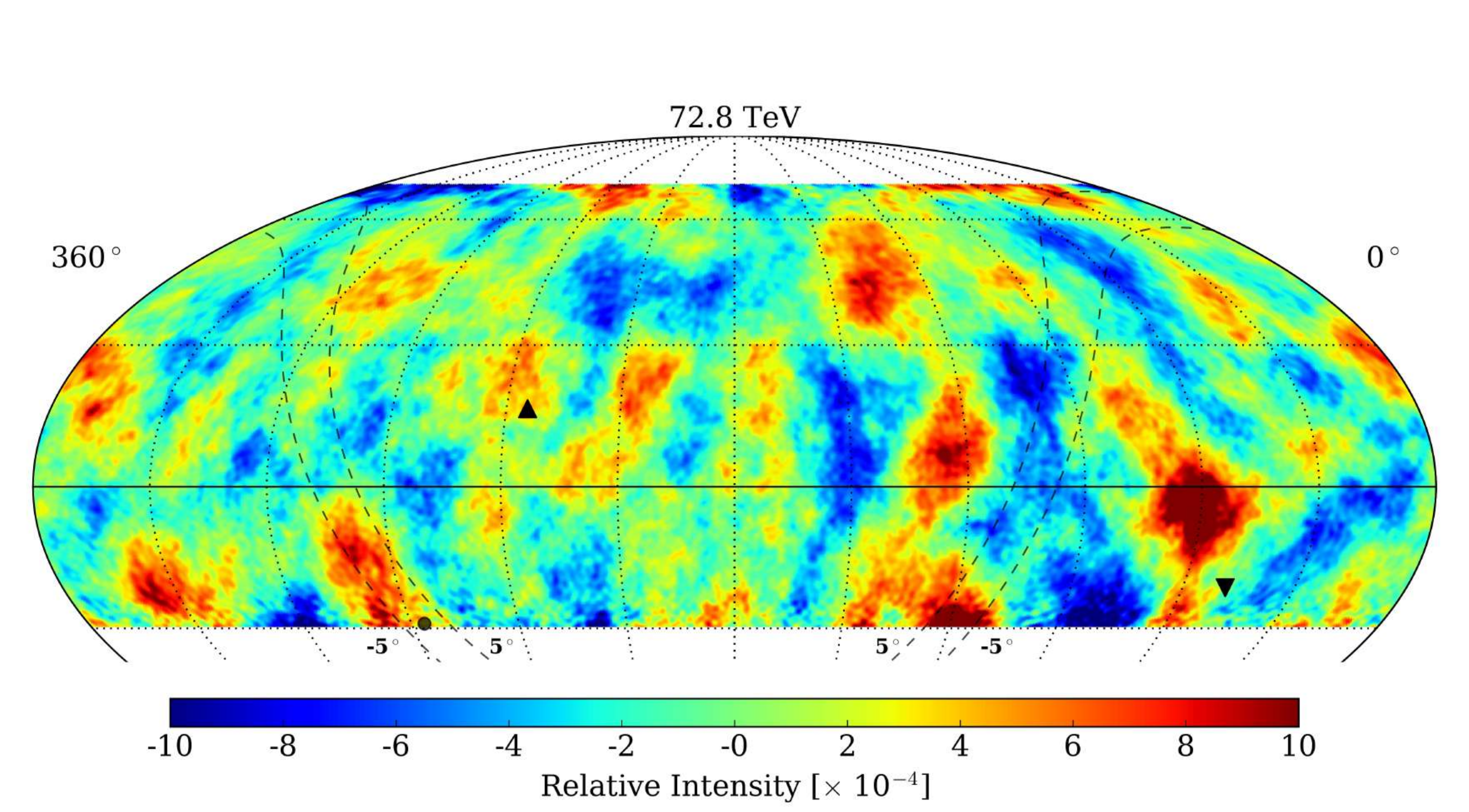}
\includegraphics[width=0.85\textwidth,trim={0 0 0 13.475cm},clip]{bin8_ell3_N256_S10p000_multipole_fit_relint.pdf}
  \end{center}
    \caption{
             Small-scale anisotropy maps in differential relative intensity $\delta I$ (smoothed $10^\circ$) for 
             eight\deleted{independent} energy bins separated by a likelihood-based reconstructed energy variable.
             A multipole fit to the moments with $\ell\leq3$ has been removed. 
             \added{The two triangle markers indicate the positive (upward-pointing) and negative (downward-pointing)
             directions of the local interstellar magnetic field $\vec{B}_{\mathrm{LIMF}}$ as inferred from
             Interstellar Boundary Explorer (IBEX) observations \citep{Zirnstein_2016}.}
             The Galactic Plane is shown with lines at $+5^\circ$ and $-5^\circ$ in Galactic latitude,
             \added{and the Galactic Center is demarcated by the solid circle}.
             }
  \label{fig:relint_ssa}
\end{figure*}

IceCube also shows a slight discrepancy in the amplitude scale of the anisotropy with other measurements, including HAWC. 
It is possible that differences in chemical compositions at detector level are the cause, as according to simulations
IceCube measures a higher-rigidity composition than IceTop \citep{Aartsen:2016ivj} and potentially other ground-based air shower detectors.
While higher-rigidity particles follow field lines more closely, an anisotropy signal
from many parsecs away might be distorted and diminished by the nearby magnetic fields of the Earth and the Sun. 
For reference, using the simulated composition described in \citep{crpaper}, we find that the fraction of events passing the selection 
criteria from proton and helium primaries decreases with energy from 91\% at 3 TeV to 78\% at 10 TeV, and to 69\% at 100 TeV.

\begin{figure*}[t]

  \begin{center}
     \includegraphics[width=0.4\textwidth,trim={0 2.7cm 0 0},clip]{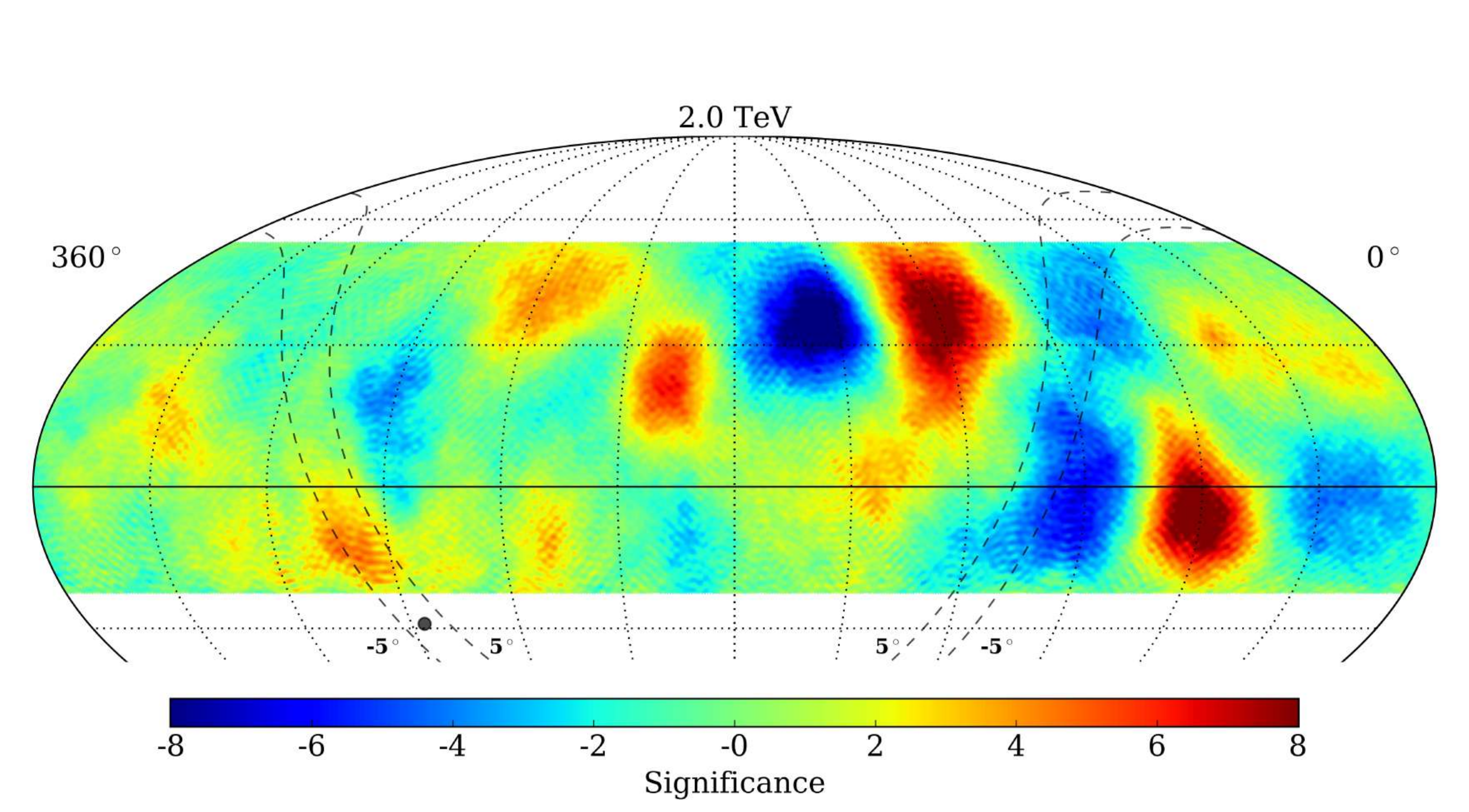}
     \includegraphics[width=0.4\textwidth,trim={0 2.7cm 0 0},clip]{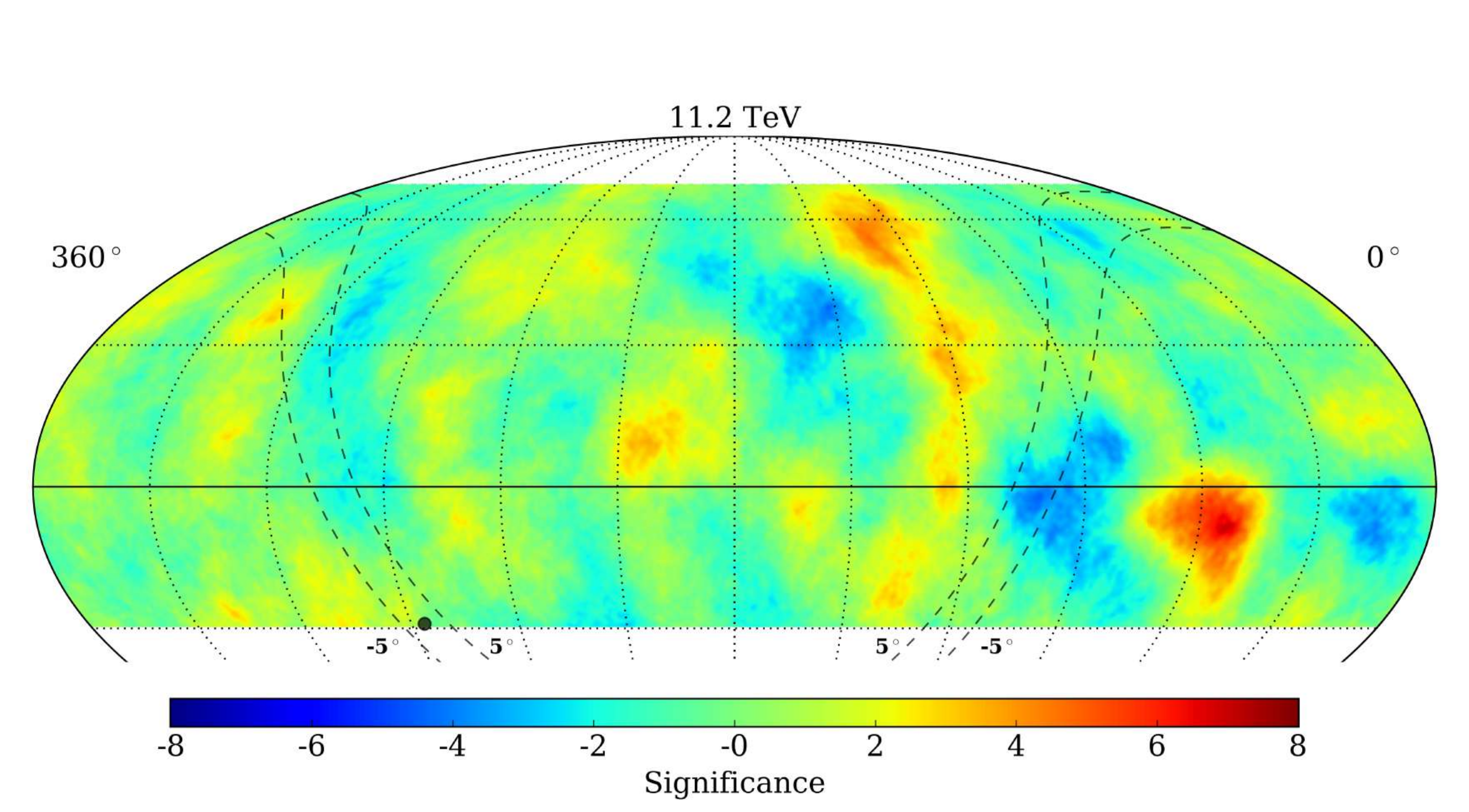}
     \includegraphics[width=0.4\textwidth,trim={0 2.7cm 0 0},clip]{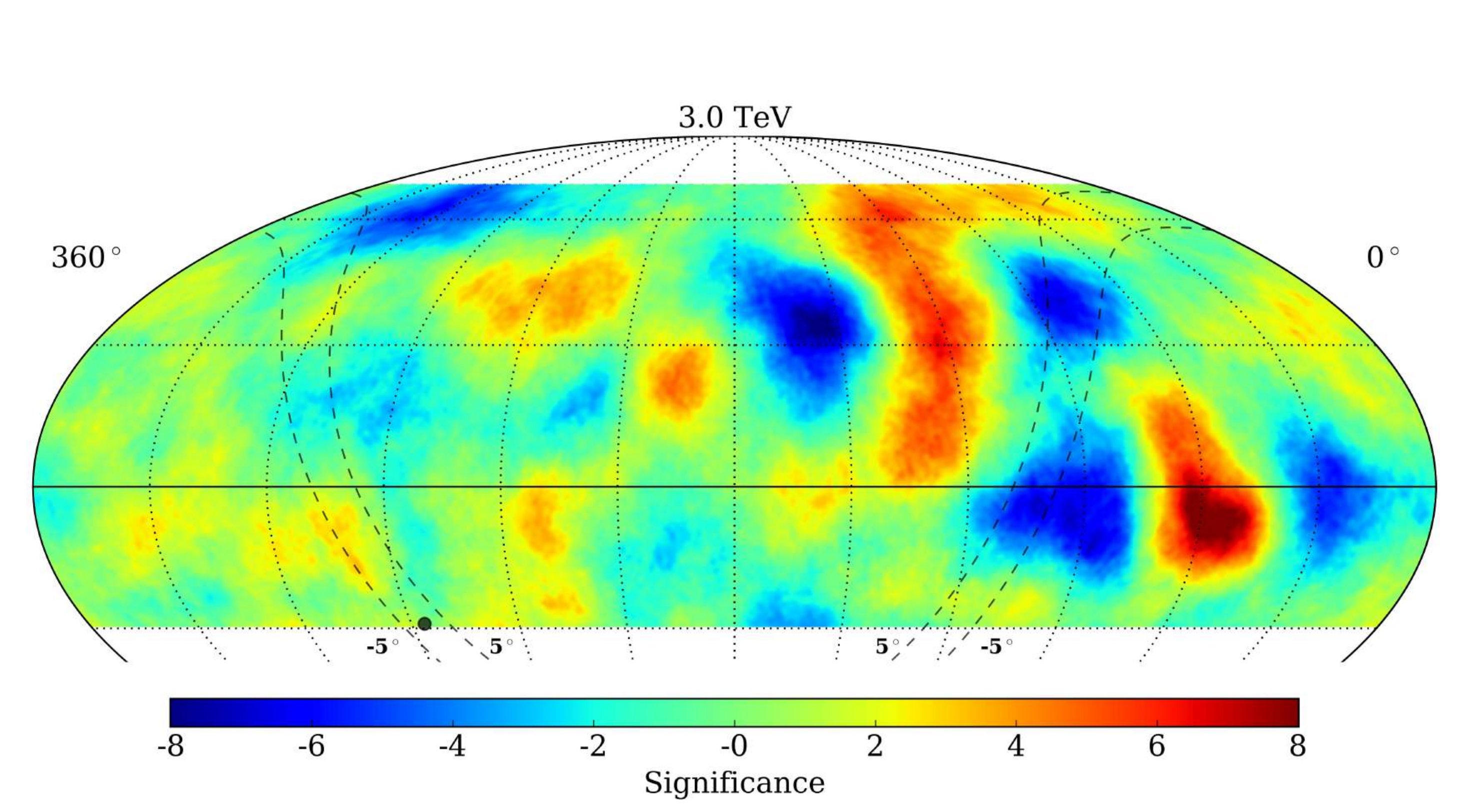}
     \includegraphics[width=0.4\textwidth,trim={0 2.7cm 0 0},clip]{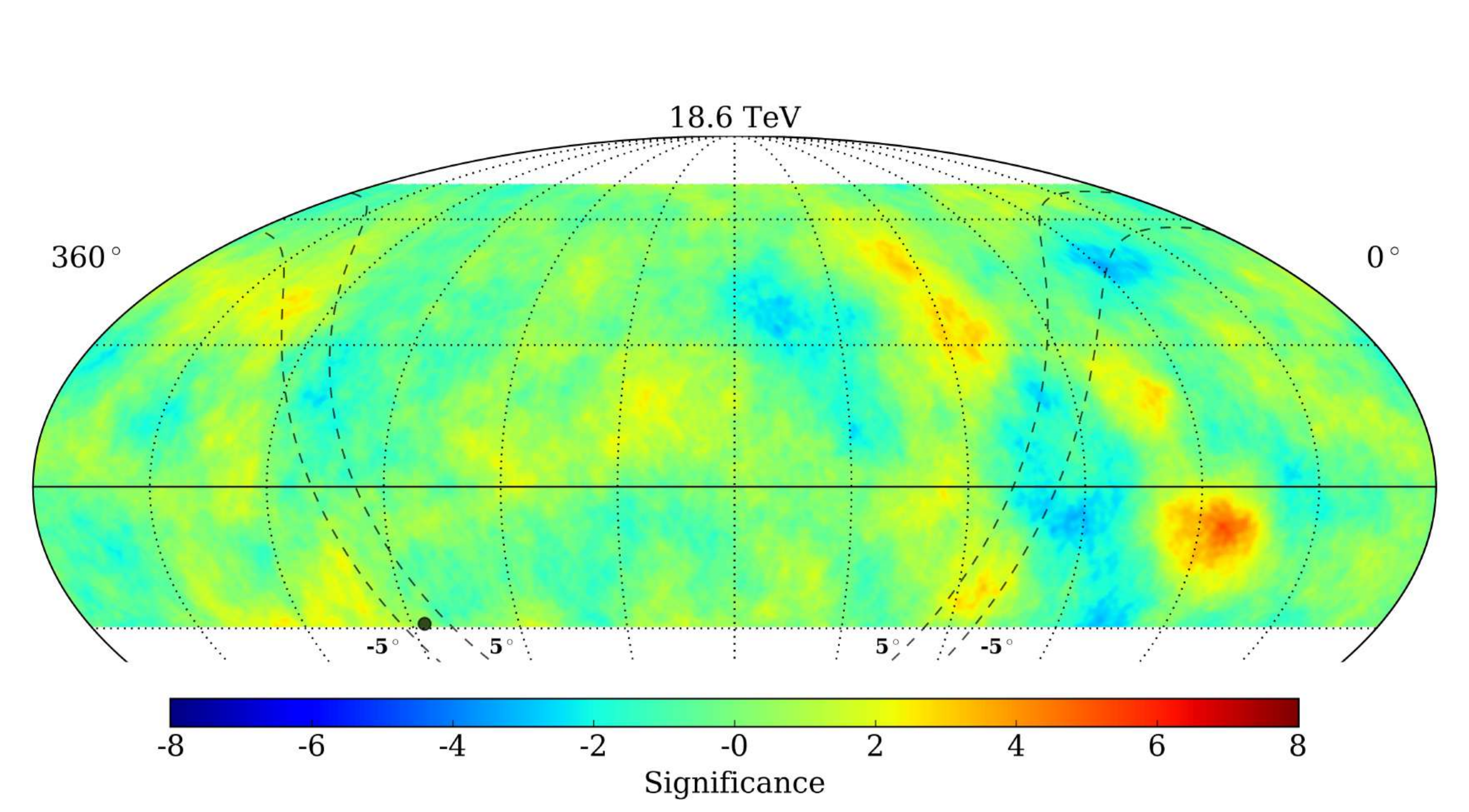}
     \includegraphics[width=0.4\textwidth,trim={0 2.7cm 0 0},clip]{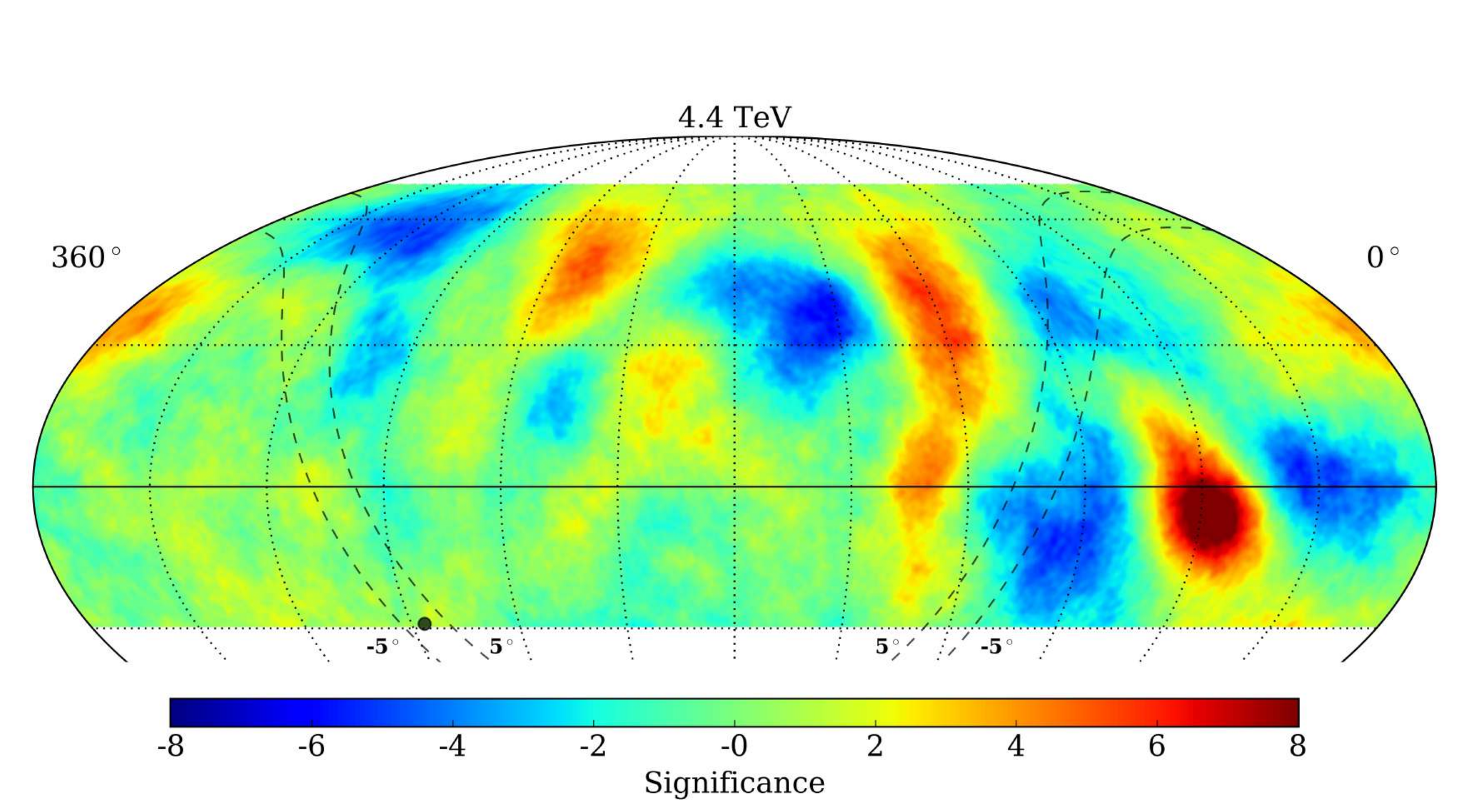}
     \includegraphics[width=0.4\textwidth,trim={0 2.7cm 0 0},clip]{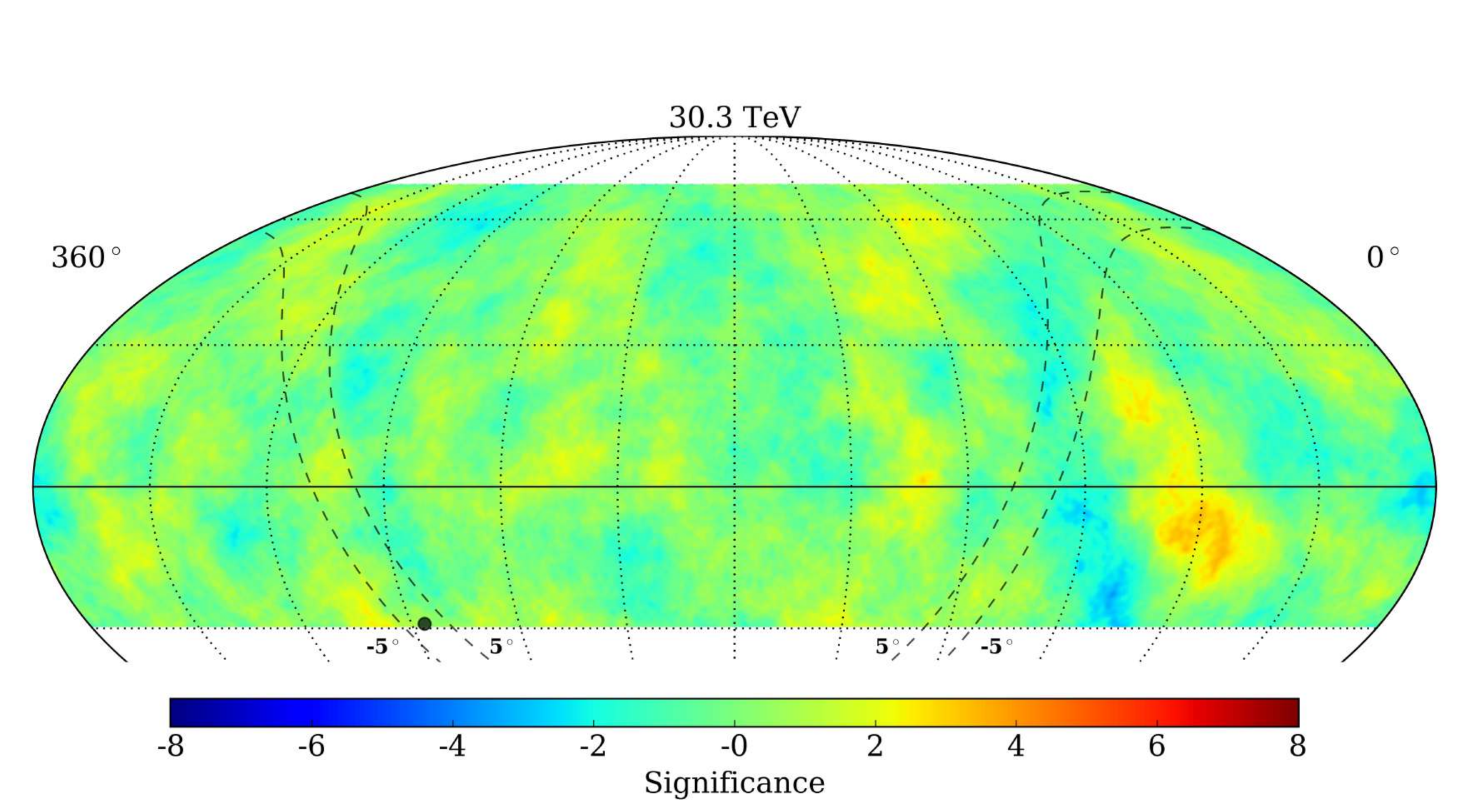}
     \includegraphics[width=0.4\textwidth,trim={0 2.7cm 0 0},clip]{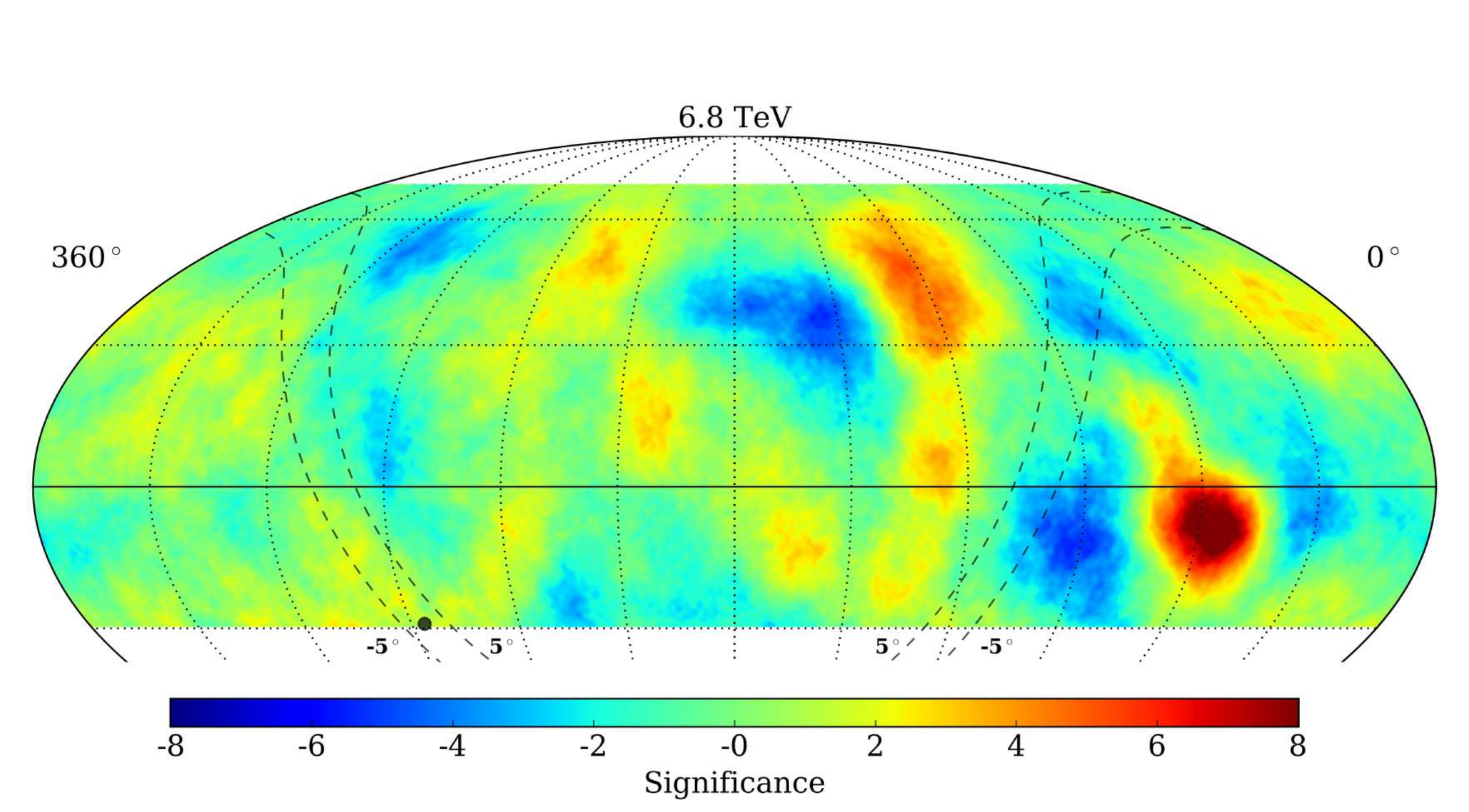}
     \includegraphics[width=0.4\textwidth,trim={0 2.7cm 0 0},clip]{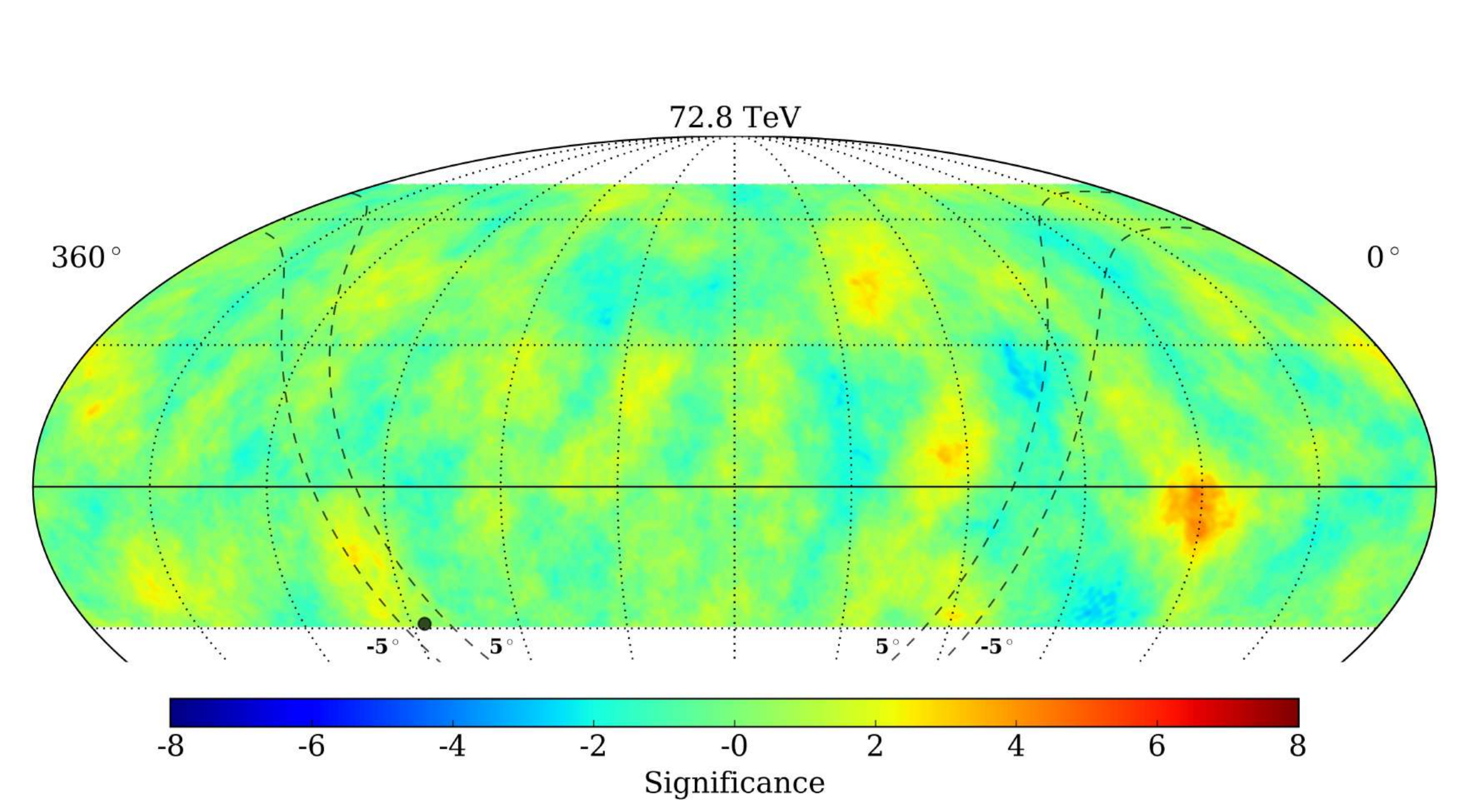}
     \includegraphics[width=0.85\textwidth,trim={0 0 0 13.475cm},clip]{bin8_ell3_N256_S10p000_multipole_fit_sig_corr.pdf}

  \end{center}
  \caption{
           Small-scale anisotropy maps in statistical significance (smoothed $10^\circ$) for 
           eight\deleted{independent} energy bins separated by a likelihood-based reconstructed energy variable.
           A multipole fit to the moments with $\ell\leq3$ has been removed.
           }
  \label{fig:signif_ssa}
\end{figure*}

In Figure~\ref{fig:dipolefits} only the HAWC measurement shown by the green squares uses a two-dimensional 
fit as described in Section~\ref{subsec:fit}.
All previous measurements represent fits of the dipole component to a Fourier series after projection of 
the relative intensity sky map onto a single declination band.
To match the method presented by the other experimental results, we also include one-dimensional fit dipole parameters (purple squares),
being between $75$--$85\%$ of the two-dimensional fit values.
This suggests that previous dipole amplitudes also are underestimated,
primarily affecting results from detectors with larger integrated fields-of-view (\textit{e.g.} HAWC, ARGO-YBJ, and Tibet)
as compared to \replaced{others}{those} with more limited fields-of-view such as IceCube.

\begin{figure*}[t]

  \begin{center}
     \includegraphics[width=0.475\textwidth,trim={0 0 0 2.75cm},clip]{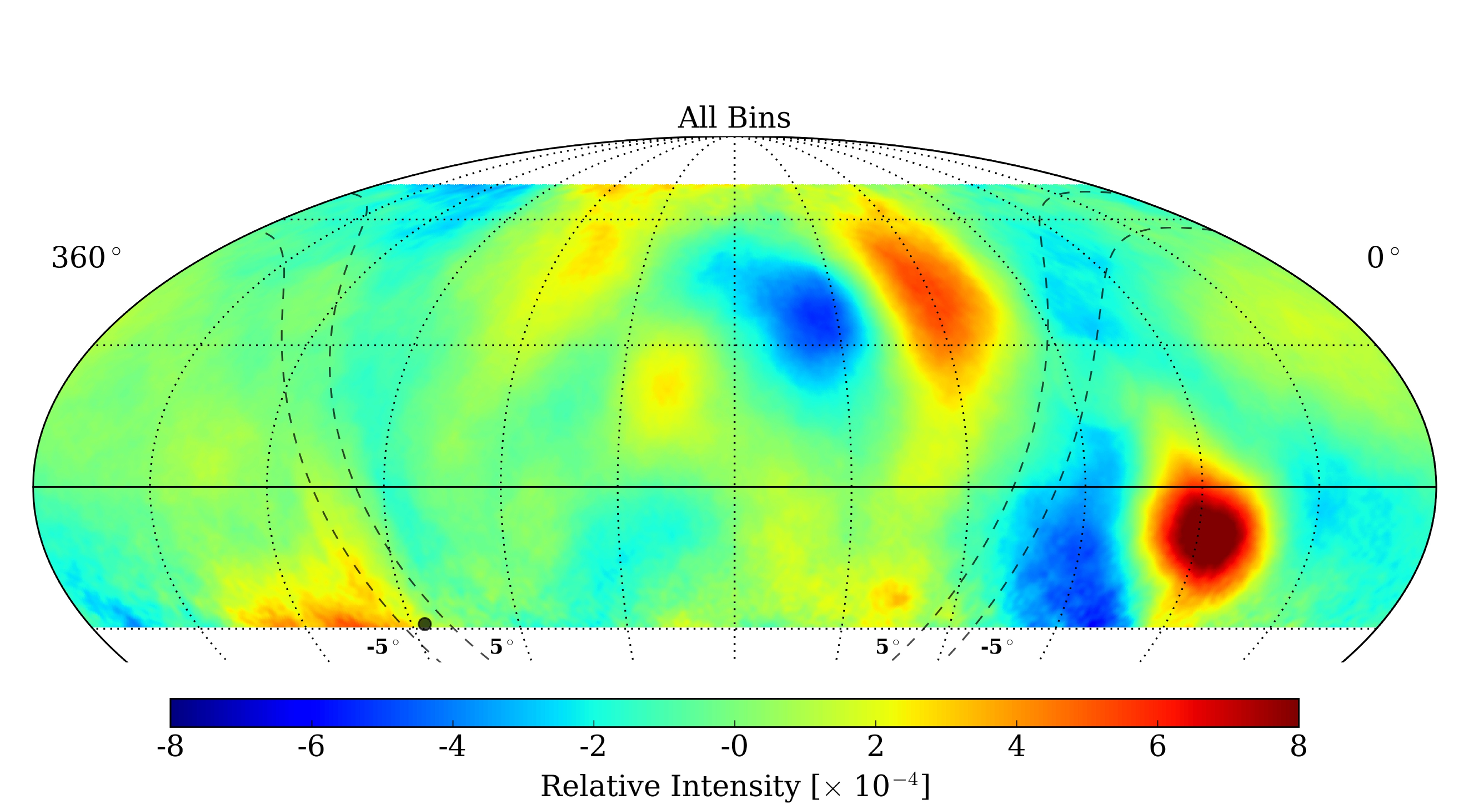}
     \includegraphics[width=0.475\textwidth,trim={0 0 0 3.5cm},clip]{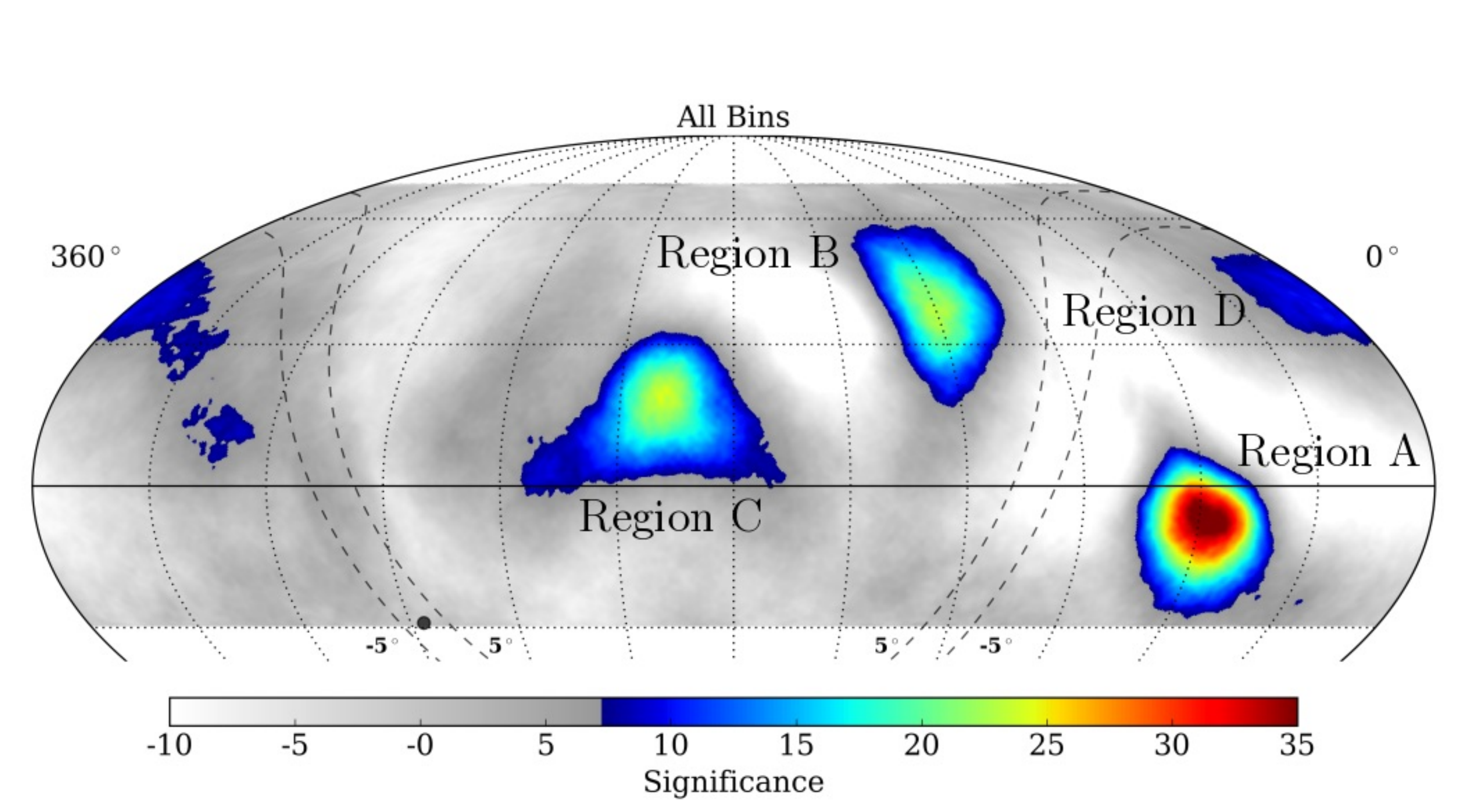}
  \end{center}
  \caption{
           Relative intensity (left) and significance (right) maps for all analysis bins combined, showing
           the locations of the most significant excesses, Regions A, B, C, and \added{the new Region} D.
           The estimated median energy of the combined map is $2.6 \left(^{-2.0}_{+9.9} \right)$ TeV,
           where the limits represent 68\% containment.
           }
  \label{fig:allbins_combined}
\end{figure*}

\begin{figure*}[!tb]

  \begin{center}
     \includegraphics[width=0.325\textwidth,trim={0.5cm 0.25cm 0.5cm 0},clip]{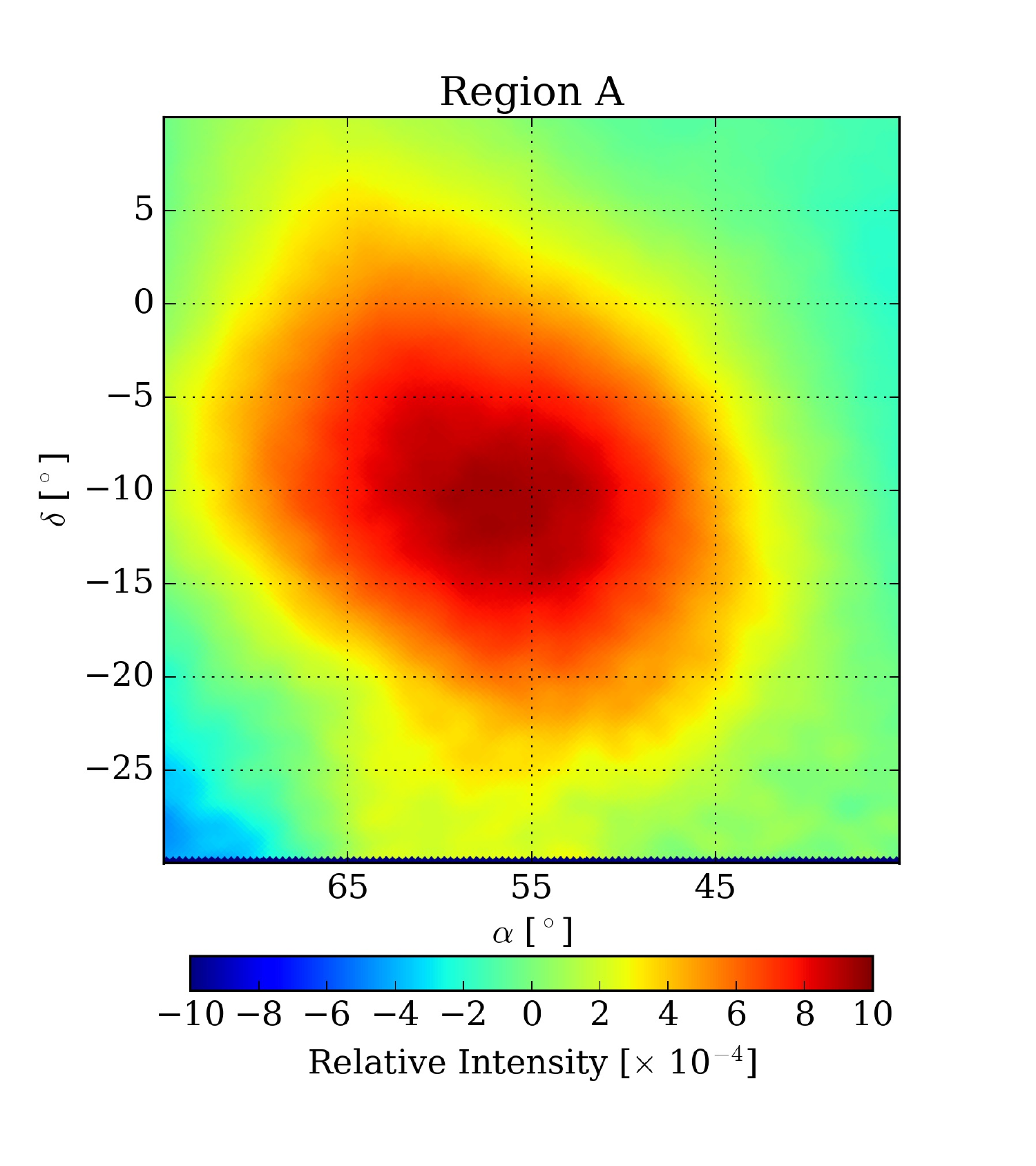}
     \includegraphics[width=0.325\textwidth,trim={0.5cm 0.25cm 0.5cm 0},clip]{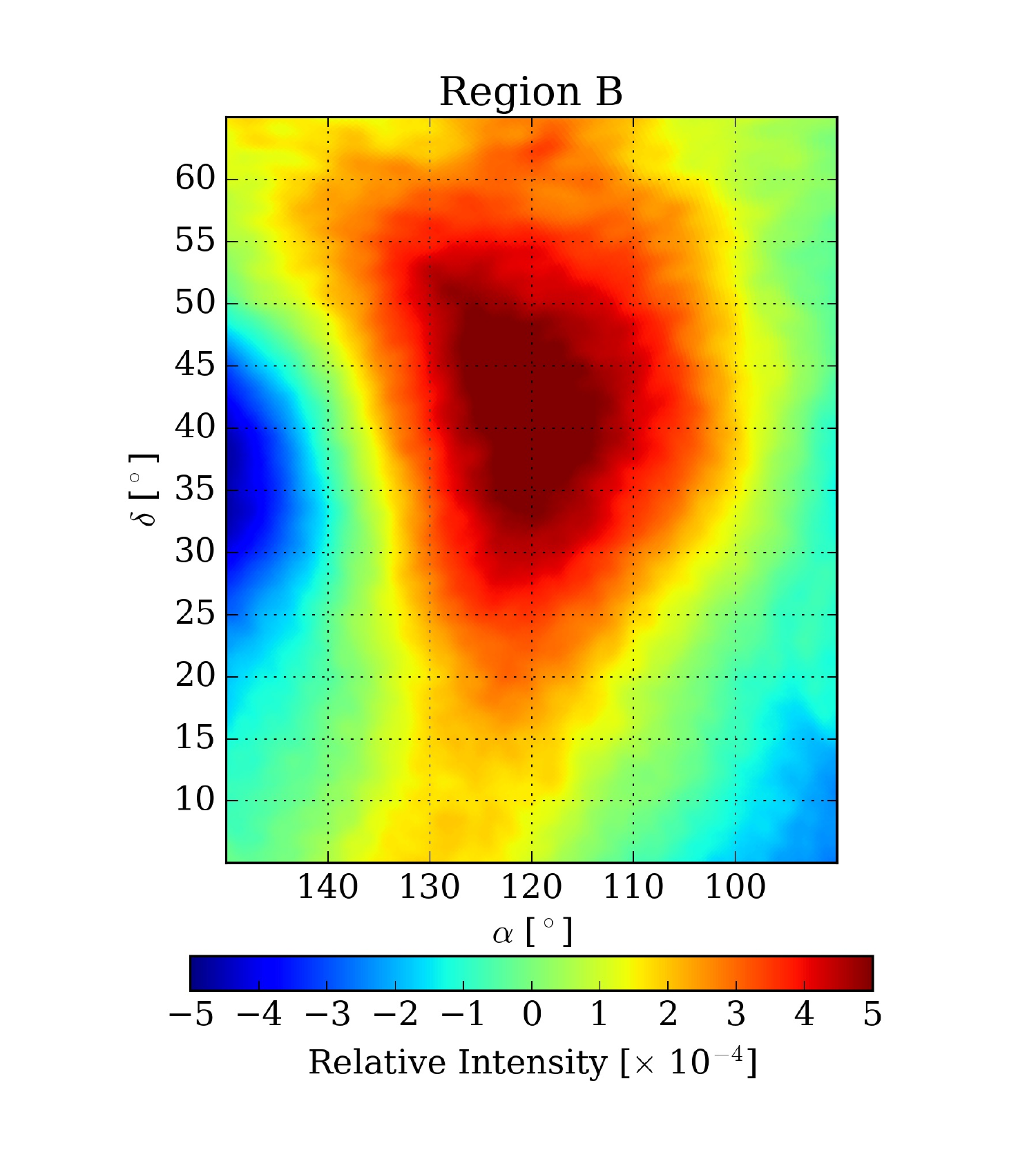}
     \includegraphics[width=0.325\textwidth,trim={0.5cm 0.25cm 0.5cm 0},clip]{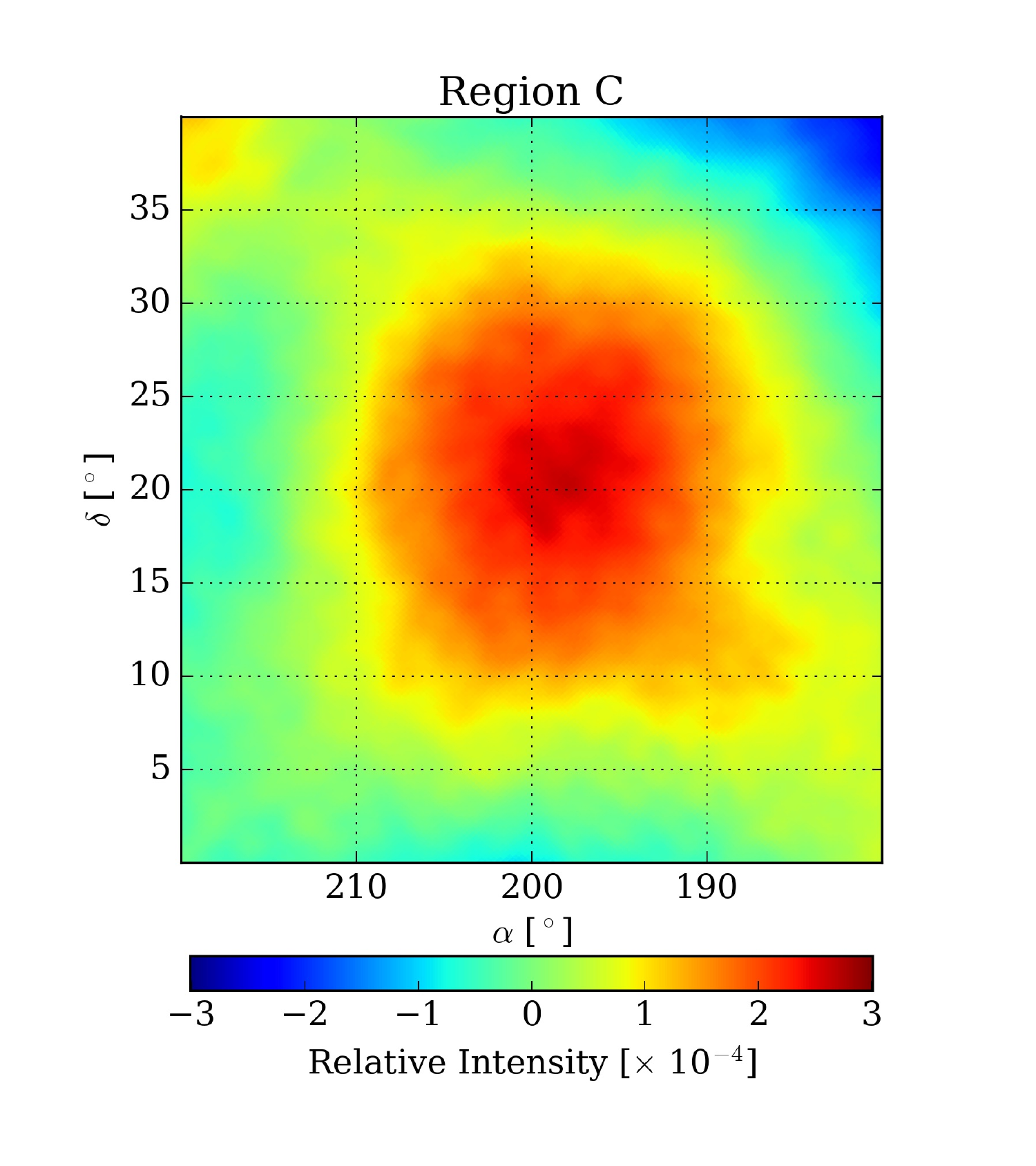}
     \includegraphics[width=0.325\textwidth,trim={0.5cm 0.25cm 0.5cm 1.69cm},clip]{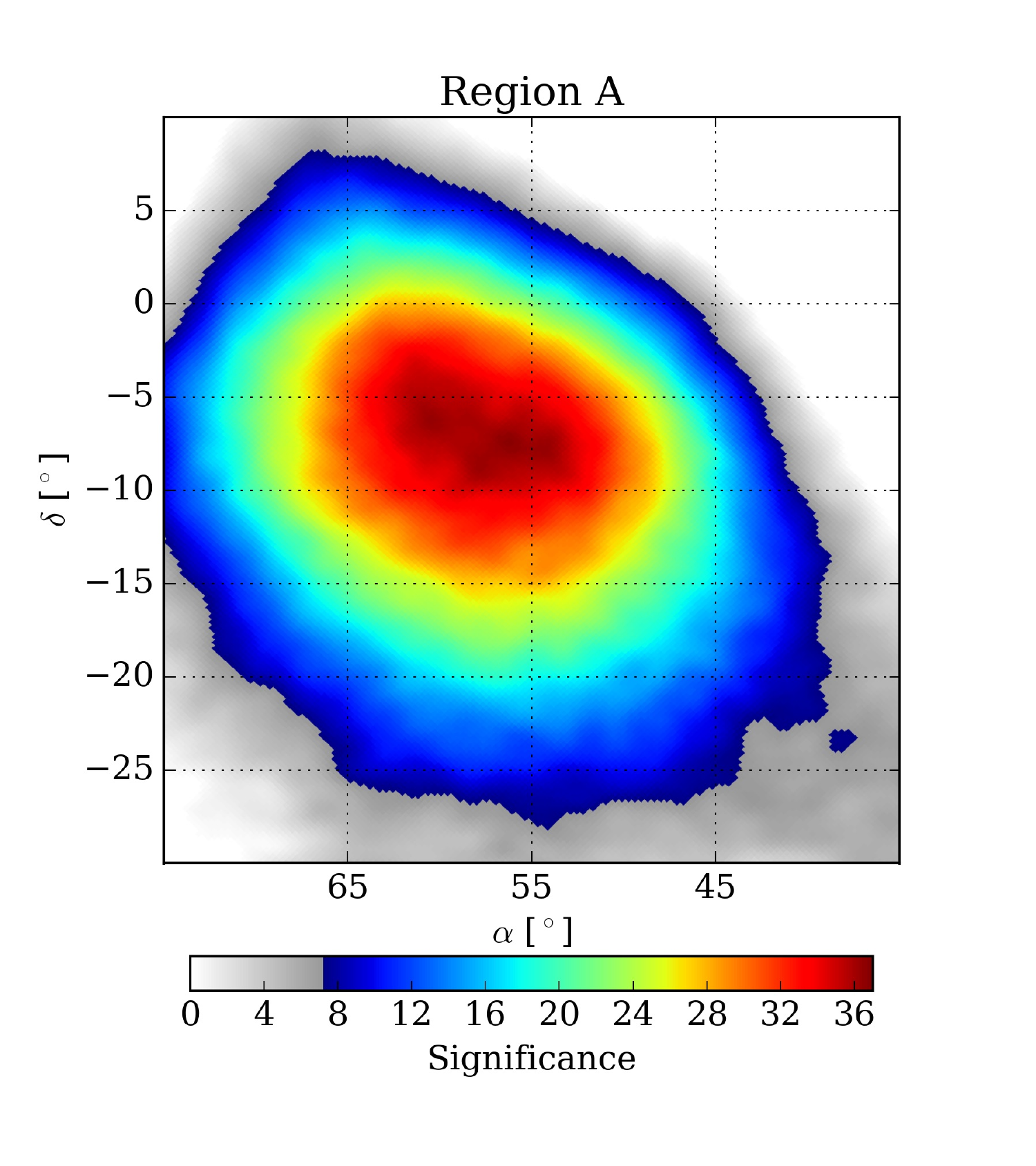}
     \includegraphics[width=0.325\textwidth,trim={0.5cm 0.25cm 0.5cm 1.69cm},clip]{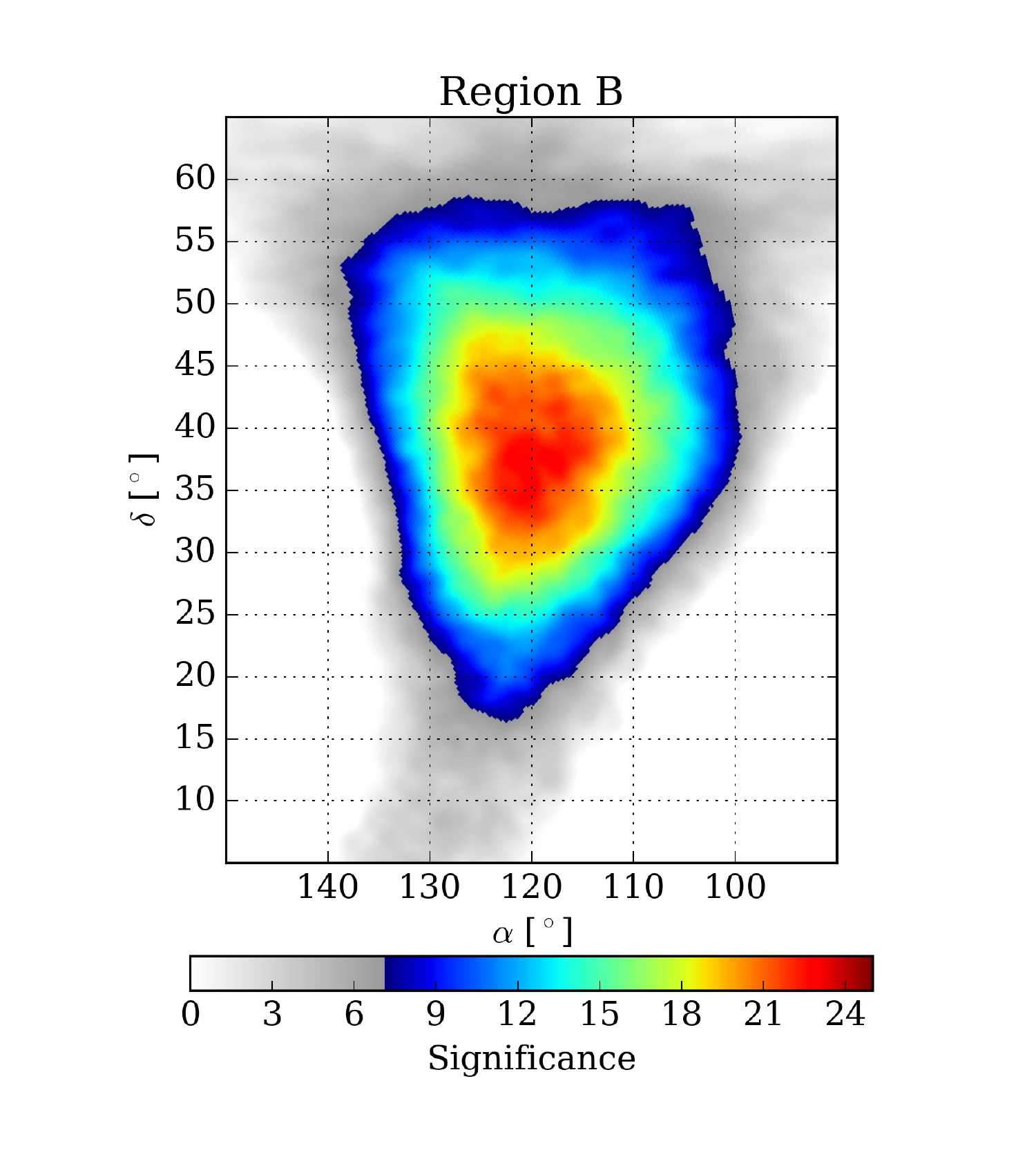}
     \includegraphics[width=0.325\textwidth,trim={0.5cm 0.25cm 0.5cm 1.69cm},clip]{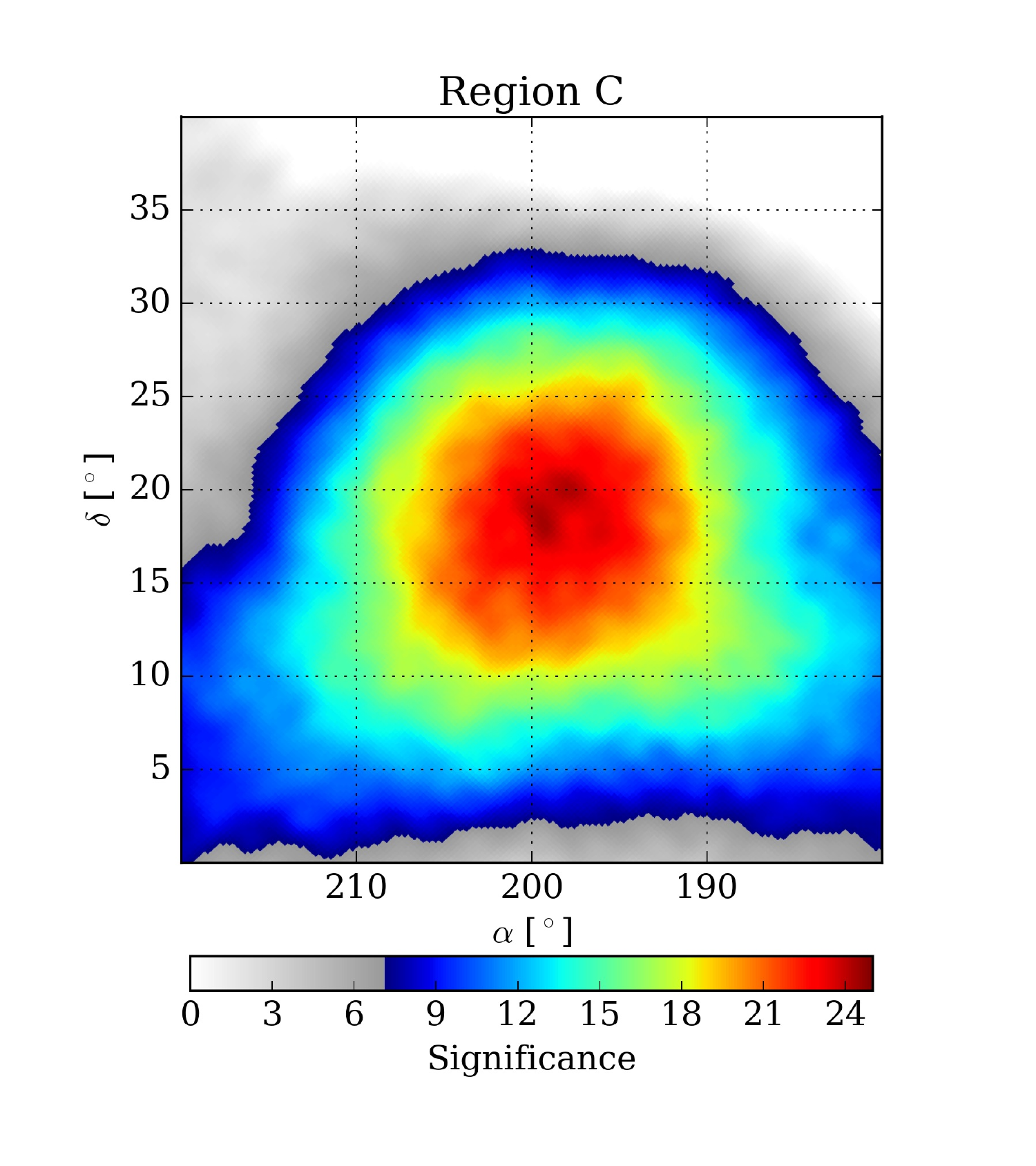}
  \end{center}
  \caption{
           Localized views of the relative intensity (top row) and significance (bottom row) of 
           Regions A (left), B (center), and C (right) having combined all energy bins into a single map.
           \replaced{The maximally significant points found for each are shown in Table~\ref{tab:ssa_max_sig}}
           {The coordinates of the maximally significant pixels found for each region are presented in Table~\ref{tab:ssa_max_sig}}.
           \added{The scales for the relative intensity and significance are different for each region.}
           }
  \label{fig:ssa_regions_abc}
\end{figure*}

\begin{figure*}[t]

  \begin{center}
     \includegraphics[width=0.49\textwidth,trim={2cm 0.75cm 2cm 1.69cm},clip]{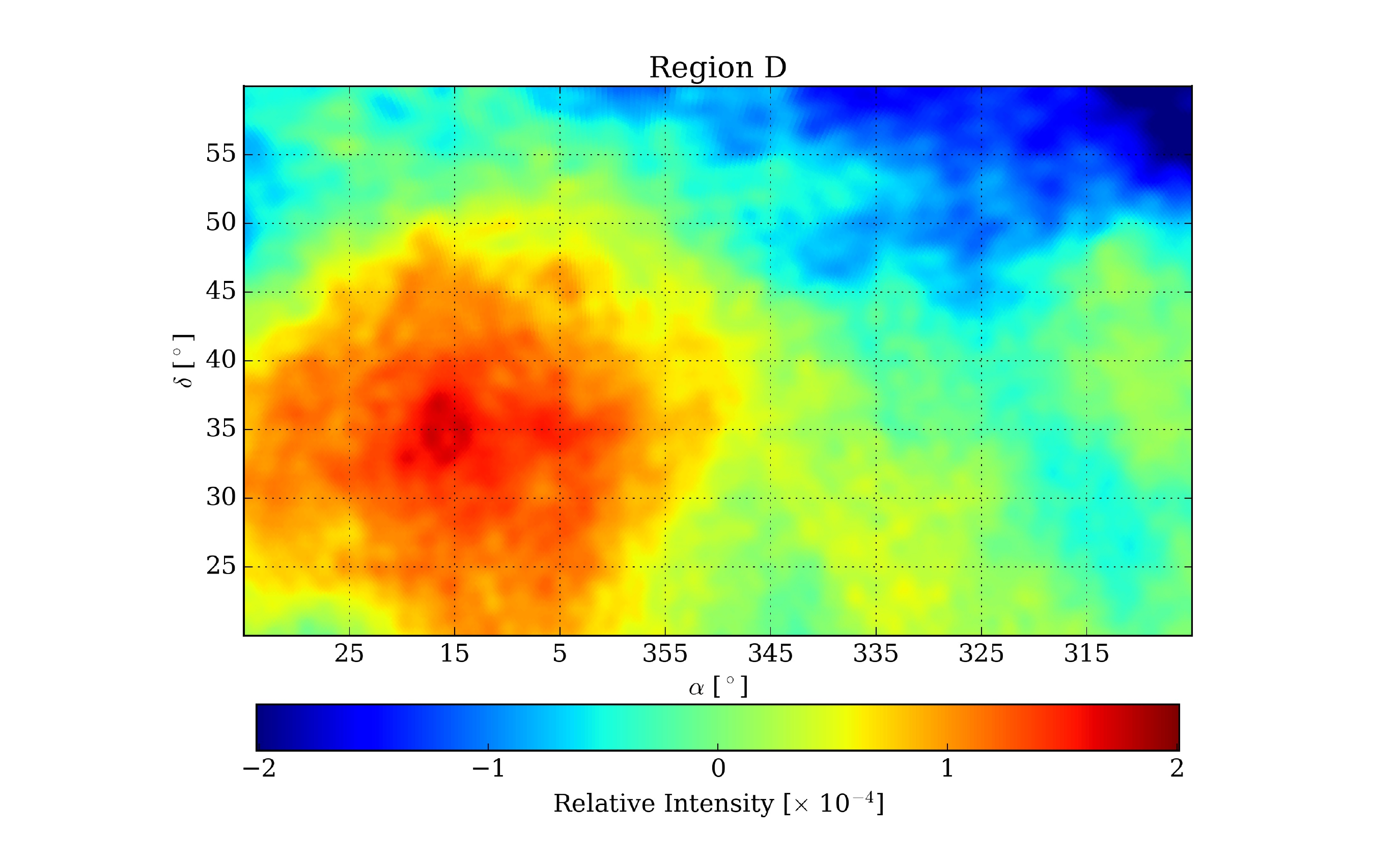}
     \includegraphics[width=0.49\textwidth,trim={2cm 0.75cm 2cm 1.69cm},clip]{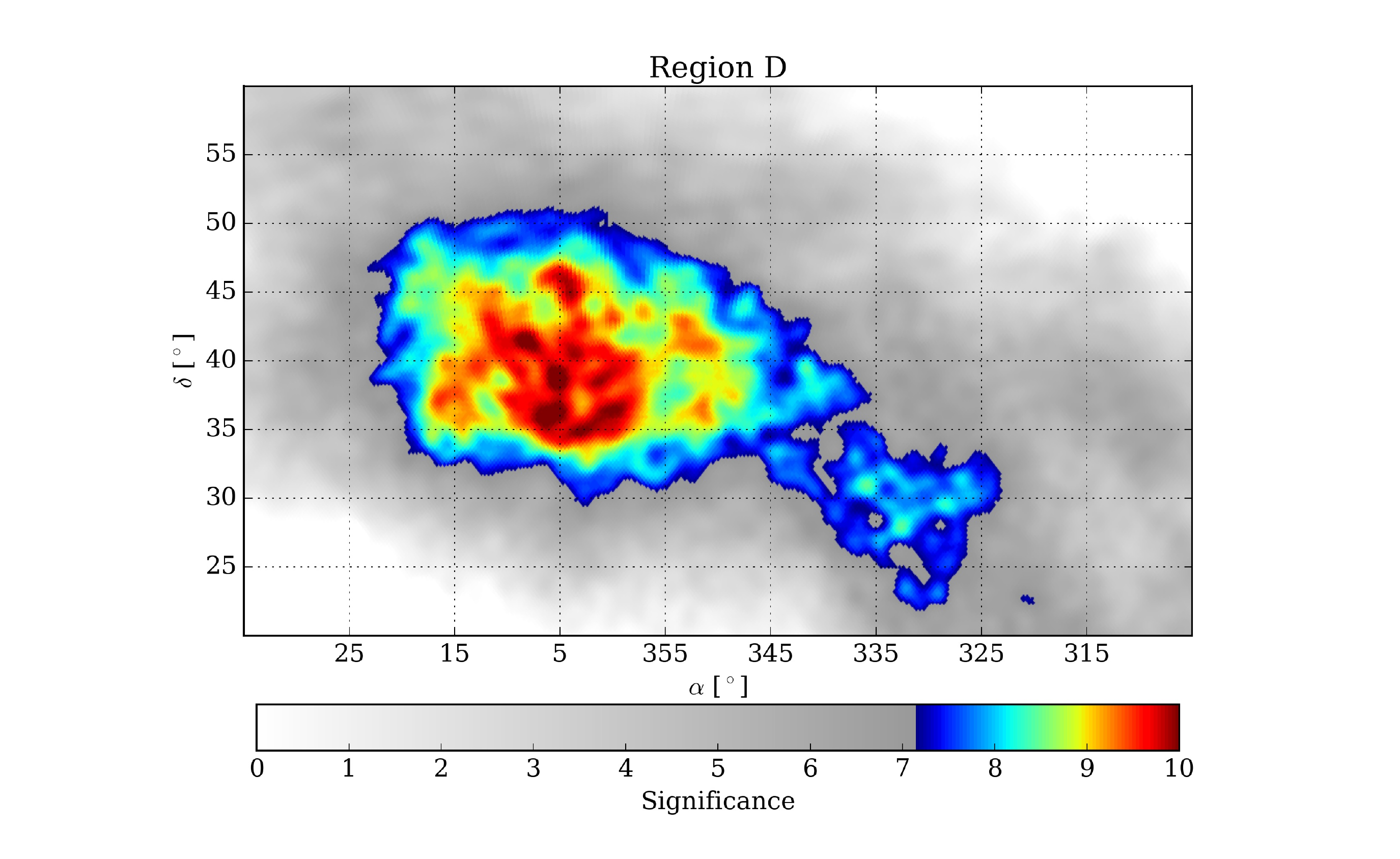}
  \end{center}
  \caption{
           Localized views of the relative intensity (left) and significance (right) of 
           Region D having combined all energy bins into a single map.
           \replaced{The maximally significant points found for each are shown in Table~\ref{tab:ssa_max_sig}}
           {The coordinates of the most significant pixel is presented in Table~\ref{tab:ssa_max_sig}}.
           }
  \label{fig:ssa_region_d}
\end{figure*}

In addition to the large-scale structure, the sky maps have significant angular power at small angular scales ($\ell\ge4$) 
as shown in the angular power spectra, and in the relative intensity and significance maps after subtraction of the $\ell\leq3$ fit multipoles,
shown in Figures \ref{fig:relint_ssa} and \ref{fig:signif_ssa}. 
The three most significant regions of excess previously observed with Milagro \citep{Abdo:2008kr} and HAWC \citep{Abeysekara:2014sna}
are Regions A, B, and C (referred to as Regions 1, 2, and 4 by ARGO-YBJ \citep{ARGO-YBJ:2013gya}).
These are\deleted{made} more apparent in the relative intensity and significance maps of Figure~\ref{fig:allbins_combined},
where all \replaced{energy bins}{events} have been combined into a single \replaced{bin}{map} having median energy 
of $2.6\left(^{-2.0}_{+9.9}\right)$ TeV to study each region's morphology.
The excess defined as Region 3 by ARGO-YBJ \citep{ARGO-YBJ:2013gya} has a maximum significance of $4.6~\sigma$ 
at $\alpha=251.2^{\circ}$, $\delta=19.5^{\circ}$ in the combined \added{HAWC} map, with a mean relative intensity for the entire region
of $(1.4 \pm 0.3)\times10^{-4}$, approximately 14\% greater than that measured by ARGO-YBJ.
Figure~\ref{fig:ssa_regions_abc} shows zoomed-in views of each\deleted{of the} regional \replaced{excesses}{excess} observed by HAWC.

Regions A and C are characterized as relatively symmetric excesses with extent of $\approx 30^{\circ}$,
as compared to Region B, which has a similar width in \replaced{$\alpha$}{right ascension} 
but is elongated by nearly a factor of two in \replaced{$\delta$}{declination}.
Similar morphology was observed for all three regions in the previous small-scale study by HAWC \citep{Abeysekara:2014sna}.
A significant feature previously not observed by HAWC nor ARGO-YBJ is labelled Region D in Figure~\ref{fig:allbins_combined}.
This new excess occupies about $25^{\circ}$ in declination while also being elongated by about a factor of two in right ascension,
though is considerably weaker than the other regions, having a maximum relative intensity of $(1.7\pm0.2)\times10^{-4}$.

Region A is the most prominent feature in $\delta I$ as seen in Figure~\ref{fig:relint_ssa}, being present at all energies,
while the shape of Region B is evident up to 18.6 TeV and that of Regions C and D are difficult to discern given the color scale.
The presence of these features in significance as a function of energy is shown in Figure~\ref{fig:signif_ssa}, 
with Regions A and B being significant up to 18.6 TeV, Region D peaking at 4.4 TeV and Region C being strongest below 4.4 TeV.

\begin{figure*}[!tb]

  \begin{center}
     \includegraphics[width=0.49\textwidth,trim={0 0 0 0},clip]{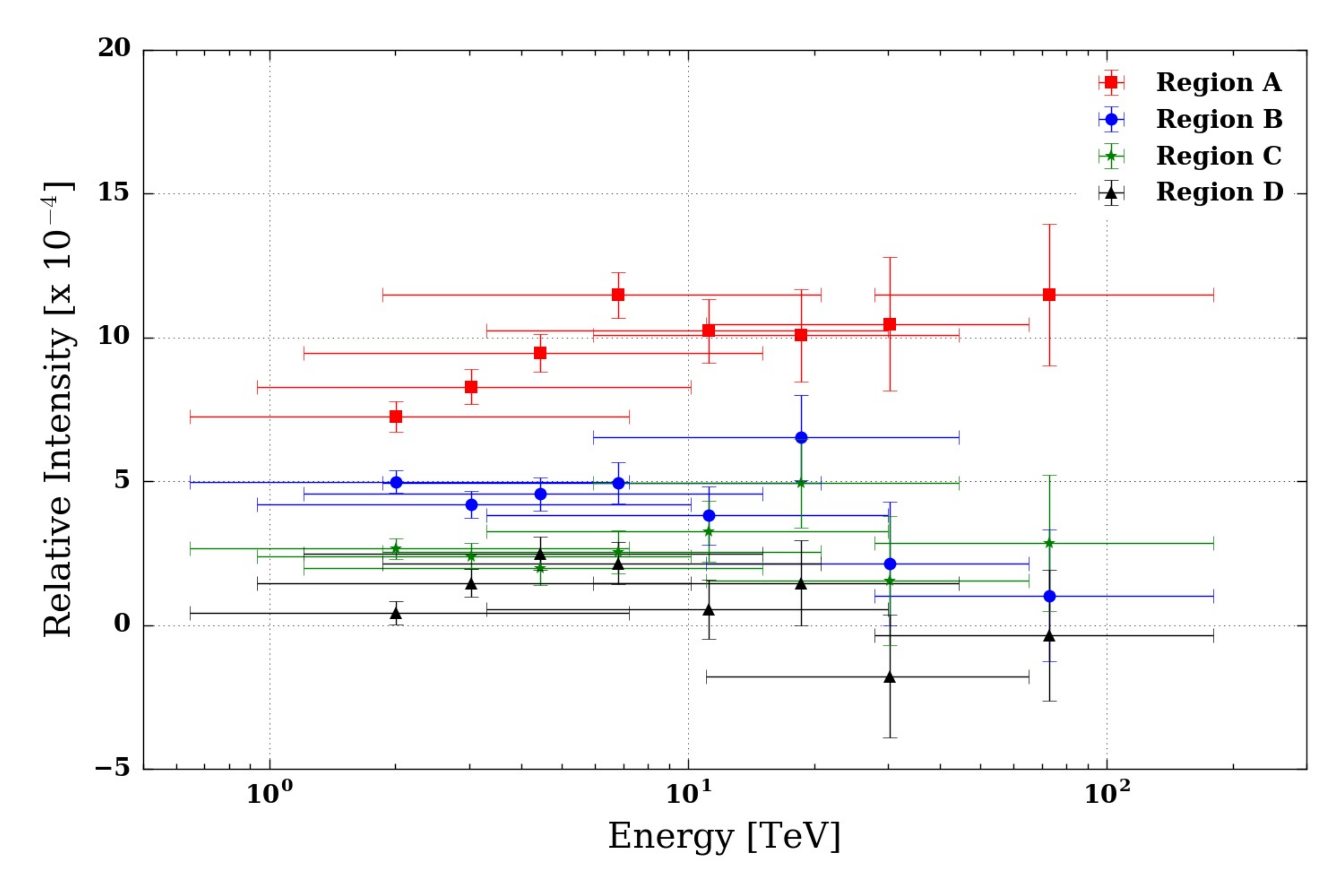}
     \includegraphics[width=0.49\textwidth,trim={0 0 0 0},clip]{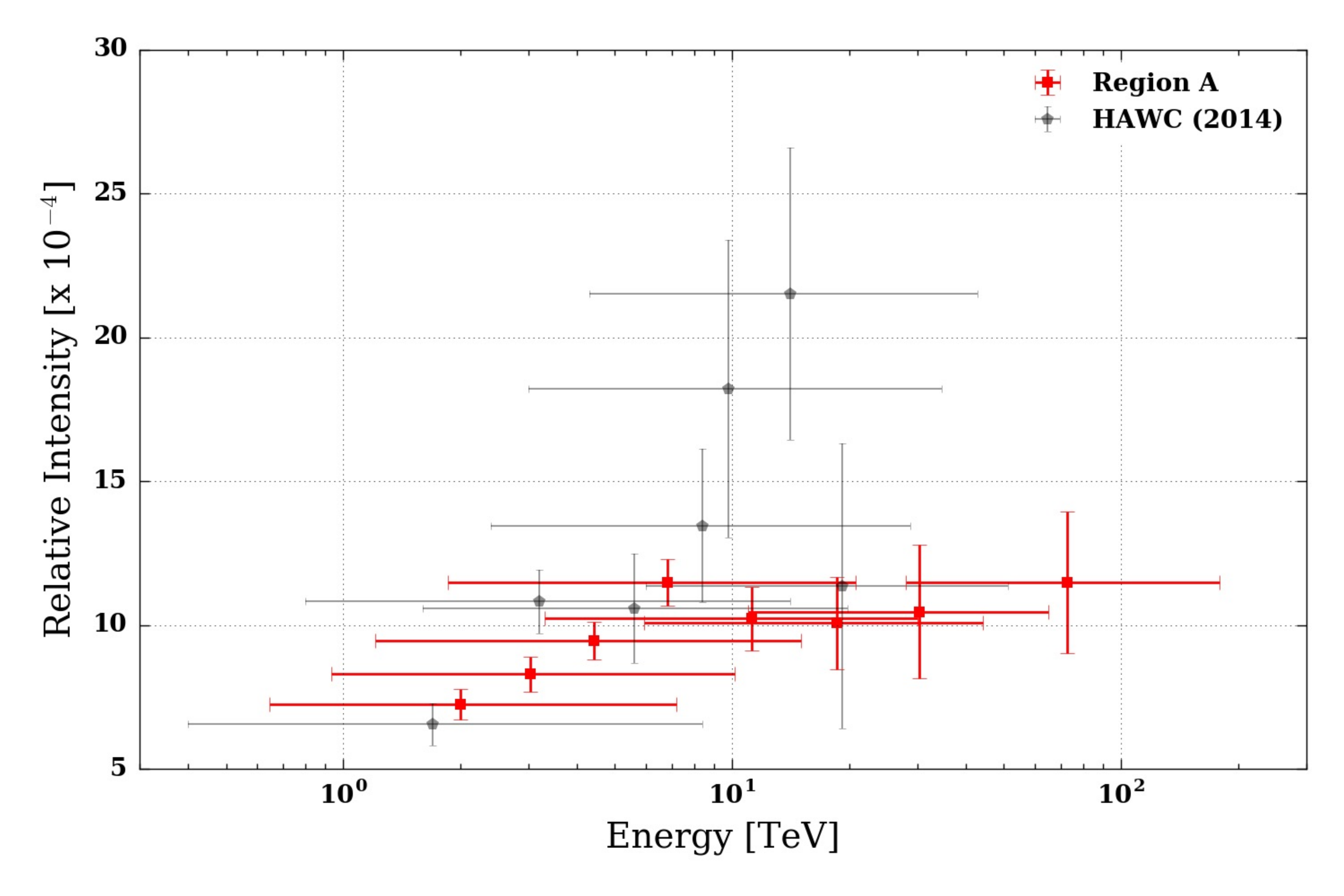}
  \end{center}
  \caption{
           Left: relative intensity spectra as a function of energy for Regions A, B, C, and D.
           Right: comparison of the relative intensity spectra for Region A from this study and the previous HAWC measurement \citep{Abeysekara:2014sna}.
           }
  \label{fig:ssa_spectra}
\end{figure*}

Table~\ref{tab:ssa_max_sig} presents for the four regions the locations in $(\alpha, \delta)$ 
of the peak significances ($\sigma_{\text{max}}$) and the corresponding $\delta I$ values found in the all-bins combined maps.
The relative intensity values are consistent with those found for Regions A, B and C at 
$\sigma_{\text{max}}$ in  \citep{Abeysekara:2014sna},
while the locations of $\sigma_{\text{max}}$ are shifted by \replaced{$O(5^{\circ})$}{$\sim 5^{\circ}$} 
from their previously identified values.
The relative intensity spectra as a function of energy extracted from the maps in Figure~\ref{fig:relint_ssa} at these locations 
are shown in Figure~\ref{fig:ssa_spectra}, including that of Region D.
Regions B and C are characterized by relatively flat spectra across the energy bins as compared to that of Region D,
which increases in $\delta I$ up to the $4.4$ TeV bin before decreasing, and Region A which has a positive slope up to $6.8$ TeV.
This is consistent with the previous spectral measurement of Region A by HAWC \citep{Abeysekara:2014sna} 
as shown in the right panel of Figure~\ref{fig:ssa_spectra}.
\replaced{, though due}{Due} to improved uncertainties from a data set an order of magnitude greater, the shape of the spectrum from this study is much
more constrained, especially for $E \geq 10$ TeV.
As was done in \citep{Abeysekara:2014sna}, we estimate the statistical significance of the spectral slope of Region A by 
comparing a linear fit of $\delta I(\log{E})$ to similar fits performed across the field of view.
The distribution of slopes across the sky (excluding points within $20^{\circ}$ of Regions A, B, C, and D) 
follows a Gaussian distribution with mean of zero and width $0.69\times10^{-4}$.
The slope fit at the location of Region A is $(1.66 \pm 0.46)\times10^{-4}$, falling $2.4~\sigma$ from the all-sky mean.

\added{
From Figures~\ref{fig:relint} and \ref{fig:relint_ssa} it appears that the large-scale structure 
as well as Region A are approximately oriented along the local 
interstellar magnetic field $\vec{B}_{\mathrm{LIMF}}$ inferred from Interstellar Boundary Explorer (IBEX) measurements.
In equatorial coordinates $(\alpha, \delta)$, this direction is $(234.43^{\circ} \pm 0.69^{\circ}, 16.3^{\circ} \pm 0.45^{\circ})$,
and $(48.5^{\circ} \pm 0.69^{\circ}, -21.2^{\circ} \pm 0.45^{\circ})$ for $-\vec{B}_{\mathrm{LIMF}}$ \citep{Zirnstein_2016}.
This alignment is consistent with local conditions playing a role in shaping the observed cosmic-ray anisotropy,
providing insight into the structure of the local interstellar medium and the heliospheric environment \citep{Desiati:2011xg, Schwadron:2014}.
}

\renewcommand{\arraystretch}{1.6}
\begin{center}
 \begin{table}[!htb]
   \caption{
            Locations and relative intensities of maximal significance found in the combined map for Regions A, B, C, and D. 
           }
   \begin{tabularx}{1.0\columnwidth}{ p{1.5cm}  p{1.25cm} p{2cm}  p{1cm}  p{1.5cm} }
     \hline
     \hline
    Region & $\sigma_{\text{max}}$  & $\delta I$ $[\times 10^{-4}]$ & $\alpha$ [$^{\circ}$] & $\delta$ [$^{\circ}$] \\
     \hline
      A & $36.6$ & $9.0\pm0.3$ & $56.4$ & $-7.5$ \\
      B & $23.0$ & $5.3\pm0.2$ & $118.0$ & $37.5$ \\
      C & $24.4$ & $2.6\pm0.2$ & $199.3$ & $18.1$ \\
      D & $10.3$ & $1.3\pm0.2$ & $5.3$ & $38.9$ \\
     \hline
   \end{tabularx}
   \label{tab:ssa_max_sig}
 \end{table}
\end{center}


\section{Conclusions} \label{sec:conclusion}

The HAWC Observatory has observed significant cosmic-ray anisotropy on both large and small scales
using $123 \times 10^9$ events comprising one of the largest TeV anisotropy data sets to date.
Implementing an energy estimation technique that has been verified using the cosmic-ray Moon shadow \citep{crpaper},
we have achieved an unprecendented energy resolution for measuring the energy-dependence of the anisotropy over eight
\replaced{energy}{analysis} bins.
Additionally, a new maximum-likelihood method was used to recover a minimally-biased estimate of the expected 
intensity of an isotropic signal.

\replaced{Using these methods, the}{The} energy dependence of the large-scale phase and amplitude
is found to be consistent with observations made by other detectors in the Northern Hemisphere.
Similarly, the morphology and relative intensity spectra of the three most significant small-scale regions
of excess are \replaced{similar to}{consistent with} previous measurements made by HAWC \citep{Abeysekara:2014sna} 
and ARGO-YBJ \citep{DiSciascio:2013cia,ARGO-YBJ:2013gya,Bartoli:2015ysa}.
\replaced{Furthermore}{Finally}, due to the increased statistics of the current data set,
it is possible to further constrain the spectrum of Region A from 
\replaced{the other measurements by Milagro and HAWC}{previous Milagro and HAWC measurements}.

\replaced{The ever-growing HAWC data set along with future optimizations of the event selection will} 
{Along with continued optimization of the event selection, the ever-growing HAWC data set will}
\replaced{further increase the ability to provide}{facilitate increasingly} accurate descriptions 
of the anisotropy as a function of energy, 
\added{providing additional insights into the nature of local accelerators and the interstellar environment}.
Furthermore, \replaced{this reconstruction}{the novel techniques used in this study} allow 
for collaboration with other observatories using
data sets consisting of targeted cosmic-ray \replaced{energies}{energy bands}.
Combination of \replaced{our}{HAWC} data with the IceCube cosmic-ray data set is an ongoing effort which
will form a nearly complete map of the \replaced{sky in cosmic rays}{cosmic-ray sky at TeV energies}.
\deleted{ and reduce mixing of angular power between the various angular scales.}

\acknowledgments

We acknowledge the support from: the US National Science 
Foundation (NSF); the US Department of Energy Office 
of High-Energy Physics; the 
Laboratory Directed Research and Development (LDRD)
program of Los Alamos National Laboratory; 
Consejo Nacional de Ciencia y Tecnolog\'{i}a (CONACyT),
M\'{e}xico (grants 271051, 232656, 260378, 179588, 239762,
254964, 271737, 258865, 243290, 132197, 281653), 
Laboratorio Nacional HAWC de rayos gamma; L`OREAL
Fellowship for Women in Science 2014; 
Belgian American Educational Foundational Fellowship;
Wallonie-Bruxelles International;
Red HAWC,
M\'{e}xico; DGAPA-UNAM (grants IG100317, IN111315,
IN111716-3, IA102715, 109916, IA102917); VIEP-
BUAP; PIFI 2012, 2013, PROFOCIE 2014, 2015; the
University of Wisconsin Alumni Research Foundation;
the Institute of Geophysics, Planetary Physics, and 
Signatures at Los Alamos National Laboratory; Polish
Science Centre grant DEC-2014/13/B/ST9/945; 
Coordinaci\'{o}n de la Investigac\'{i}on Cient\'{i}fica 
de la Universidad Michoacana.
We thank Markus Ahlers for help with the implementation of the maximum likelihood method.
We gratefully acknowledge Scott DeLay for his dedicated 
efforts in the construction and maintenance of the HAWC experiment. 
Additional thanks to Luciano D\'{i}az and Eduardo 
Murrieta for technical support.

\added{\software{HEALPix \citep{Gorski:2004by}}}
\explain{Added software citation for HEALPix, per lead editor's request.}

\bibliography{lsa}               

\end{document}

%% file: authors.tex
\author{A.U.~Abeysekara}\affiliation{Department of Physics and Astronomy, University of Utah, Salt Lake City, UT, USA}
\author{R.~Alfaro}\affiliation{Instituto de F{\'i}sica, Universidad Nacional Aut{\'o}noma de M{\'e}xico, Mexico City, Mexico}
\author{C.~Alvarez}\affiliation{Universidad Aut{\'o}noma de Chiapas, Tuxtla Guti{\'e}rrez, Chiapas, Mexico}
\author{R.~Arceo}\affiliation{Universidad Aut{\'o}noma de Chiapas, Tuxtla Guti{\'e}rrez, Chiapas, Mexico}
\author{J.C.~Arteaga-Vel{\'a}zquez}\affiliation{Universidad Michoacana de San Nicol{\'a}s de Hidalgo, Morelia, Mexico}
\author{D.~Avila Rojas}\affiliation{Instituto de F{\'i}sica, Universidad Nacional Aut{\'o}noma de M{\'e}xico, Mexico City, Mexico}
\author{H.A.~Ayala Solares}\affiliation{Department of Physics, Pennsylvania State University, University Park, PA, USA}
\author{A.~Becerril}\affiliation{Instituto de F{\'i}sica, Universidad Nacional Aut{\'o}noma de M{\'e}xico, Mexico City, Mexico}
\author{E.~Belmont-Moreno}\affiliation{Instituto de F{\'i}sica, Universidad Nacional Aut{\'o}noma de M{\'e}xico, Mexico City, Mexico}
\author{S.Y.~BenZvi}\affiliation{Department of Physics \& Astronomy, University of Rochester, Rochester, NY, USA}
% ...
